%% file: main-bn_arxiv.tex
\documentclass[fleqn,usenatbib]{mnras}
\usepackage{newtxtext,newtxmath}
\usepackage[T1]{fontenc}
\DeclareRobustCommand{\VAN}[3]{#2} 
\let\VANthebibliography\thebibliography
\def\thebibliography{\DeclareRobustCommand{\VAN}[3]{##3}\VANthebibliography}
\usepackage{graphicx}	
\usepackage{amsmath}	
\usepackage{xspace}
\usepackage[dvipsnames, table]{xcolor}
\input{defs}

\begin{document}
\large
\input{body.tex}
\bibliographystyle{mnras}
\bibliography{bn.bib}
\input{appendix.tex}
\bsp \label{lastpage} 
\end{document}

%% file: defs.tex
\def\rf{\par\vspace{1pt plus 1pt minus 1pt}\noindent \hangindent 13pt}
\newcommand{\equ}[1]{eq.~(\ref{eq:#1})}

\newcommand{\se}[1]{\S \ref{sec:#1}}
\newcommand{\fig}[1]{Fig.~\ref{fig:#1}}

\newcommand{\Fig}[1]{Figure~\ref{fig:#1}}

\newcommand{\tab}[1]{Table~\ref{tab:#1}}
\newcommand{\tabAppOne}[1]{Table~\ref{tab_app:#1}}
\newcommand{\tabApp}[1]{Table~\ref{tab_app:#1}}
\newcommand{\seApp}[1]{\S \ref{sec_app:#1}}
\newcommand{\figApp}[1]{Fig.~\ref{fig_app:#1}}
\newcommand{\FigApp}[1]{Figure~\ref{fig_app:#1}}

\def\ul{\underbar}
\newcommand{\be}{\begin{equation}}
\newcommand{\ee}{\end{equation}}
\newcommand{\ba}{\begin{align}}
\newcommand{\ea}{\end{align}}

\def\la{\langle}

\newcommand{\no}{\noindent}

\newcommand{\msun}{\,\rm M_\odot}
\newcommand{\Msun}{\,\rm M_\odot}

\newcommand{\ifm}[1]{\relax\ifmmode#1\else$\mathsurround=0pt #1$\fi}
\newcommand{\kms}{\ifmmode\,{\rm km}\,{\rm s}^{-1}\else km$\,$s$^{-1}$\fi}

\newcommand{\cMpc}{\,{\rm cMpc}}
\newcommand{\Mpc}{\,{\rm Mpc}}
\newcommand{\kpc}{\,{\rm kpc}}
\newcommand{\pc}{\,{\rm pc}}
\newcommand{\Gyr}{\,{\rm Gyr}}

\newcommand{\Myr}{\,{\rm Myr}}

\newcommand{\ergs}{\,{\rm erg}\,{\rm s}^{-1}}

\newcommand{\cmc}{\,{\rm cm}^{-3}}
\newcommand{\cms}{\,{\rm cm}^{-2}}

\newcommand{\ssim}{\!\sim\!}
\newcommand{\simi}{\sim\!}
\newcommand{\ssimeq}{\!\simeq\!}
\newcommand{\simeqi}{\simeq\!}
\newcommand{\seq}{\!=\!}
\newcommand{\sgt}{\!>\!}
\newcommand{\slt}{\!<\!}
\newcommand{\sgeq}{\!\geq\!}
\newcommand{\sleq}{\!\leq\!}
\newcommand{\sdash}{\!-\!}
\newcommand{\ltsima}{$\; \buildrel < \over \sim \;$}
\newcommand{\lsim}{\lower.5ex\hbox{\ltsima}}
\newcommand{\gtsima}{$\; \buildrel > \over \sim \;$}
\newcommand{\gsim}{\lower.5ex\hbox{\gtsima}}
\newcommand{\sprop}{\!\propto\!}
\newcommand{\dd}{\rm d}

\newcommand{\const}{\rm const.}

\newcommand{\BNs}{BN$\star$\xspace}

\def\sy{\,M_\odot\, {\rm yr}^{-1}}

\def\omm{\Omega_{\rm m}}
\def\oml{\Omega_{\Lambda}}

\def\Mv{M_{\rm v}}

\def\Rv{R_{\rm v}}
\def\Vv{V_{\rm v}}
\def\Tv{T_{\rm v}}

\def\Mg{M_{\rm g}}
\def\Ms{M_{\rm s}}
\def\M1s{M_{\rm s1}}

\def\Re{R_{\rm e}}
\def\Resf{R_{\rm e, SFR}}
\def\Ve{V_{\rm e}}

\def\fg{f_{\rm g}}

\def\Vrot{V_{\rm rot}}
\def\vrot{V_{\rm rot}}
\def\vcirc{V_{\rm circ}}
\def\Vcirc{V_{\rm circ}}

\def\slos{\sigma_{\rm los}}
\def\sigr{\sigma_{\rm r}}

\def\thub{t_{\rm hub}}

\def\brho{\bar\rho}

\def\bmu{\mu}

\def\vela{{\sc vela}\xspace}
\def\velaC{{\sc vela3}\xspace}
\def\velaF{{\sc vela6}\xspace}
\def\nh{{\sc NewHorizon}\xspace}
\def\nihao{{\sc nihao}\xspace}

\def\Sig1{\Sigma_{1\kpc}}
\def\Sige{\Sigma_{\rm e}}
\def\abn{a_{\rm BN}}
\def\aabn{a/a_{\rm BN}}
\def\Mbn{M_{\rm BN}}
\def\abng{a_{\rm BNg}}
\def\abns{a_{\rm BN\star}}

\def\Mdm{M_{\rm DM}}

\def\DMS{\Delta_{\rm MS}}
\def\tdep{t_{\rm dep}}
\def\fgs{f_{\rm gs}}
\def\rd{r_{\rm d}}
\def\Mcrit{M_{\rm crit}}
\def\Mbn{M_{\rm BN}}

\def\sigg{\sigma_{\rm g}}
\def\sigs{\sigma_{\rm s}}
\def\Sigg{\Sigma_{\rm g}}
\def\Sigs{\Sigma_{\rm s}}
\def\Sigsf{\Sigma_{\rm SFR}}
\def\Sg1{\Sigma_{{\rm g},1\kpc}}
\def\Ss1{\Sigma_{{\rm s},1\kpc}}
\def\vos{v_{\rm rot}/\sigma}
\def\vosg{v/\sigma_{\rm g}}

\def\lambdag{\lambda_{\rm g}}
\def\alphav{\alpha_{\rm v}}
\def\alpharho{\alpha_{\rho}}

\def\fdm{f_{\rm DM}}
\def\Zg{Z_{\rm g}}
\def\rp{\vert r_{\rm P}\vert}

\def\sersic{S\'{e}rsic }

\colorlet{TabCol}{SkyBlue!30!}

%% file: body.tex

\title[Blue Nuggets]
{Wet Compaction to a Blue Nugget: a Critical Phase in Galaxy Evolution}

\author[S. Lapiner et al.]
{\parbox[t]{\textwidth}
{Sharon Lapiner$^{1}$\thanks{Email: sharon.lapiner@mail.huji.ac.il},
Avishai Dekel$^{1,2}$,
Jonathan Freundlich$^{3}$,
Omri Ginzburg$^{1}$,
Fangzhou Jiang$^{4,5}$,
Michael Kretschmer$^{6}$,
Sandro Tacchella$^{7,8}$,
Daniel Ceverino$^{9,10}$,
Joel Primack$^{2,11}$
}
\\ \\
$^{1}$Center for Astrophysics and Planetary Science, Racah Institute of Physics,
The Hebrew University, Jerusalem 91904, Israel\\
$^{2}$SCIPP, University of California, Santa Cruz, CA 95064, USA\\ 
$^{3}$Universit\'e de Strasbourg, CNRS UMR 7550, Observatoire astronomique de Strasbourg, 67000 Strasbourg, France\\
$^{4}$Carnegie Observatories, 813 Santa Barbara Street, Pasadena, CA 91101, USA \\
$^{5}$TAPIR, California Institute of Technology, Pasadena, CA 91125, USA\\
$^{6}$Institute for Computational Science, Universität Zürich, Winterthurerstrasse 190, CH-8057 Zürich, Switzerland\\
$^{7}$Kavli Institute for Cosmology, University of Cambridge, Madingley Road, Cambridge, CB3 0HA, UK\\
$^{8}$Cavendish Laboratory, University of Cambridge, 19 JJ Thomson Avenue, Cambridge, CB3 0HE, UK\\
$^{9}$Departamento de Fisica Teorica, Facultad de Ciencias, Universidad Autonoma
de Madrid, Cantoblanco, 28049 Madrid, Spain\\
$^{10}$CIAFF, Facultad de Ciencias, Universidad Autonoma de Madrid, 28049 Madrid, 
Spain\\ 
$^{11}$Department of Physics, University of California, Santa Cruz, CA 95064, USA
}
\pagerange{\pageref{firstpage}--\pageref{lastpage}} \pubyear{2023}
\maketitle
\label{firstpage}
\begin{abstract}
We utilize high-resolution cosmological simulations to reveal that high-redshift galaxies tend to undergo a robust `wet compaction' event when near a `golden' stellar mass of $\simi\!10^{10}\msun$. This is a gaseous shrinkage to a compact star-forming phase, a `blue nugget' (BN), followed by central quenching of star formation to a compact passive stellar bulge, a `red nugget' (RN), and a buildup of an extended gaseous disc and ring. Such nuggets are observed at cosmic noon and seed today's early-type galaxies. The compaction is triggered by a drastic loss of angular momentum due to, e.g., wet mergers, counter-rotating cold streams, or violent disc instability. The BN phase marks drastic transitions in the galaxy structural, compositional and kinematic properties. The transitions are from star-forming to quenched inside-out, from diffuse to compact with an extended disc-ring and a stellar envelope, from dark matter to baryon central dominance, from prolate to oblate stellar shape, from pressure to rotation support, from low to high metallicity, and from supernova to AGN feedback. The central black hole growth, first suppressed by supernova feedback when below the golden mass, is boosted by the compaction, and the black hole keeps growing once the halo is massive enough to lock in the supernova ejecta. 
\end{abstract}
\begin{keywords}
{
galaxies: evolution --
galaxies: formation --
galaxies: high-redshift --
galaxies: haloes --
galaxies: starburst --
galaxies: interactions. 
}
\end{keywords}
\section{Introduction}
\label{sec:intro}

\begin{figure*} 
\includegraphics[width=0.95\textwidth]
{"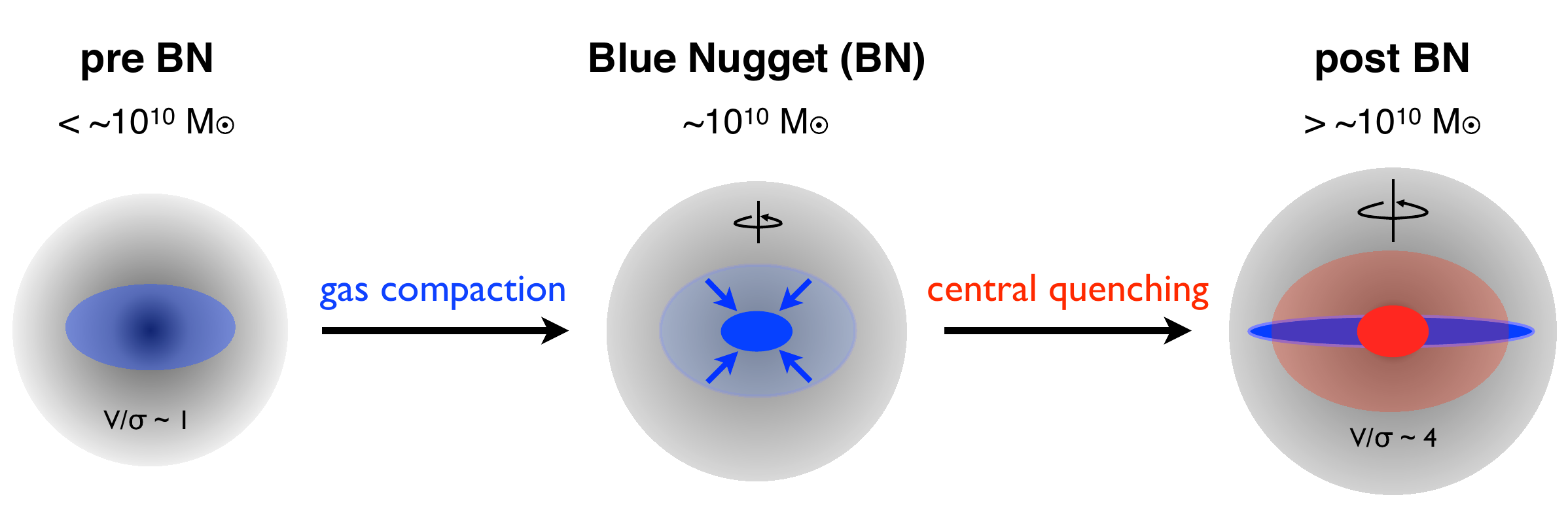"}
\caption{
A cartoon illustrating the phases with respect to the BN at $\Ms\ssim10^{10}\msun$. The pre-BN system is elongated, diffuse and perturbed, only marginally supported by rotation and with a DM-dominated core. Gas compaction leads to a compact star-forming BN, which is baryon dominated. Post-BN, the gas is depleted from the centre leaving behind a compact passive nugget, while a new extended rotation-supported gas disc/ring and/or a stellar envelope develop.
}
\label{fig:cartoon}
\end{figure*}

About half the massive galaxies at $z\ssim2$ are observed to be compact and already quiescent, thus dubbed `{\it red nuggets}' \citep[hereafter RNs,][]{trujillo07, vandokkum08, buitrago08, damjanov09, newman10, vandokkum10,damjanov11, whitaker12, bruce12, vandokkum14, vandokkum15}. While star-forming discs of $\simi10^{10-11}\msun$ can extend to several kpc \citep{genzel06,genzel08}, the typical quenched systems of a similar mass have smaller effective radii of order $1\kpc$. Most of these RNs are likely to be the progenitors of the central regions of today's early-type galaxies, and some may remain as fossil naked RNs \citep{saulder15,yildirim17}.

\smallskip
What is the origin of the compactness of the RNs? It may partly reflect the high density of the Universe at early times when the RN had formed \citep[e.g.,][]{lilly16}. This is provided that the RNs formed, quenched and established their compactness much earlier than the time when they were observed. If angular momentum is conserved during the gas contraction from the dark matter (DM) halo virial radius $\Rv$ to the central galaxy of an effective, half-mass radius $\Re$, then one expects $\Re\ssim\lambda \Rv$ where $\lambda\ssim0.04$ is the halo spin parameter that is expected to be drawn from the same distribution at all times and masses \citep[e.g.,][]{white84,bullock01_j}. The virial radius of the halo of a given galaxy evolves on average as $\Rv\sprop \Mv^{1/3} (1+z)^{-1}$, where the virial mass grows in time as $\Mv \sprop \exp(-\alpha z)$ with $\alpha \!\lsim\! 1$ slowly varying with mass \citep[][where it is derived analytically in the EdS regime at $z\sgt1$]{wechsler02,dekel13}. Thus, if angular momentum is conserved for the gas inflowing to the galaxy, the average effective radius of an evolving galaxy is expected to {grow as} $\Re\ssim0.04\Rv$, namely $\Re \sprop \exp(-\alpha z/3)\, (1+z)^{-1}$.

\smallskip
We refer to any further dissipative contraction, which must be associated with drastic angular-momentum (AM) loss, as {\it `wet compaction'} \citep{db14,zolotov15}. It leads to typical sizes for the forming RNs that are smaller than $0.04 \Rv$ and predicts the existence of {\it `blue nuggets'} (hereafter BNs\footnote{Most of the BNs may actually be red, by dust. They typically have lower U-V colours but higher V-J colours with respect to the RNs.}), namely compact, gas-rich, star-forming systems that are the immediate progenitors of the young RNs. Wet compaction is naturally expected at high $z$, where the accretion-merger rate is high {\citep{neistein08_m,dekel13}} and where the gas fraction is high {\citep{daddi10_gas, tacconi10, tacconi13, tacconi18, tacconi20, freundlich19}}. These permit frequent events of angular-momentum loss \citep{danovich15} and allow the gas to fall in prior to its conversion into stars \citep{db14}.

\smallskip
Indeed, BNs are convincingly observed at $z\ssim2\sdash 3$, with consistent masses, structure, kinematics and abundance for being the immediate progenitors of the RNs \citep{barro13,barro14_bn_rn,barro14_kin,williams14,vandokkum15, williams15,barro16_kin,barro16_alma,barro17_uni}. The observed sizes of the compact galaxies at $z\ssim2$ are on the order of $1\kpc$ \citep{barro16_alma,tadaki16}, which is typically $\simi0.01\Rv$, a factor of four smaller than that predicted without wet compaction associated with AM loss. This is a factor of $\simi64$ in density. We conclude that the BNs must have formed by wet compaction events, thus explaining the compactness of both the BNs and the subsequent RNs.

\smallskip
Our cosmological simulations reveal a generic sequence of events in the evolution of galaxies at moderately high redshift, with dramatic consequences for the galaxy properties, as follows.
\rf
a. A typical high-redshift and low-mass galaxy is a gas-rich, star-forming, highly perturbed and possibly rotating system, fed by intense streams from the cosmic web, including multiple wet mergers which cause frequent spin flips \citep{dekel20_flip}.
\rf
b. When the stellar mass is in the ballpark of $\simi10^{10}\msun$, the galaxy undergoes a major, last, wet compaction into a BN, starting with a compact gaseous star-forming system that rapidly turns into a compact stellar system.
\rf
c. A central gas-depletion process starts immediately {after} compaction, which leads to inside-out quenching into an RN.
\rf
d. Post compaction, there is {sometimes} 
a (partial) rejuvenation of a fresh extended, gas-rich, clumpy disc or ring about the compact stellar system and/or growth of an elliptical envelope by dry mergers.
 
\no 
\Fig{cartoon} illustrates some of the properties of the three stages of evolution, pre-BN, the BN itself, and post-BN.

\smallskip
The BN phase thus marks major transitions in galaxy properties, which are key to understanding {massive} galaxies. These can be translated to observable property dependencies on mass, redshift, position relative to the main sequence of star-forming galaxies, and galactocentric distance within each galaxy. {In this paper, we explore these transitions using cosmological simulations.} \tab{transitions} lists many of the transitions of galaxy properties at the BN phase, to be elaborated on below, and listed with some more detail in \se{conc}.

\begin{table}
\centering
\resizebox{\columnwidth}{!}{%
\begin{tabular}{@{}c|ccc}
\hline
Property & Pre BN & BN & Post BN  
\\
\hline
\hline
$\Ms$       & $<\!\Mbn$ & $\Mbn\ssim10^{10}\msun$ &      $>\!\Mbn$\\
$\Mv$        & $<\!\Mbn$ & $\Mbn\ssim10^{12}\msun$ & $>\!\Mbn$\\
\hline
$\Re$     & slow rise           & contracted $\lsim 1\kpc$  & rapid rise \\
$\Ss1$       & low                 & $>\!10^{9.2}\msun\kpc^{-3}$ & same high \\
$\Sg1$       & low                 & $>\!10^{8.7}\msun\kpc^{-3}$ & very low \\
\hline
sSFR    & MS                  & top of MS           & quenching \\
$\tdep$      & short               & short               & long  \\
$\fgs$       & high                & higher              & low   \\
$\Zg$       & rising              & shoulder            & plateau \\
\hline
centre mass  & dark matter          & mixed               & baryon  \\
shape        & prolate             & compact disc        & oblate  \\
\hline
$V/\sigg$    & $\simi1$            & rising              & $\simi4$ \\
$V/\sigs$    & $\simi1$            & $V,\sigs$ rising    & $\simi1$ \\
Jeans equil.  & crude                &                    & valid \\
spin       & normal              & low                 & high \\
\hline
$\Sigg(r)$   & diffuse              & compact             & disc/ring\\
$\Sigs(r)$   & diffuse              & compact             & compact\\
sSFR$(r)$    & flat                & declining           & rising \\
$Z(r)$       & shallow decline        &                     & steep decline \\
\hline
BH growth    & slow by SN          & activated, rapid           & rapid{\small$\rightarrow$}self regul. \\
feedback     & supernova           &                     & AGN    \\
\hline\hline
\end{tabular}}
\caption{
Transitions in properties at the BN phase. {The critical BN mass is marked as $M_{\rm BN}$. Galaxy-scale properties are measured within $0.1\Rv$. {\bf Rows:} 
(1) $\Ms$ is the galaxy's stellar mass. 
(2) $\Mv$ is the virial mass. 
(3) $\Re$ is the 3D stellar half mass radius. 
(4) $\Ss1$ is the surface density of stars within $1\kpc$. 
(5) $\Sg1$ is the surface density of gas within $1\kpc$. 
(6) sSFR is the specific SFR. 
(7) $\tdep$ is the depletion time ($\Mg/SFR$ assuming that the SFR remains constant after the time of measurement). 
(8) $\fgs$ is the gas fraction ($\Mg/\Ms$). 
(9) $\Zg$ is the gas-phase metallicity.
(10) `centre mass' represents the dominant component (dark matter or baryons), see $\fdm(r<\Re)$. 
(11) `shape' is the shape of the stellar component within $\Re$. 
(12) $V/\sigg$ is the gas rotational velocity over velocity dispersion. 
(13) $V/\sigs$ is the stellar rotational velocity over velocity dispersion. 
(14) `Jeans equil.' represents whether or not Jeans equilibrium is valid in each phase. 
(15) `spin' is the gas spin $\lambdag= (J_{\rm g}/M_{\rm g})/(\sqrt{2}\Vv\Rv)$. 
(16) $\Sigg(r)$ is the surface density of the gas as a function of radius. 
(17) $\Sigs(r)$ is the surface density of the stars as a function of radius. 
(18) sSFR$(r)$ is the specific SFR as a function of radius. 
(19) $\Zg(r)$ is the gas-phase metallicity as a function of radius. 
(20) `BH growth' represents the mode of BH mass growth (suppressed or active accretion). 
(21) `feedback' represents the most effective feedback in each phase.}
}
\label{tab:transitions}
\end{table}


\smallskip
As pointed out in \citet{zolotov15} and demonstrated in \citet{tacchella16_ms}, the evolution of a galaxy through episodes of compaction and quenching is associated with a characteristic evolution with respect to the universal main sequence (MS) of star-forming galaxies (SFGs), as defined in the plane of specific star-formation rate (sSFR) versus stellar mass ($\Ms$). Early oscillations about the MS ridge eventually bring the galaxy to a BN phase in the upper region of the MS (high sSFR), which triggers quenching of {star-formation rate (SFR)} towards the lower parts of the MS and down to the quenched regime, typically once more massive than $\Ms \ssim 10^{10}\msun$. This translates the transitions of galaxy properties through the phases of evolution to observable gradients along and across the MS as a function of $\Ms$ and sSFR, respectively. 

\smallskip
The simulations reproduce the observed characteristic mass scale for galaxies at a stellar mass of $\Ms\ssim10^{10-10.5}\msun$, or a dark matter halo mass of $\simi10^{11.5-12}\msun$, roughly the same mass at all times. {This mass scale is known to mark a peak in the efficiency of galaxy formation, as seen from the stellar-to-halo mass ratio \citep{moster10, moster13, behroozi13, rodriguez17, moster18, behroozi19}. It is imprinted as a peak time in the cosmic star-formation efficiency at $z\ssim1\sdash2$ \citep{madau14} when the typical halo has a comparable mass \citep{press74}, and it marks a bimodality in many galaxy properties \citep[e.g.,][]{db06}.} We learn that the BN phase typically occurs in the ballpark of this mass scale, and the transitions associated with the BN phase are associated with the observed bimodality. {Indications of a similar characteristic mass were also seen in observed galaxies from the HST CANDELS survey using machine learning models \citep{huertas18}, revealing a preference for BNs near a stellar mass of $\simi 10^{10}\msun$. The rough constancy of this critical BN mass in time allows the useful advantage of translating the time evolution of galaxies through the BN phase to an observable mass dependence of galaxy properties. Several physical processes may give rise to this magic scale \citep{dekel19_gold},} in particular the upper limit for effective supernova-driven outflows at a virial velocity (or potential depth) of $\Vv\ssim100\kms$ \citep{ds86}, and the threshold halo mass for virial shock heating (due to slow-enough cooling) of the circum-galactic medium (CGM) \citep{bd03,db06}. We will discuss how the processes responsible for this magic mass scale can combine to define the BN scale and a threshold for rapid black hole (BH) growth in the galaxy centre \citep{lapiner21} and, thus, an onset for AGN feedback, which in turn, helps quench the galaxy above this mass scale. We will discuss the mechanisms for the quenching process following the BN event, those triggered by the BN and maintained by the hot CGM and possibly AGN feedback. {However, they deserve a detailed study using simulations incorporating black holes.}

\smallskip
{Several mechanisms could trigger the AM loss, which leads to a wet compaction event.} These include major mergers, multiple minor mergers, counter-rotating streams, galactic fountains, violent disc instability, and more. Compaction is thus a generic process whose origin could be in a major merger but not necessarily so. We describe below preliminary results concerning the possible origins of compaction.

\smallskip
{Our purpose in this paper is to present the characteristic details of the transitions in galaxy properties due to the wet compaction events based on the \vela cosmological simulations. We will seek their origin and summarize observable predictions.}

\smallskip
The paper is organized as follows.
In \se{sims}, we present the simulations and discuss their limitations.
In \se{phases}, we describe the phases of evolution as seen in the simulations, the way we identify the BN phase, and the associated characteristic mass.
In \se{MS}, we connect the phases of evolution through the BN phase with the evolution of galaxies along and across the main sequence of SFGs and relate to observable gradients across the MS.
In \se{radius}, we refer to the evolution of effective radii of galaxies.
In \se{shape}, we address the transition in the overall shape of the galaxy as a result of the transition from dark matter to baryon central domination.
In \se{kinematics}, we describe the transition in the kinematic properties of the galaxy and the validity of Jeans equilibrium for estimating dynamical masses.
In \se{metallicity}, we study the evolution of metallicity through the BN phase.
In \se{profiles}, we summarize the evolution of profiles within the galaxy, including the DM fraction.
In \se{SN_BH}, we address the interplay between supernova feedback, black hole growth and the trigger of rapid black hole growth and AGN activity by compaction at the critical mass.
In \se{trigger}, we address the origins of compaction.
In \se{abundance}, we estimate the abundance of BNs.
In \se{disc}, we refer to results from other simulations.
In \se{conc}, we summarize our results and conclusions.

\smallskip
In several sections, following the presentation of theory results, we refer to certain relevant observational results, but with {little elaboration and no attempt to be complete.} Detailed comparisons with observations are beyond the scope of this paper and are deferred to other papers by us and others. {The same is true for comparisons with other simulations, which are discussed only briefly below.}

\section{Simulations}
\label{sec:sims}

\subsection{Method of Simulation and Analysis}

We use zoom-in hydro-cosmological simulations of 32 moderately massive galaxies that comprise the \vela simulation suite \citep{ceverino14,zolotov15}. This is the generation-3 suite, in which radiation-pressure feedback has been added to earlier versions which were analysed in previous work. We have used this suite to study a variety of issues related to high-redshift galaxy evolution, including violent disc instability and clumpiness \citep{inoue16,mandelker17}, and the issues of compaction and quenching most relevant to the current paper \citep{zolotov15,ceverino15_shape,tomassetti16,tacchella16_ms,tacchella16_prof}. We provide here a brief overview of the key aspects of the simulations, while more details are provided in \seApp{sims_app} and in \citet{ceverino14}.

\smallskip
The \vela simulations utilize the {\sc adaptive refinement tree (art)} code {\citep{krav97,ceverino09}}, which accurately follows the evolution of a gravitating $N$-body system and the Eulerian gas dynamics, with an AMR maximum resolution of $\simi25\pc$ in physical units at all times. The dark matter particle mass is $8.3\times 10^4 \msun$, and the stellar particles have a minimum mass of $10^3 \msun$. This resolution is higher than in most current cosmological simulations, allowing a proper resolution of the central, compact $\simi1\kpc$ central regions of galaxies at $z\ssim 1\sdash5$, which are the subject of the current investigation.

\smallskip
Beyond gravity and hydrodynamics, the code incorporates the physics of gas and metal cooling, UV-background photoionization, stochastic star formation, gas recycling, stellar winds and metal enrichment. The main feedback processes are thermal supernova feedback \citep{ceverino10,ceverino12} and radiative pressure from massive stars \citep{ceverino14}. AGN feedback is not included in the current simulations. The overall strength of the feedback in these simulations is on the moderate-low side compared to several other cosmological simulations that adopt stronger supernova feedback \citep[e.g.,][]{wang15,dubois15} or stronger radiative feedback \citep[e.g.,][]{hopkins14_fire}. 

\smallskip
The initial conditions for the simulations are based on dark matter haloes drawn from dissipationless $N$-body simulations at lower resolution in larger cosmological boxes. We assume the standard $\Lambda$CDM cosmology with the WMAP5 cosmological parameters, namely $\Omega_{\rm m}\seq 0.27$, $\Omega_{\rm\Lambda}\seq 0.73$, $\Omega_{\rm b}\seq 0.045$, $h\seq 0.7$ and $\sigma_8\seq 0.82$ \citep{WMAP5}. The haloes were randomly selected from the haloes of a desired virial mass at $z\seq 1$ in the range $\Mv\seq 10^{11.3-12.3}\msun$ {uniformly distributed in $\log \Mv$} with a median of $5.6\times10^{11}\msun$. The median halo mass at $z\seq 0$ would be $\gsim\!10^{12}\msun$, including massive galaxies and groups.

\smallskip
More than half the sample evolved to $z\seq 1$ (or later), and all but five evolved to $z\!\le\!2$. The simulation outputs were stored and analysed at times separated by fixed intervals of $\Delta a\seq 0.01$, where $a\seq (1+z)^{-1}$ is the cosmological expansion factor, which at $z\seq 2$ corresponds to timesteps of about $100\Myr$. Galaxies are found inside the haloes using the \texttt{AdaptaHOP} group finder \citep{tweed09,colombi13} on the stellar particles. The central galaxy is identified in the final available output, and its main progenitor is traced back in time until it contains less than 100 stellar particles, typically at $a\ssim0.11$ ($z\ssim8$).

\smallskip
The virial mass $\Mv$ is defined as the total mass within a sphere of radius $\Rv$ that encompasses a given overdensity $\Delta(z)$ relative to the cosmological mean mass density, $\Delta(z)\seq (18\pi^2\sdash 82\oml(z)\sdash 39\oml(z)^2)/\omm(z)$, where $\omm(z)$ and $\oml(z)$ are the cosmological density parameters of mass and cosmological constant at $z$ \citep{bryan98}.

{
\smallskip
\tabAppOne{sample_app} lists halo and galaxy properties at $z\seq 2$, except for the five galaxies (marked by $*$) that were stopped at a higher redshift $z_{\rm fin}$ (specified in the table), for which the properties are quoted at $z_{\rm fin}$. For certain purposes, we divide the sample of galaxies into two sub-samples, a {\it high-mass} sub-sample and a {\it low-mass} sub-sample, according to the galaxy stellar mass at $z\seq 2$ being above and below the median {value, $10^{10.06}\msun$,} of the whole sample at that time. 

\smallskip
The stellar mass $\Ms$ is the instantaneous mass in stars, typically within $0.1\Rv$, accounting for past stellar mass loss. The SFR is obtained from the initial mass in stars younger than $\simi60\Myr$, thus serving as a proxy for H$_\alpha$-based SFR measurements (while UV-based measurements are sensitive to stars younger than $\simi100\Myr$). The specific SFR is ${\rm sSFR}\seq {\rm SFR}/\Ms$, the gas mass $\Mg$ is measured within the same volume, and the gas fraction is $\fg\seq \Mg/(\Mg\!+\!\Ms)$. 
}

\subsection{Limitations of the Current Simulations}
\label{sec:limitations}

The cosmological simulations used in this paper are state-of-the-art in terms of high-resolution AMR hydrodynamics and the treatment of key physical processes at the subgrid level, highlighted above. These simulations successfully trace the cosmological streams that feed galaxies at high redshift, including mergers and smooth flows \citep[e.g.][]{danovich12,danovich15,goerdt15,ceverino16_drops}. They properly resolve the violent disc instability that governs high-$z$ disc evolution and bulge formation \citep{mandelker14,mandelker17}, they reproduce galactic outflows \citep{ceverino16_outflow}, and they capture the phenomenon of wet compaction on $1\kpc$ scales and its implications \citep{zolotov15,tacchella16_ms,tacchella16_prof}.

\smallskip
However, like other simulations, they are not perfect in their treatment of star formation and feedback processes. While the SFR recipe was calibrated to reproduce the Kennicutt-Schmidt relation and a realistic SFR efficiency per free-fall time \citep{ceverino14}, the code does not yet follow in detail the formation of molecules and the effect of metallicity on SFR \citep{kd12}. The resolution does not capture the Sedov-Taylor adiabatic phase of supernova feedback, where most of the energy and momentum is deposited in the ISM, which introduces significant uncertainty in the feedback strength \citep{gentry17}. The adopted radiative stellar feedback assumes low infrared trapping, in the spirit of the low trapping advocated by \citet{dk13} based on \citet{krum_thom13}. However, there is an ongoing controversy concerning the level of trapping \citep{murray10} and the counter effect of photon leakage through a porous medium \citep{hopkins11}, which led others to adopt stronger radiative feedback \citep{hopkins12b,hopkins12c,genel12,hopkins14_fire,oklopvcic17}. Finally, AGN feedback and feedback associated with cosmic rays and magnetic fields are not incorporated in the current \vela simulations. 

\smallskip
The resultant star formation rates, gas fractions, and stellar-to-halo mass ratios are in the ballpark of the estimates deduced from observations based on abundance matching {\citep{moster10, moster13, behroozi13, rodriguez17, moster18, behroozi19}}, though with apparent mismatches by factors of order two. These offsets are comparable to the observational uncertainties, which can be estimated from the fact that other measurements, based on the kinematics of $z\ssim0.6\sdash2.8$ galaxies \citep[][{Fig.~7}]{burkert16}\footnote{{We note that these galaxies do not constitute a statistical sample ($\simi 360$), and were selected to be SF discs.}}, reveal larger stellar-to-halo mass ratios for galaxies in the relevant mass range $\Mv\slt 10^{12}\msun$ \citep[see][Fig. 3]{mandelker17}. These uncertainties in accurately matching certain observations may lead to quantitative inaccuracies in the estimated characteristic masses and redshifts associated with certain events. Still, one can assume that these simulations properly capture the qualitative features of these events.

\smallskip
The details of the phenomena addressed in the current paper may depend on the strength of feedback effects and the resultant gas fraction. While one may assume that the \vela simulations qualitatively capture the effects of compaction and the associated transitions in galaxy properties, it will be important to study the effects of feedback on these phenomena in simulations that incorporate the feedback effects in different ways. For this purpose, we briefly refer in \se{disc} to {results from the \nh simulations \citep{dubois20_NH}, which incorporate strong supernova feedback as well as AGN feedback, showing similar compaction events \citep{lapiner21} as seen in this work. In addition, we discuss preliminary results showing signatures of compaction events in the \nihao simulations \citep{wang15} and in the \velaF simulations \citep{ceverino22_vela6}, both of which incorporate stronger supernova feedback.}

\begin{figure*} 
\centering
\includegraphics[width=0.95\textwidth]
{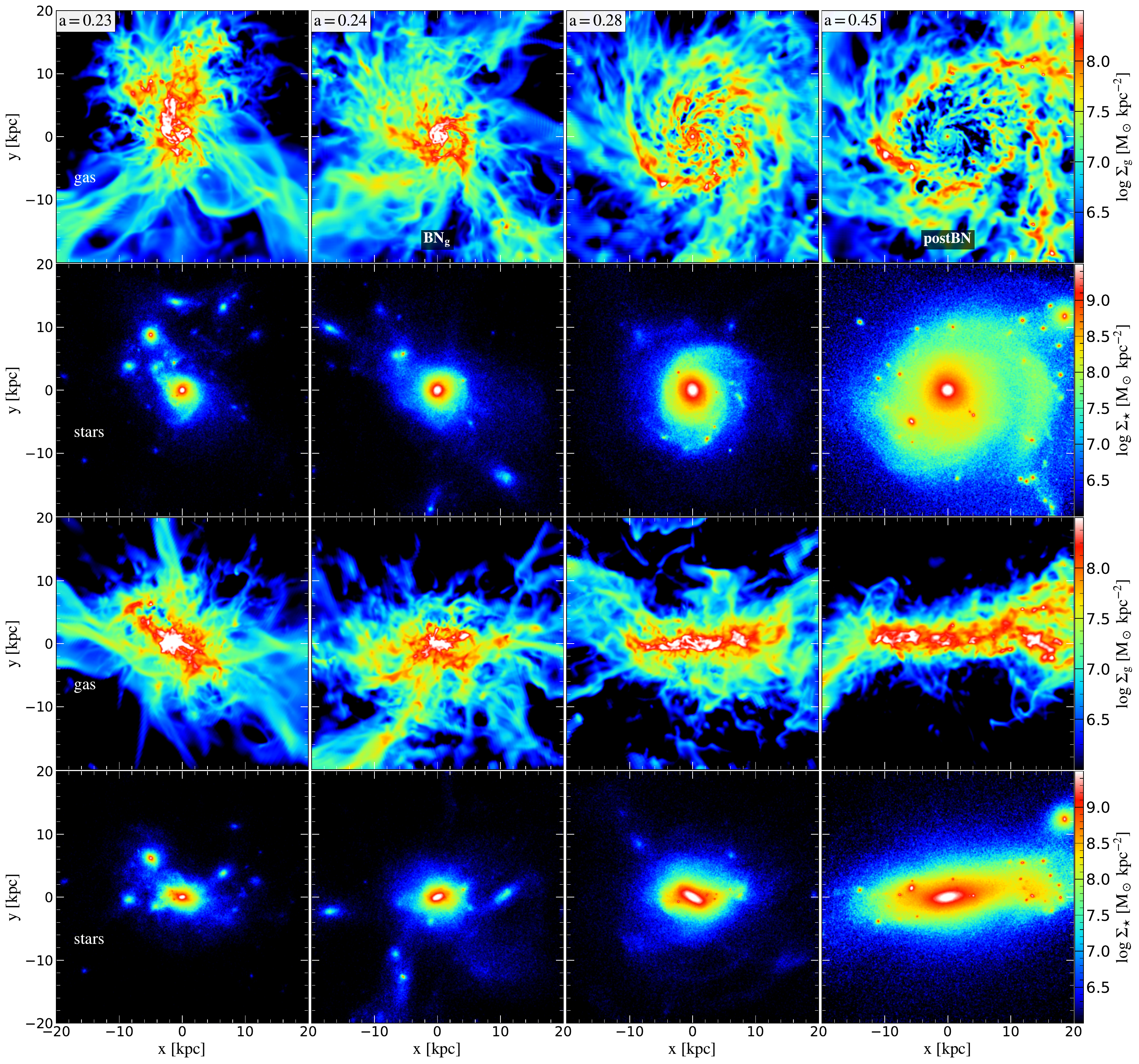}
\caption{
The evolution of V07 through the BN phase in pictures. Shown are the projected densities in {gas} (first and third row) and stars (second and fourth row), face-on (top two rows) and edge-on (bottom two rows). Gas compaction forms a compact, star-forming BN that leads to a compact stellar core. The latter remains compact while the gas depletes from the core that passively turns into a RN. The newly accreted gas forms an extended star-forming clumpy ring about the RN, and a stellar envelope grows by dry minor mergers. An additional example is shown in \figApp{pics_V12}.
}
\label{fig:pics_V07}
\end{figure*}

\begin{figure*} 
\includegraphics[width=0.8\textwidth]{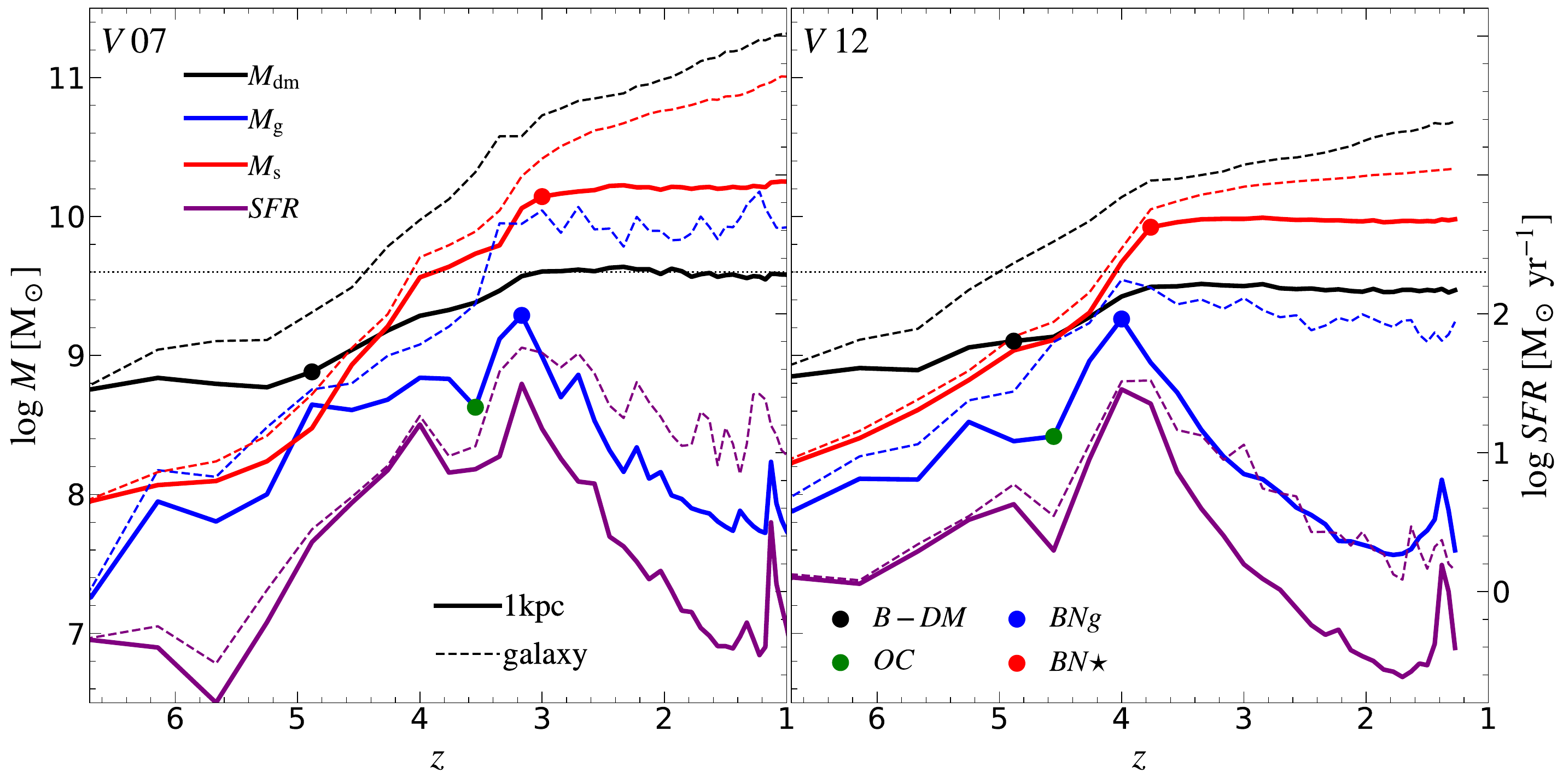}
\caption{
Examples (V07, V12) of the evolution of mass and its rate of change within $1\kpc$ (solid) and for the whole galaxy within $0.1\Rv$ (dashed), as a function of expansion factor $a$. {Shown are the masses of gas (blue), stars (red), dark matter (black) and SFR (purple).} The BN is identified by the sharp peak in gas mass {(or SFR)} within $1\kpc$ and the subsequent saturation of the compact stellar mass to its high value. The transition from central dominance of dark matter to baryon dominance occurs in association with the BN.
}
\label{fig:M_V07_V12}
\end{figure*}

\begin{figure*} 
\includegraphics[width=0.8\textwidth]{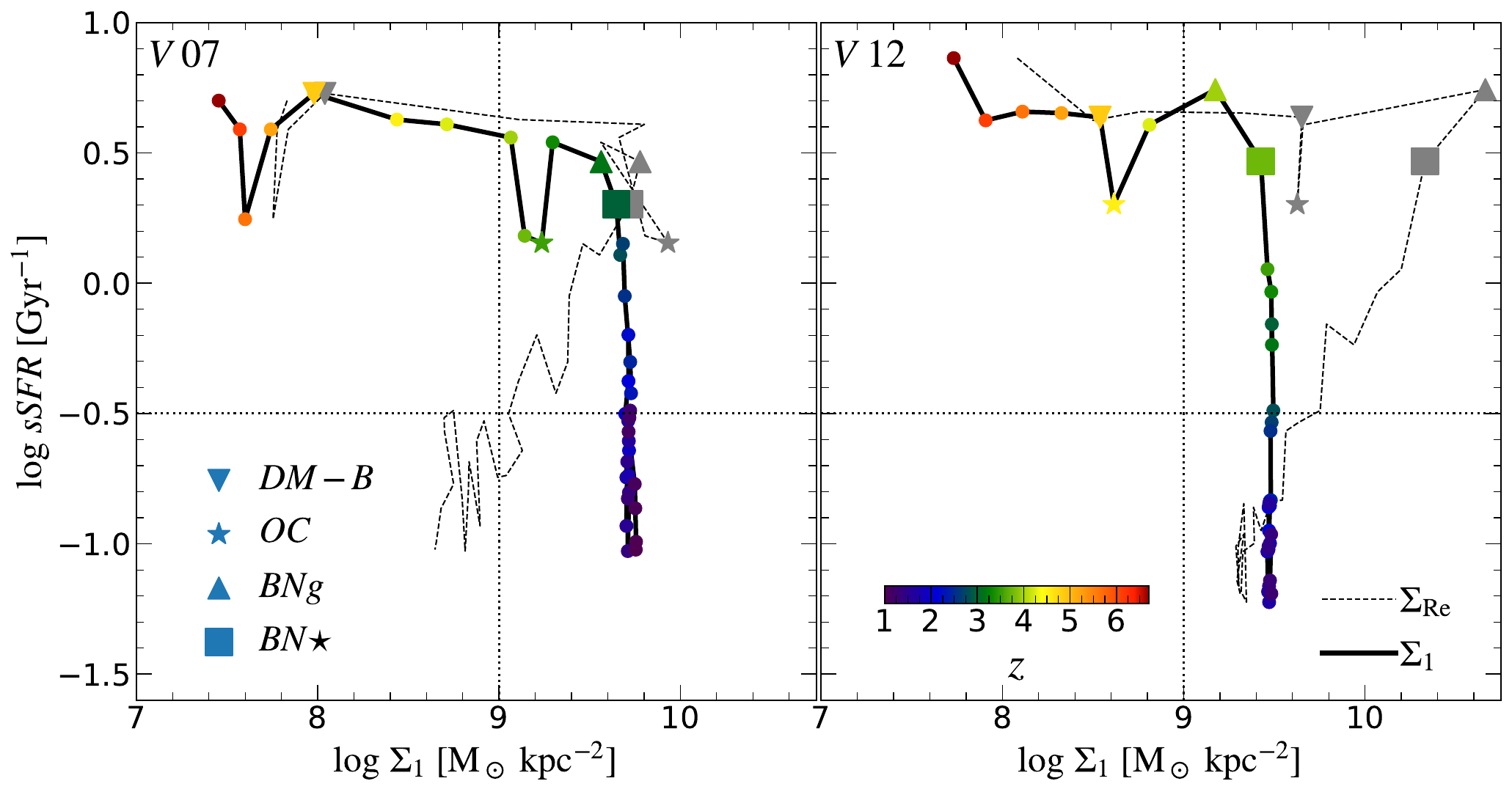}
\caption{
Examples (V07, V12) of evolution tracks in the plane of sSFR and stellar compactness, as measured by $\Sig1$ (solid) or $\Sige$ (dashed). The colour marks redshift. The typical track has an L shape, starting with a compaction phase at a constant sSFR and turning at the \BNs ({at $z\ssim 3, 3.8$}) into quenching at a constant $\Sig1$. 
}
\label{fig:L_V07_V12}
\end{figure*}
%

\section{Phases of Evolution through the BN: Central Masses and SFR}
\label{sec:phases}

The generic pattern of evolution revealed by our simulations, from a diffuse galaxy through compaction to a BN and the following quenching, has been first pointed out in \citet{zolotov15}. Here we use the whole sample of simulations to explore the systematic evolution through a BN phase and demonstrate the associated characteristic mass.

\subsection{Example Galaxies}

\smallskip
\Fig{pics_V07} shows images of one galaxy, V07, at several stages of evolution. {Shown are gas and stellar mass density, face-on (top) and edge-on (bottom). The first snapshot, at $a\seq 0.23$, is during the gas compaction phase, associated with intense multiple streams and minor mergers. The second snapshot, $a\seq 0.24$, is at the peak of gas density within the inner $1\kpc$, which marks the gas BN phase, BNg. The third and fourth snapshots, at $a\seq 0.28$ and much later at $a\seq 0.45$,} demonstrate central gas depletion and the growth of an extended clumpy gaseous disc, eventually a ring, fed by in-spiralling fresh streams. The spatial distribution of SFR and young stars is similar to that of gas above a certain threshold density (see \figApp{pics2D_gas_vs_sfr}). In terms of the stellar distribution, a compact stellar core emerges soon after the formation of a gas core, to be identified as the stellar BN phase, \BNs. The stellar core remains compact at a rather constant density as it passively evolves into a compact RN. A stellar envelope develops around it in a disc/ring following the gas and in a spheroidal envelope due to dry minor mergers. Images of another galaxy, V12, are shown in \figApp{pics_V12}.

\smallskip
\Fig{M_V07_V12} shows the time evolution of masses and SFR for two example galaxies, V07 and V12. {Shown are the DM, gas, stellar mass and the SFR} as a function of $z$. These quantities refer to either the inner $1\kpc$ (solid) or the whole galaxy out to $0.1\Rv$ (dashed). The four events associated with the BN, as defined within $1\kpc$, are marked: {the peak of gas density (BNg, at $z\ssim 3.2, 4$ respectively), the shoulder of stellar density (\BNs, $z\ssim3, 3.8$), the onset of compaction (OC, $z\ssim3.5, 4.6$) and the transition from DM to baryon dominance (DM-B, $z\ssim 5, 5$).} The compaction phase is seen as a drastic increase in central gas density and SFR, by order of magnitude, from OC to BNg. Immediately following is a gradual central gas depletion associated with the quenching of SFR. The central stellar mass is growing rapidly during the increase in central gas. It continues to grow after the peak of central gas and SFR until it flattens off at \BNs, after which it keeps a rather constant stellar density within the inner $1\kpc$.

\smallskip
The evolution of the same quantities for the whole galaxy shows a more gradual quenching post-BN, associated with the formation of an extended disc or ring from fresh gas. 

\smallskip
\Fig{L_V07_V12} shows the evolution tracks of V07 and V12 in the plane of overall sSFR and an inner measure of compactness, $\Sig1$, the stellar surface density within $1\kpc$ (solid) or $\Sige$ within $\Re$ (dashed). It shows an L-shape evolution pattern, where a phase of compaction at a roughly constant sSFR abruptly turns into a phase of quenching at a roughly constant $\Sig1$. The turning point marks the \BNs event, at which the galaxy spends a rather short time.

\smallskip
The evolution track with respect to $\Re$ is similar pre-BN when $\Re$ is comparable to $1\kpc$. Post-BN, the quenching is associated with a gradual decrease in $\Sige$ due to the increase in $\Re$ with time (\se{radius}).

\smallskip
Certain other galaxies in the simulations show a similar L-shape track which is preceded by several oscillatory episodes of minor compaction and quenching attempts. These oscillations {eventually end with a} more drastic compaction to a major BN phase, followed by fuller quenching, typically once the halo mass exceeds a threshold of $\simi10^{11.5}\msun$ \citep[][Fig. 21]{zolotov15}. The characteristic mass is demonstrated below in \fig{char_M-a_event}.

\subsection{Identifying the BN}
\label{sec:identify}
 
\smallskip
We try to identify one major BN phase for each galaxy, the one that leads to significant central gas depletion and SFR quenching (second column in \tabApp{comp_time_app}, $a_{\rm comp^1: BNg}$). {The most physical way to define it is at the highest peak of central gas density, as identified in \fig{M_V07_V12} using the inner $1\kpc$\footnote{{See \seApp{physical} for the choice of $1\kpc$ in the identification of compaction.}}, as long as a significant, long-term decline follows it in central gas mass and SFR.} We term this event BNg. This event coincides with the peak in SFR, and it would be the observable BN if it is based on peak central SFR. The curve of central gas density also allows us to identify the onset of compaction, termed OC, as the start of the steep rise prior to the BNg.

\smallskip
An alternative for identifying the BN is by the shoulder where the inner stellar density in \fig{M_V07_V12}, or $\Sig1$ in \fig{L_V07_V12}, {reaches its} plateau of maximum long-term compactness, slightly after the BNg. We term this event \BNs and note that it may be closer to the BN phase as identified observationally based on the compactness of the stellar component.

\Fig{M_V07_V12} also reveals another characteristic feature associated with the onset of the BN phase, namely a transition of the core from being dominated by dark matter to being dominated by baryons, mostly stars. We refer to this event as the DM-B transition, to be discussed in \se{DM-B}. Finally, a fifth way to identify the BN phase is by its kinematic characteristics, to be described in \se{kinematics}.

\subsection{Systematic Evolution through the BN}

\begin{figure*} 
\includegraphics[width=0.95\textwidth]{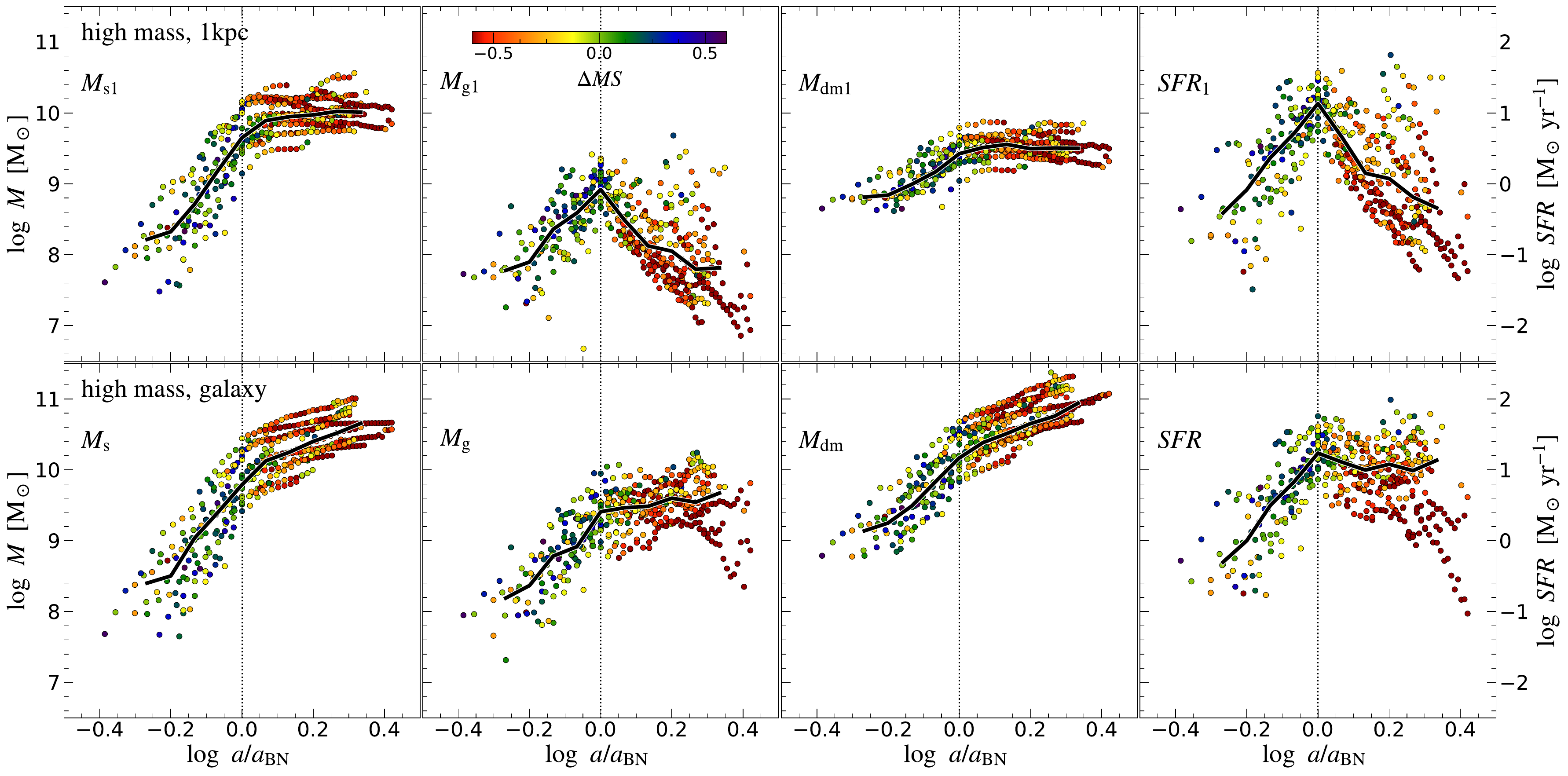}
\caption{
Evolution of global properties. All snapshots for the high-mass subsample are stacked such that the time is with respect to the BN {(BNg), $a/\abng$}. The properties refer to the inner $1\kpc$ (top, marked with a subscript `1') or the whole galaxy within $0.1\Rv$ (bottom). Shown are stellar mass, gas mass, dark matter mass and SFR. Colour marks the deviations of sSFR from the universal main sequence, $\DMS$ {(see \equ{DMS})}. The median in equal log bins is marked (black line). The BN is best identified by a peak in $M_{\rm g1}$ or SFR$_1$, followed by a shoulder in $M_{\rm s1}$. \FigApp{M-a_lm} shows the same for the low-mass subsample.
}
\label{fig:M-a_hm}
\end{figure*}

\begin{figure*} 
\includegraphics[width=0.8\textwidth]{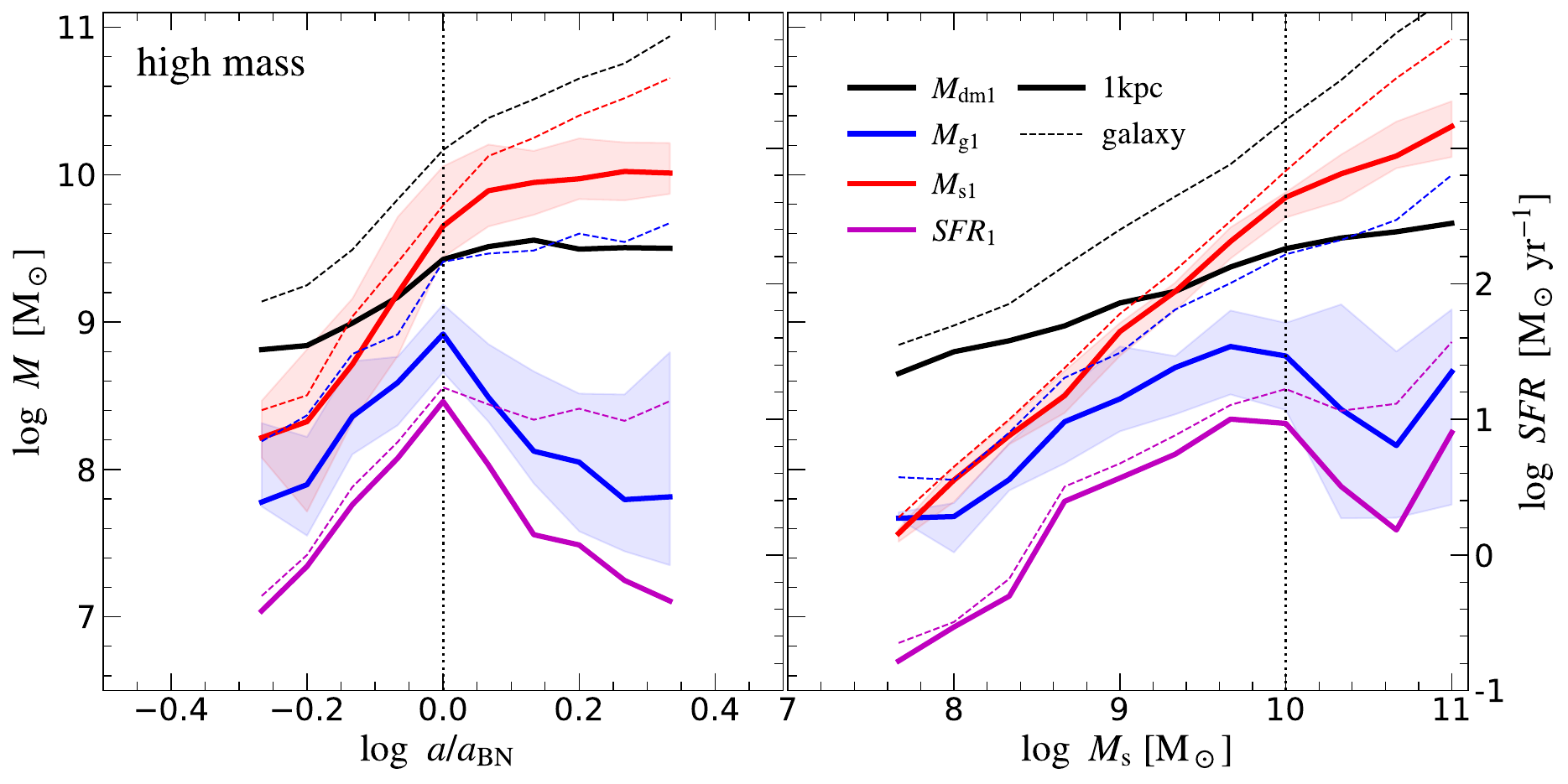}
\caption{
The medians for the properties shown in \fig{M-a_hm} and \figApp{M-Ms_hm} are put together here for the inner $1\kpc$ (solid) and the whole galaxy (dashed). The shaded regions indicate the $1\sigma$ scatter about the median for the gas and stellar mass within $1\kpc$ (see \figApp{M-a_med_lm} for the low-mass subsample). The BN is characterized by a peak in $M_{\rm g1}$ and SFR$_1$ and by a transition from central dark matter dominance to baryon (mostly stellar) dominance. 
}
\label{fig:M-a_med}
\end{figure*}

\begin{figure*} 
\includegraphics[width=0.8\textwidth]{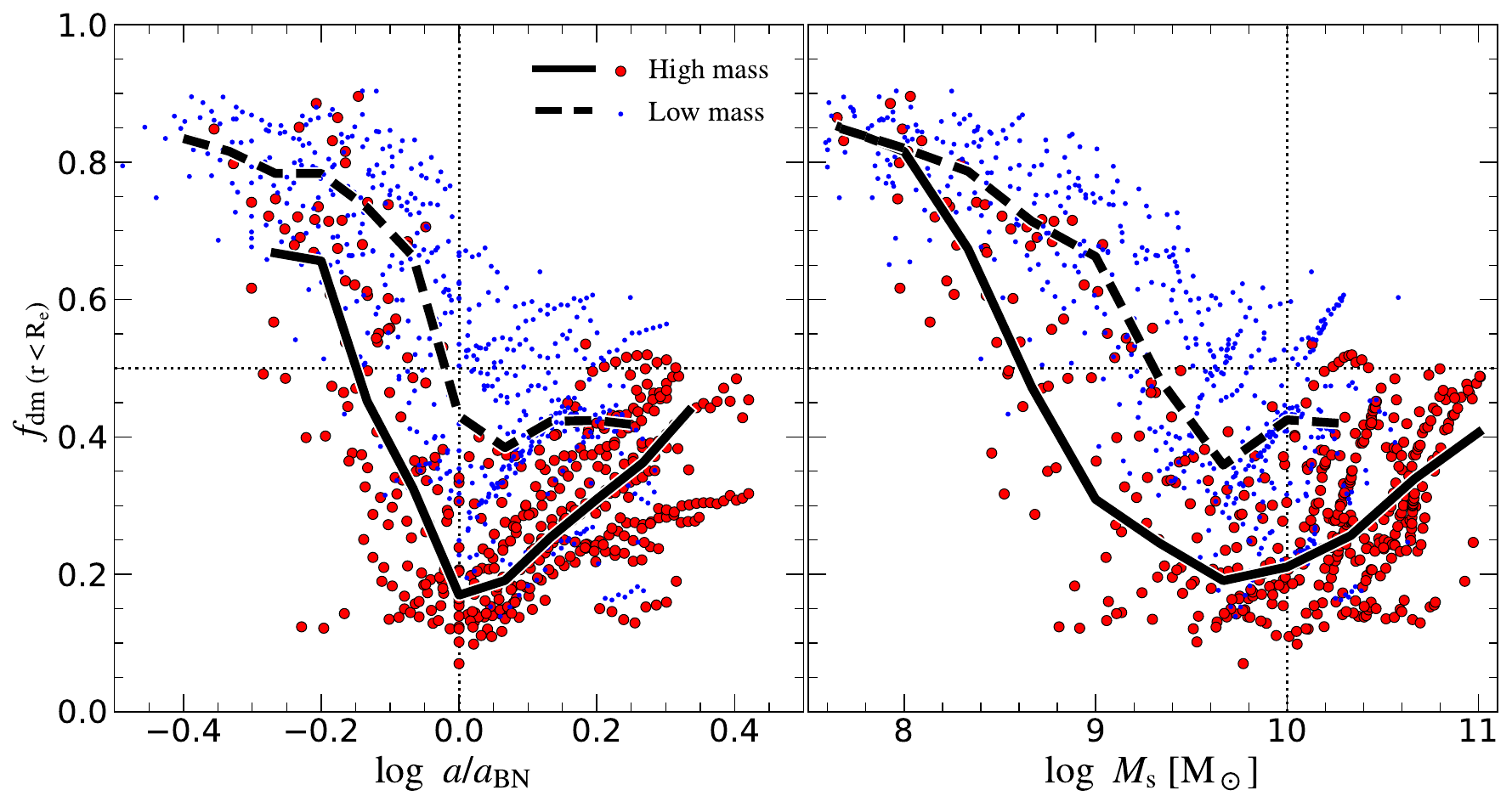}
\caption{
Dark matter fraction within $\Re$ as a function of $a/\abn$ and $\Ms$. {Shown are all snapshots divided at $z\ssim2$ to a high-mass (red, median in solid black line) and a low-mass (blue, dashed line) subsamples. The dark matter fraction is high pre-BN and drops to low values during compaction. Post-BN, $\fdm$ is mildly rising, partly due to an increase in $\Re$. The qualitative similarity between the two panels demonstrates that the evolution with respect to the BN phase (left) can be translated to evolution with respect to the critical mass (right).}
}
\label{fig:fdm_Re}
\end{figure*}

\begin{figure*} 
\includegraphics[width=0.8\textwidth]{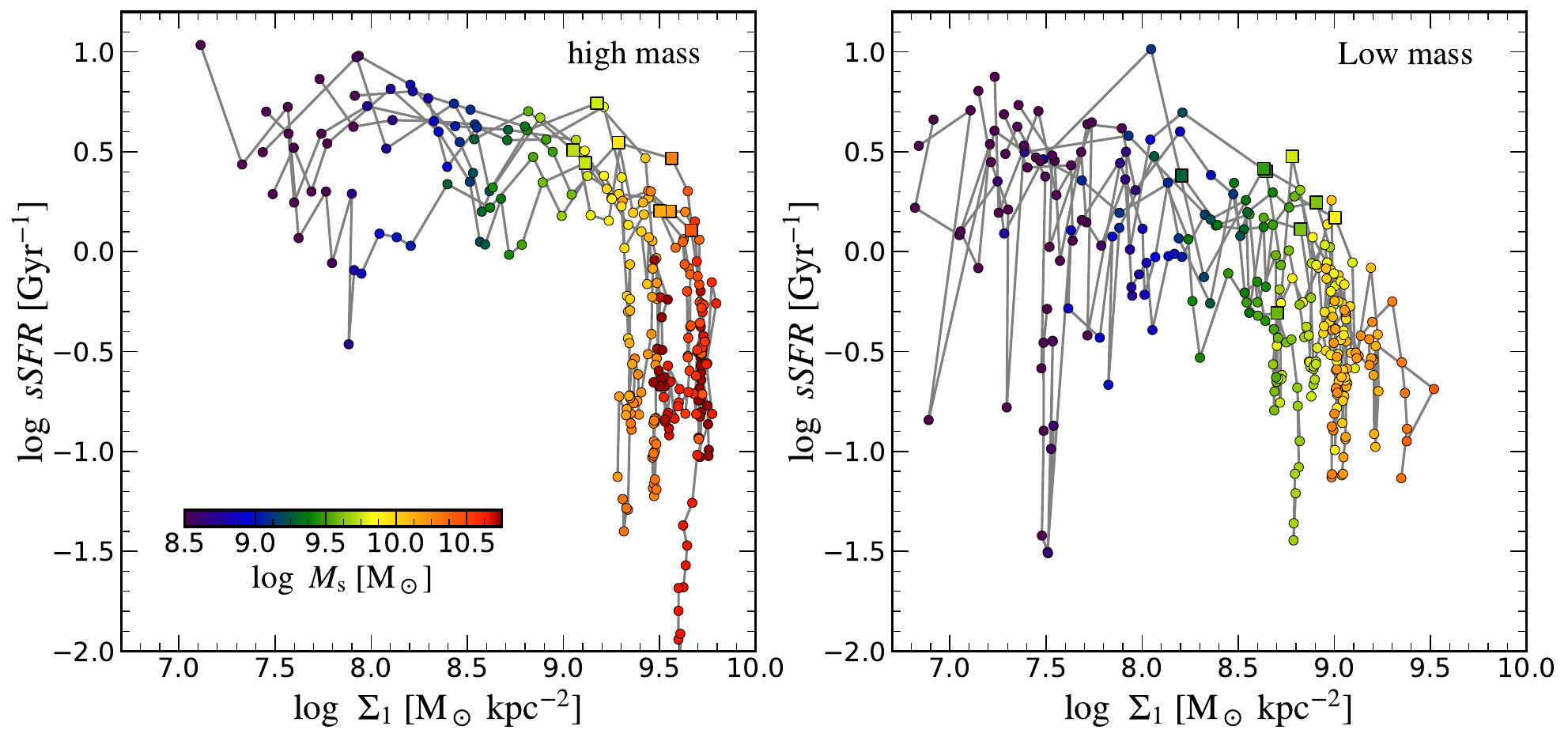}
\caption{
Typical L-shape evolution tracks in sSFR versus $\Sig1$. Stellar mass is indicated by colour, starting from $\Ms \sgt 10^{7.5} \msun$. {\bf Left:} Eight galaxies from our high-mass subsample, where the BN occurs at high redshift. The characteristic track is L-shaped: an early evolution as an SFG at a constant or slowly declining sSFR that changes abruptly at the BN phase (marked by squares) to quenching at a constant $\Sig1$. This resembles the observed universal L-shape pattern in Fig.~9 of \citet{barro17_uni}. {\bf Right:} Eight galaxies from our low-mass subsample, where the BN occurs at a lower redshift. In the pre-BN phase, the oscillations in sSFR are larger, but the transition at the BN into quenching along an L-shape track is still apparent, with a somewhat lower $\Sig1$ during the quenching. { We find an indication for stronger major compaction events in the high-mass subsample compared to the low-mass subsample (see \seApp{physical}).}
}
\label{fig:L-shape}
\end{figure*}

In \fig{M-a_hm} (and \figApp{M-Ms_hm})
we consider {all} the timesteps of the high-mass subsample of galaxies (with stellar mass above the median at $z\seq 2$). The same but for the low-mass subsample is shown in \figApp{M-a_lm}  (and \figApp{M-Ms_lm}). In each figure, the top row refers to the inner sphere of radius $1\kpc$ and the bottom row to the whole galaxy out to $0.1\Rv$. The four panels in each row refer to the quantities stellar mass ($\Ms$), gas mass ($\Mg$), dark matter mass ($\Mdm$) and star-formation rate (SFR). The colour denotes $\DMS$, the sSFR with respect to the universal main sequence ridge (see \se{MS} below). In \fig{M-a_hm}, these quantities are shown versus the cosmological expansion factor $a\seq (1+z)^{-1}$ {and stacked (normalized)} with respect to the {peak of the major compaction event as measured by the central gas density, namely, the BNg event at $\abng$. The curves mark the median in bins of $a/\abng$. Here, and in similar calculations throughout the paper, before measuring the median in bins, we first calculate the median of each galaxy within each bin, thus preventing the possibility that one galaxy with many timesteps in one bin will dominate the median value.} These median curves are displayed together in \fig{M-a_med}.

\smallskip
We see a universal evolution pattern shared by almost all the {higher mass} galaxies with a relatively small scatter about the median. We focus first on the inner $1\kpc$ shown in the top row. The mass within a sphere of $1\kpc$ serves as a proxy for the surface density within a circle of $1\kpc$ {(where we approximate the stellar surface density by $\Sigma(\slt r)\ssim M_{\rm s}(\slt r)/(\pi r^2)$}, e.g., ${\Ms}_{1}$ approximates the observable central stellar surface density $\Sig1$ \citep{zolotov15}, which is commonly used to identify compact BNs and RNs in observations.

\smallskip
A wet compaction phase is easily identified in the gas evolution (second column), where the central gas density rises steeply by an order of magnitude from the onset of compaction at $\simi0.63\,\abng$ up to $\abng$. Following a sharp peak at $\abng$, the central gas density steeply declines immediately thereafter. The central SFR follows the central gas density, reflecting the Kennicutt-Schmidt law, SFR$\,\propto\!\Sigma_{\rm gas}^{1.5}$. There is a sharp peak of SFR at the BN event, immediately followed by a central quenching process. A steep rise in central stellar density follows the rising gas and SFR density during the gas compaction phase, as gas efficiently turns into stars. This rise continues till $\abns\ssim1.35\abng$, where the stellar density sharply flattens to a plateau. The central stellar density remains constant afterwards as the core quenches to a passive RN.

\smallskip
Before \BNs, $\Sig1$ rises at a roughly constant sSFR, as indicated by comparing the SFR to $\Ms$ for the whole galaxy or by the colours of the symbols referring to $\DMS$. Post-\BNs, the sSFR (or $\DMS$) is declining at a rather constant $\Sig1$. This is responsible for the universal L-shape in the sSFR-$\Sig1$ diagram, as seen in \fig{L_V07_V12} {and in \se{L-shape} below}.

\smallskip
Turning to the galaxy as a whole, for the massive subsample, in the bottom row of \fig{M-a_hm}, the BN event is still well defined. The pre-BN evolution is similar to the evolution in the inner $1\kpc$ because the effective radius is comparable to $1\kpc$ and because most of the star formation occurs there during the compaction phase. Post-BN, the total gas mass remains roughly constant; and even rises slowly, reflecting the formation of a new extended disc or ring while obeying the predictions of the bathtub model in a quasi-steady state for a constant gas mass \citep{dm14}. The overall SFR remains roughly constant for a while until $\simi2 \abng$, after which it drops. Therefore, the overall stellar mass keeps rising gradually in the post-BN phase. As a result, the sSFR is declining, eventually driving the galaxy below the MS, which reflects an overall quenching of star formation.

\smallskip
The low-mass sample shown in \figApp{M-a_lm} tends to have the BN event at a similar critical mass (see below), hence at later times, when the gas fraction is lower. As a result, the central gas compaction is slower, it grows only by a factor of $\simi4$ in density, the peak central gas density is lower, and so are the central SFR and the central stellar mass density $\Sig1$ at the BN event. Nevertheless, the total stellar mass at $\abng$ is comparable for the low-mass and massive galaxies.

\smallskip
\FigApp{M-Ms_hm} and \figApp{M-Ms_lm} show the same quantities as in \fig{M-a_hm} and \figApp{M-a_lm} but against the total stellar mass $\Ms$ instead of $a/\abng$, now without any {x-axis} scaling of the individual galaxies. The right panels of \fig{M-a_med} show the corresponding medians in comparison to each other. The stacked evolution tracks are qualitatively similar to the tracks when plotted against $a/\abng$. The peak of central gas density and SFR always occurs near a critical total stellar mass of $\Ms\ssim10^{9.5-10}\msun$, with the sharpness of the peak somewhat smeared. Pre-BN, the rise of the core gas density as a function of $\Ms$, is similar to the rise as a function of $a/\abng$, but the post-BN drop as a function of $\Ms$ is steeper because the growth rate of $\Ms$ is suppressed in this quenching phase. The low-mass galaxies in our sample tend to reach the BN phase toward $z\ssim1$ when they reach the critical mass. This is when many of our simulations stop, which makes it harder to analyse the BN phenomenon in some of these low-mass galaxies.

\subsection[Central Dark Matter to Baryon Dominance]{{Central} Dark Matter to Baryon Dominance}
\label{sec:DM-B}

\smallskip
One of the important transitions due to the compaction process is a transition from central dark matter dominance to baryon dominance. This transition and its association with the BN phase is clearly identified in \fig{M-a_med} {close to the crossing of the dark matter and stellar mass growth curves (black and red curves).} The DM-B transition in the inner $1\kpc$ occurs near the same critical mass that characterizes the BN. For the whole galaxy, within $0.1\Rv$, the dark matter always dominates over the baryons, but only by a factor of $\simi2-3$.

\smallskip
\Fig{fdm_Re} shows the dark matter fraction within $\Re$ as a function of $\aabn$ and as a function of $\Ms$. It demonstrates the sharp transition due to the gas compaction from DM dominance pre-BN, at a level of $\fdm\ssim0.8$, to baryon dominance post-BN, with $\fdm(r\slt \Re)\ssim0.2$ for the high-mass subsample, and some cases as low as $\simi0.1$. 
{During the post-BN phase, the central dark matter fraction is somewhat increasing. This is partly caused by the post-BN increase in the effective radius $\Re$ (see \fig{Re-Ms}). The transition in the central DM fraction is associated with a transition in shape, as discussed in \citet{tomassetti16,ceverino15_shape} and \se{shape}.}{Within $2\Re$ (\figApp{fdm_2Re}) we find a slightly higher $\fdm$, $\simi0.3$ for the high-mass subsample.}

\smallskip
{The profiles of dark matter fraction, $\fdm(r)$, are shown in \fig{profiles_re} at four evolutionary phases: pre-BN, BN, post-BN and a late post-BN quenching phase (see \se{profiles}). We see a significant decrease in the central dark matter fraction during the transition from the pre-BN phase to the peak of compaction at the BN and only mild central evolution in the post-compaction phases.}

\subsection{L Shape}\label{sec:L-shape}

\smallskip
\Fig{L-shape} puts together typical evolution tracks of galaxies in the plane of sSFR versus $\Sig1$. Stellar mass is indicated by colour, starting from $\Ms \sgt 10^{7.5} \msun$. Shown on the left are eight galaxies from our high-mass subsample, where the BN tends to occur at high redshift, $z\ssim4-2$. These galaxies demonstrate the characteristic L-shaped track, {where pre-compaction, the galaxies are initially diffuse with low $\Sig1$. They then evolve with a rather constant and slowly declining sSFR as they shrink into a BN once $\Ms\ssim10^{9-9.5}\msun$.} This is the trigger for an abrupt change in the evolution track, where the galaxies start their quenching process in which the sSFR is dropping while $\Sig1$ remains roughly constant at $\Sig1\ssim10^{9.5} \msun\kpc^{-2}$.

\smallskip
As discussed in \se{sims}, in most of the simulated galaxies, the quenching is not complete by the end of the simulations at $z\ssim2-1$ (partly due to the moderate-strength stellar/supernova feedback incorporated and the lack of AGN feedback). {Still, the early stages of the quenching process are clearly identified.}

\smallskip
The right panel of \fig{L-shape} shows eight galaxies from our low-mass subsample, where the BN tends to occur at lower redshifts, $z\ssim2-1$. {Before the major compaction event, which occurs above the critical BN mass, these galaxies may show bigger oscillations in sSFR, indicating episodes of compaction and quenching attempts followed by rejuvenation \se{MS}.} Nevertheless, the transition at the BN into quenching along an L-shape track is still clear. It occurs above the same critical BN mass as for the high-mass subsample, but with the quenching at a somewhat lower value of $\Sig1\ssim10^9 \msun\kpc^{-2}$. {The slow decline of the critical $\Sig1$ with time is related to both the cosmological decrease in density and the decline of gas fraction with time, making the galaxy less susceptible to instabilities \citep[as discussed in, e.g.,][]{db14}, and resulting in a less dissipative contraction and smaller collapse factor.}

\begin{figure*} 
\centering
\includegraphics[width=0.592\textwidth]{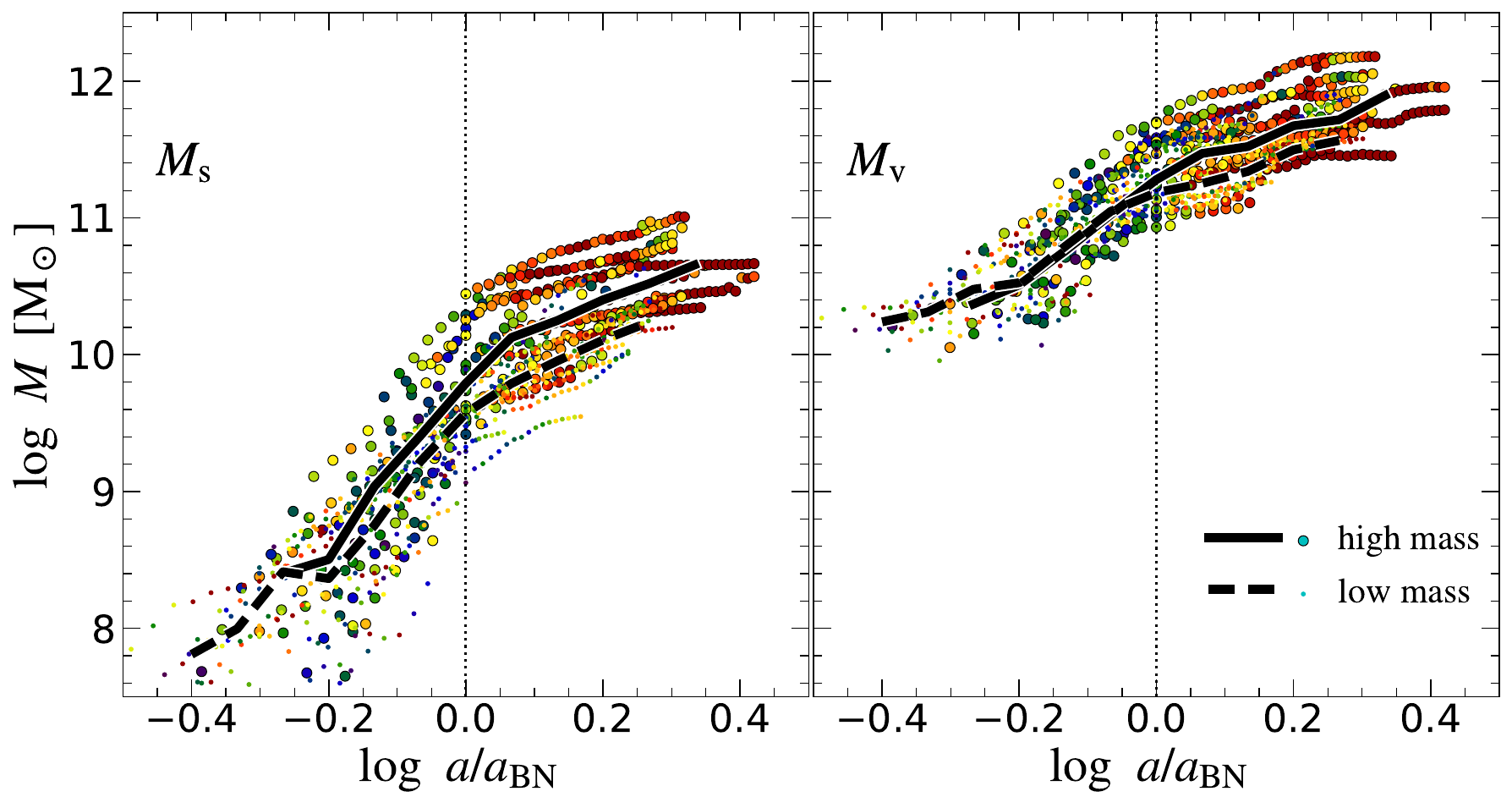}
\includegraphics[width=0.33\textwidth]{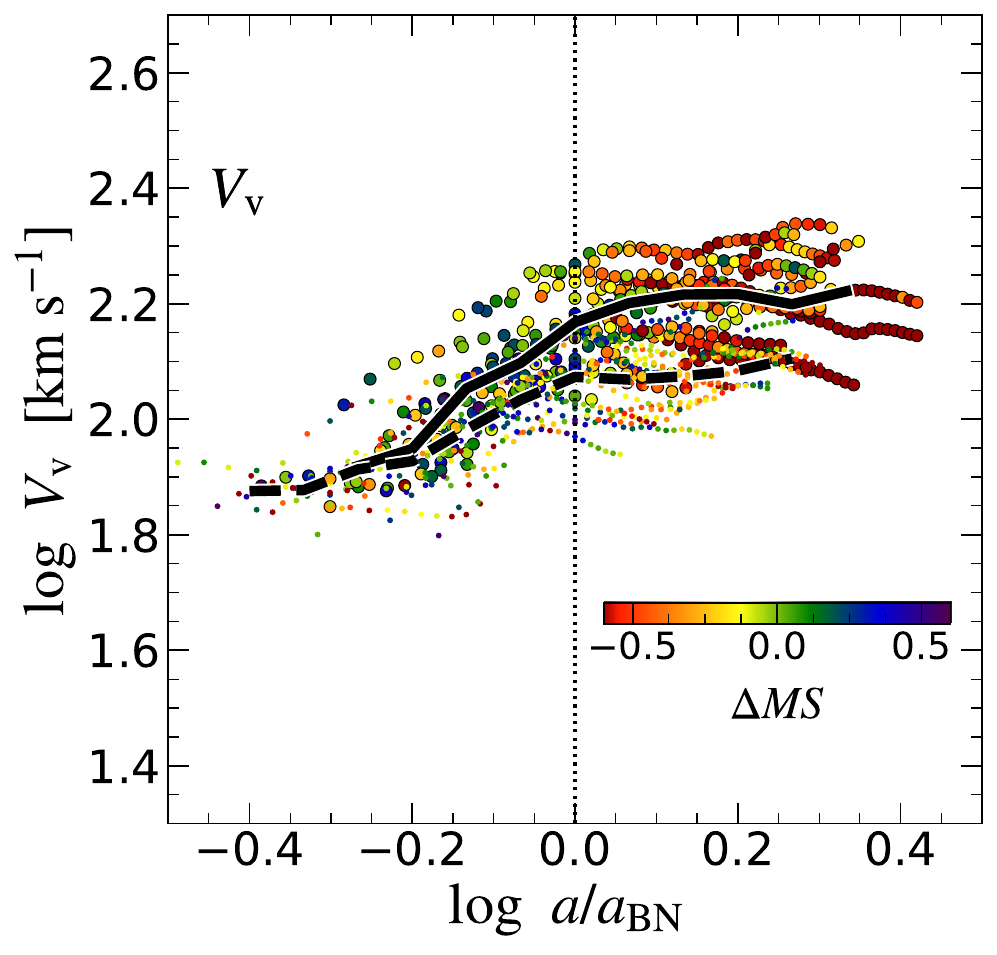}
\caption{
Characteristic BN mass. {Shown are the stellar mass, virial mass and the virial velocity, as a function of time (expansion factor), stacked with respect to the BN time ($\abn$). The medians are shown for the high-mass (solid) and low-mass (dashed) subsamples. The colour marks the deviation from the MS. The BN occurs at characteristic values of $\Ms$ and $\Mv$, with a small scatter. The value of $\Vv$ is somewhat lower for the low-mass subsample (a later BN).}
}
\label{fig:char_M-a}
\end{figure*}

\begin{figure*} 
\includegraphics[width=0.95\textwidth]{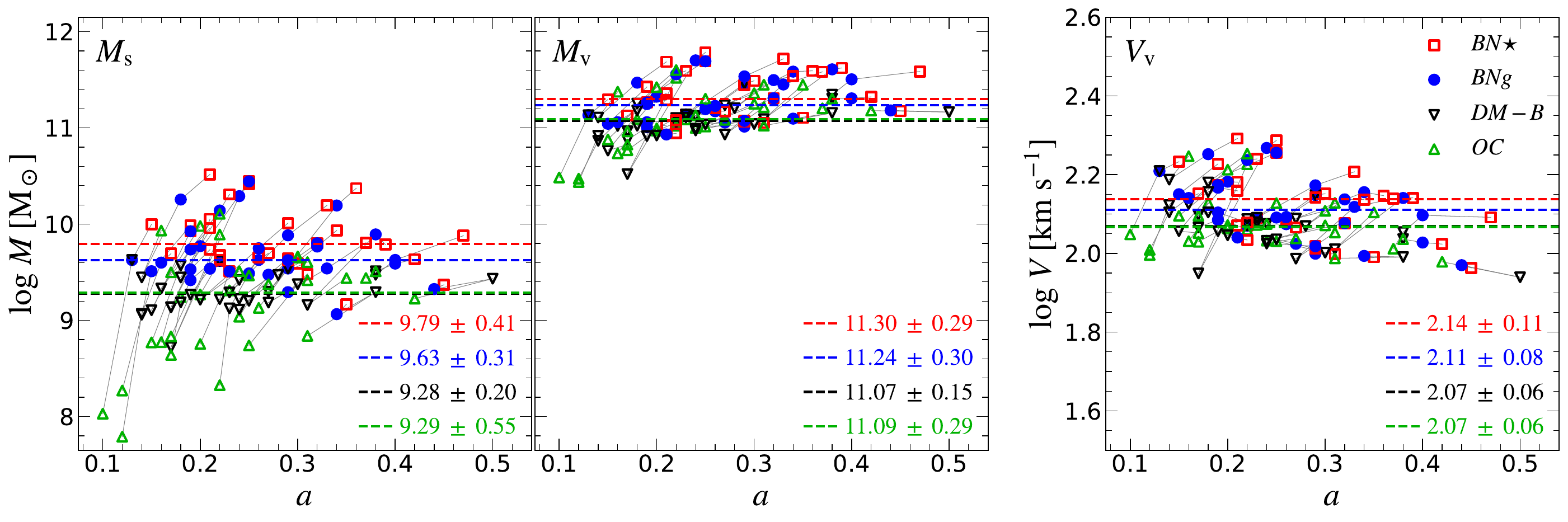}
\caption{
Characteristic BN mass {at four events associated with the compaction phase. Shown from left to right are the stellar mass, virial mass and virial velocity as a function of the expansion factor. We show the entire sample where each galaxy is marked four times, once for each compaction-associated event. The four events are as follows: 
(1) OC, the onset of compaction (green), (2) DM-B, transition from central DM to baryon dominance (black), (3) BNg, the peak of compaction measured in central gas density (blue), (4) \BNs, the shoulder of a plateau in central stellar density (red). The median masses (and velocities) are shown in dashed horizontal lines, with values of the medians and the $1\sigma$ scatter quoted for each event. The BN occurs at a characteristic mass or velocity with negligible time dependence. The halo quantities, $\Mv$ and $\Vv$ show a smaller scatter, displaying a similar scatter given the virial relation $\Vv\ssim \Mv^{1/3}$.}
}
\label{fig:char_M-a_event}
\end{figure*}

\subsection{A Characteristic BN Mass} \label{sec:cmass}
{The characteristic mass (or velocity) of the BN phase {in the simulations} is demonstrated in \fig{char_M-a} and \fig{char_M-a_event}, which refer to the global galaxy properties of stellar mass, virial mass and virial velocity. In \Fig{char_M-a}, these quantities are plotted as a function of $a/\abng$ for all snapshots of all galaxies in the sample. One can see that the halo mass is growing continuously, following on average $\Mv \sprop \exp (-\alpha z)$ with $\alpha\ssim0.8$, as predicted in the EdS regime at $z\sgeq1$ \citep{dekel13}. At $\abng$, it has a characteristic value of $\Mv\seq 10^{11.24\pm 0.3}\msun$, the same for the massive and low-mass subsamples. The stellar mass grows at a similar pace before the BN event, and its growth is flattened off soon after this event. The critical value of the stellar mass at $\abng$ for the whole sample is $\Ms\seq 10^{9.6\pm 0.3}\msun$, the same for the massive and low-mass subsamples.}
{\citet{huertas18} identified a consistent preferred characteristic stellar mass of $\simi 10^{10}\msun$ for BNs, by applying deep learning models on observed galaxies from the HST CANDELS survey.}

\smallskip
{The corresponding characteristic virial velocity at $\abng$ is $\simi125\pm25\kms$. The virial velocity post-BN is lower for the low-mass subsample. This is consistent with the virial relation $\Vv\sprop \Mv^{1/3} a^{-1/2}$ and the fact that the BN occurs at later times for low-mass galaxies. At the time of the BN, the scatter in $\Vv$ is similar to the scatter in $\Mv^{1/3}$, with the median $\Vv$ of the low-mass subsample lower by $0.1$dex than that of the massive subsample.}

\smallskip
{In figure \figApp{char_Ve-a} we show the circular velocity, $\Ve$,  measured {within} $\Re$. Although we see similar qualitative behaviour in $\Ve$, as seen in $\Vv$, the central circular velocity shows a significant scatter, with a factor of $\simi2$ difference between the low-mass and high-mass subsample around the time of compaction. This is an indication of the weaker compactions experienced by the low-mass galaxies, which results in lower concentration (see \seApp{cmass_app}).}

\smallskip
\Fig{char_M-a_event} shows the same global properties for each galaxy at the slightly different events associated with the BN phase as a function of the time $a$ when the event occurs. The events, referring to the inner $1\kpc$, are the DM-B transition, the onset of compaction OC, the peak of gas density and onset of central quenching BNg, and the shoulder of peak stellar density \BNs where it turns into a plateau. The log y range of the velocities is one-third of that of the masses to reflect the virial relation $V \sprop M^{1/3}$ with a characteristic density for all haloes at a given time.

\smallskip
All the quantities show a weak or no systematic variation with time. The trends with the time of the virial mass and velocity are related through $\Vv\sprop \Mv^{1/3} a^{-1/2}$, which explains why the velocities are constant or slowly declining with time while the masses are constant or slowly growing.
 
\smallskip
{The virial quantities $\Mv$ and $\Vv$ tend to show a smaller scatter than the galaxy quantity $\Ms$, possibly indicating that the driving of compaction is more strongly associated with the halo properties.}

\smallskip
{While there are promising indications of a similar critical BN stellar mass in observed galaxies \citep{huertas18}, it should be noted that the value found in simulations could be subject to the implemented feedback. One may assume that the simulations capture a qualitative representation of the compaction process and its effect on the galaxy, keeping in mind their limitations (e.g., the deviation from the {stellar-to-halo mass relation}, see \se{limitations}). When the compaction process is examined in simulations with a given feedback recipe, we expect the BN mass scale will maintain a critical value with little to no time dependence. We will assume throughout the paper that the critical BN stellar mass is in the ballpark of $\simi 10^{10}\msun$ and a halo mass of $\simi 10^{12}\msun$.
To further study the effect of feedback, we briefly discuss in \se{disc} the occurrence of compaction events \citep{lapiner21} in the \nh simulations \citep{dubois20_NH}, and preliminary results from the \nihao simulations \citep{wang15} and the \velaF simulations \citep{ceverino22_vela6}.}

\smallskip
The DM-B transition tends to occur soon after the onset of compaction OC, followed by the BNg, which the \BNs immediately follow. The duration from OC to \BNs is typically $\Delta a/a \ssim 0.2$, corresponding to $\Delta t/t \ssim 0.3$ \citep[][Fig.~16]{zolotov15}. {During the compaction, while the halo mass typically grows by 0.2dex, the stellar mass within $1\kpc$ typically grows by 0.5dex.}

\smallskip
In summary, the events associated with the BN phase occur at a critical mass to within a factor of two, rather independent of redshift. The medians of the {OC, DM-B, BNg and \BNs events are respectively at $\log \Ms/\msun \ssim 9.3, 9.3, 9.6, 9.8$, $\log \Mv/\msun \ssim 11.1, 11.1, 11.2, 11.3$, and $\log \Vv/\kms \ssim 2.07, 2.07, 2.10, 2.14$. Therefore the mass or velocity are proxies for $a/\abng$.}

\subsection{Comparison to Observations}
\label{sec:phases_obs}

The L-shape evolution tracks in \fig{L-shape} are remarkably similar to the observed population of galaxies in the plane of sSFR versus $\Sig1$. Figure~6 of \citet{barro17_uni} shows this for CANDELS galaxies in redshift bins in the range $z\seq 0.5\sdash 3.0$. Figure~7 of that paper shows the galaxies of all redshifts stacked, where the weak redshift dependence of the threshold $\Sig1$ for nuggets (determined by the RNs) is scaled out, demonstrating that the L-shape is universal. This is a version of earlier presentations, e.g. Fig.~2 of \citet{barro13} also spanning $z\seq 0.5\sdash 3.0$, \citet{cheung12} based on the DEEP/AGEIS survey at $z\seq 0.5\sdash 0.8$, and low-redshift SDSS galaxies in \citet{fang13} and \citet{woo17}. Similar results are shown in Figs.~13-15 of \citet{mosleh17} for $z\seq 1-2$.

\smallskip
The top-right panel of \figApp{M-Ms_hm} shows a tight correlation between $\Sig1$ and $\Ms$, with a close to linear slope for SFGs, and a flatter slope for quiescent galaxies at high $\Sig1$ and $\Ms$. This is observed at high redshifts {\citep{barro17_uni,saracco12,tacchella15_sci, suess21}} and at low redshifts \citep{cheung12,fang13,tacchella17_zens}, as well as in different environment \citep{saracco17}. \citet{tacchella16_prof} demonstrated that the simulations agree both qualitatively and quantitatively with these observations of the $\Sig1-\Ms$ relation.

\Fig{fdm_Re} predicts that compaction leads to a low fraction of DM within $\Re$ in BN and post-BN galaxies, where $\Ms\!\geq\!10^{10}\msun$, reaching values as low as $\fdm\ssim0.1$. This is indeed consistent with the DM fractions deduced from observations for massive galaxies at $z\ssim2$ \citep{price16, wuyts16, genzel17, genzel20, liu23}, typically $\fdm\ssim0.1$. However, one should bear in mind the large uncertainties associated with estimating dynamical masses from observations, as discussed in \se{jeans}. {Observation by \citet{genzel20} of rotation curves in massive galaxies show a significant percentage of galaxies with low central DM fraction, $\fdm(\slt\Re)\slt0.3$, which is associated with surprisingly extended DM cores ($\simi10\kpc$). Higher prevalence of both low $\fdm$ and the existence of extended DM cores are shown in the high-z subsample. While we find low $\fdm$ for post-compaction massive galaxies, as seen in \fig{fdm_Re}, such extended DM cores in massive galaxies are not reproduced by cosmological simulations to date. \citet{dekel21_core} suggested a \emph{two-step} scenario for core formation in massive haloes $\Mv\!\geq\!10^{12}\msun$. In the first step, the inner DM halo is heated by dynamical friction due to a merger with a post-compaction satellite. Once the DM is sufficiently pre-heated, namely closer to the escape velocity from the cusp, outflows by AGN activity may have a greater effect causing further expansion and creating extended cores. Using a combination of an analytical toy model, a semi-analytical model {SatGen} \citep{jiang21} and results from the {\vela} simulations shown in this work, it was estimated that the pre-heating would be efficient for a satellite which went through a wet compaction event, with preference to high redshifts. Using the {CuspCore} model \citep{freundlich20_cuspcore}, it was shown that pre-heated cusps develop a DM core in response to the removal of half the gas mass by AGN feedback.}

\smallskip
\Fig{M-a_hm} and \figApp{M-Ms_hm} show that at the BN phase the SFR within $1\kpc$ is at a peak. ALMA observations of CO and dust reveal that the SFR in blue nuggets is indeed enhanced in a sub-kpc core {\citep{barro16_alma,tadaki16,tadaki17,tadaki20,barro17_alma,elbaz18}}.

\smallskip
\Fig{char_M-a} and \fig{char_M-a_event} indicate a characteristic mass for the BN phase. \citet{huertas18} used deep learning to search for BNs as defined in the simulations in the HST CANDELS survey. The machine was trained using mock images of simulated galaxies and successfully identified BNs among the observed galaxies. This clearly revealed a preferred stellar mass for {BNs in the range of $10^{9.2-10.3}\msun$,} as predicted. While an attempt has been made to eliminate the direct information concerning the total luminosity (and hence mass) from the training set, it is yet to be convincingly demonstrated that the classification as a BN was dominated by the compact young-SFR nature of the galaxy rather than by its mass.

\section{Phases of Evolution with respect to the Main Sequence}
\label{sec:MS}

\subsection{Galaxy Properties Along the Main Sequence}

\begin{figure*} 
\includegraphics[width=0.95\textwidth]{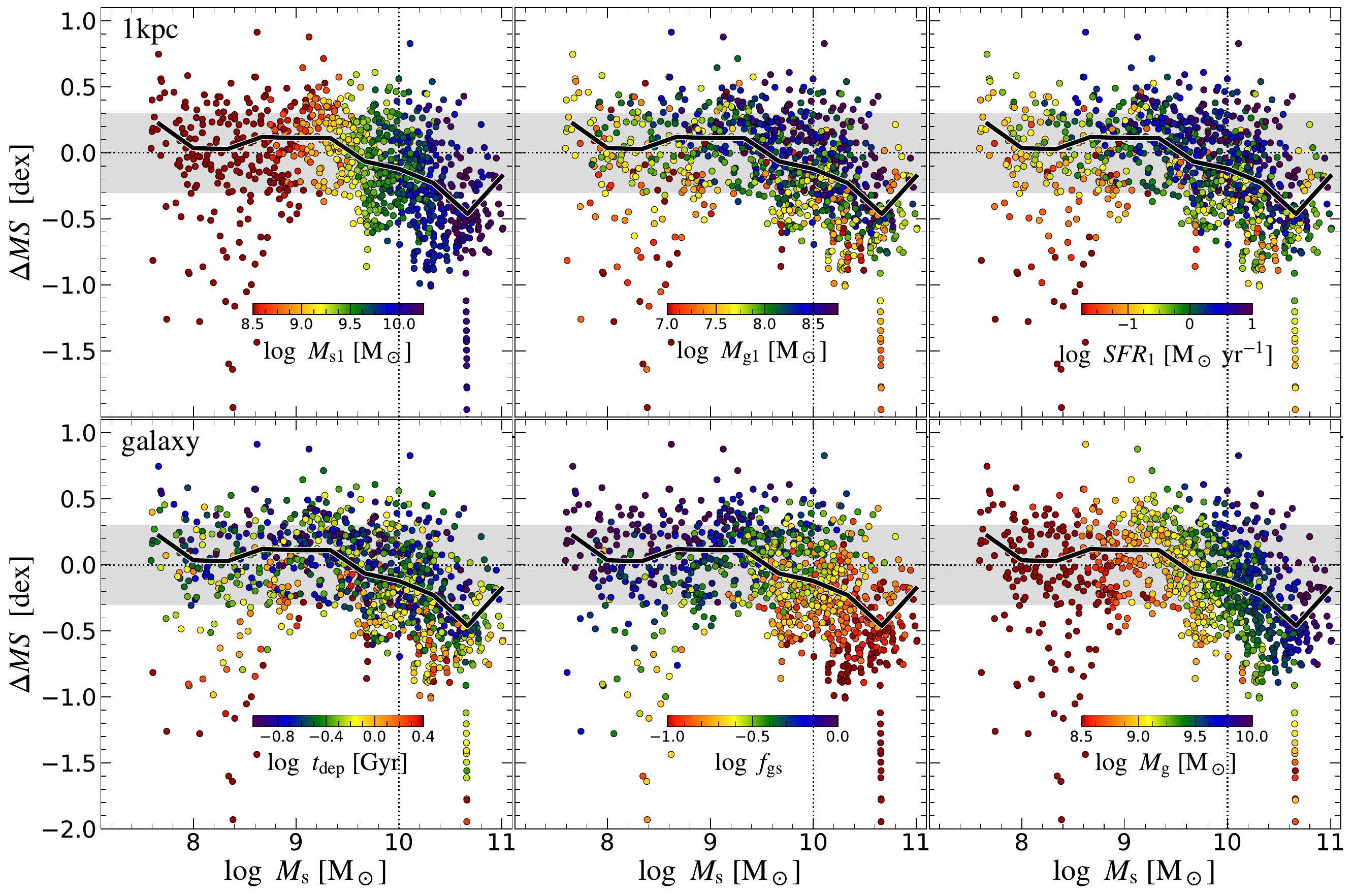}
\caption{
Galaxy properties with respect to the MS, in the plane of $\DMS$ versus $\Ms$. The value of the desired property is indicated by colour. {\bf Top:} Stellar mass, gas mass and SFR within $1\kpc$ (left to right). {\bf Bottom:} depletion time, gas-to-star fraction and gas mass for the whole galaxy.
}
\label{fig:dMS-Ms}
\end{figure*}

The evolution through compaction and quenching phases is naturally translated to evolution along and across the universal Main Sequence (MS) of SFGs in the diagram of $\DMS$ versus stellar mass. The quantity $\DMS$ is the log deviation from the universal MS ridge, 
\be
\DMS = \log {\rm sSFR}- \log {\rm sSFR}_{\rm MS} \, 
\label{eq:DMS}
\ee
with sSFR$_{\rm MS}$ defining the MS ridge and scaled by $1/(1+z)^{5/2}$ to reflect the systematic redshift dependence of the sSFR \citep[following the systematic redshift dependence of the specific accretion rate,][]{dekel13}. This has been discussed based on the same simulations in \citet{zolotov15} and \citet{tacchella16_ms}. Pre-BN galaxies evolve along the universal MS while oscillating across the MS ridge, and they climb to the top of the MS as they {shrink to a compact BN. Then, once they reach the} critical BN mass, they gradually {transition into a quenched phase \citep{pandya17}, moving} down across the MS through the `Green Valley' at the bottom of the MS, aiming at the quenched region well below the MS.

\begin{figure*} 
\includegraphics[width=0.95\textwidth]{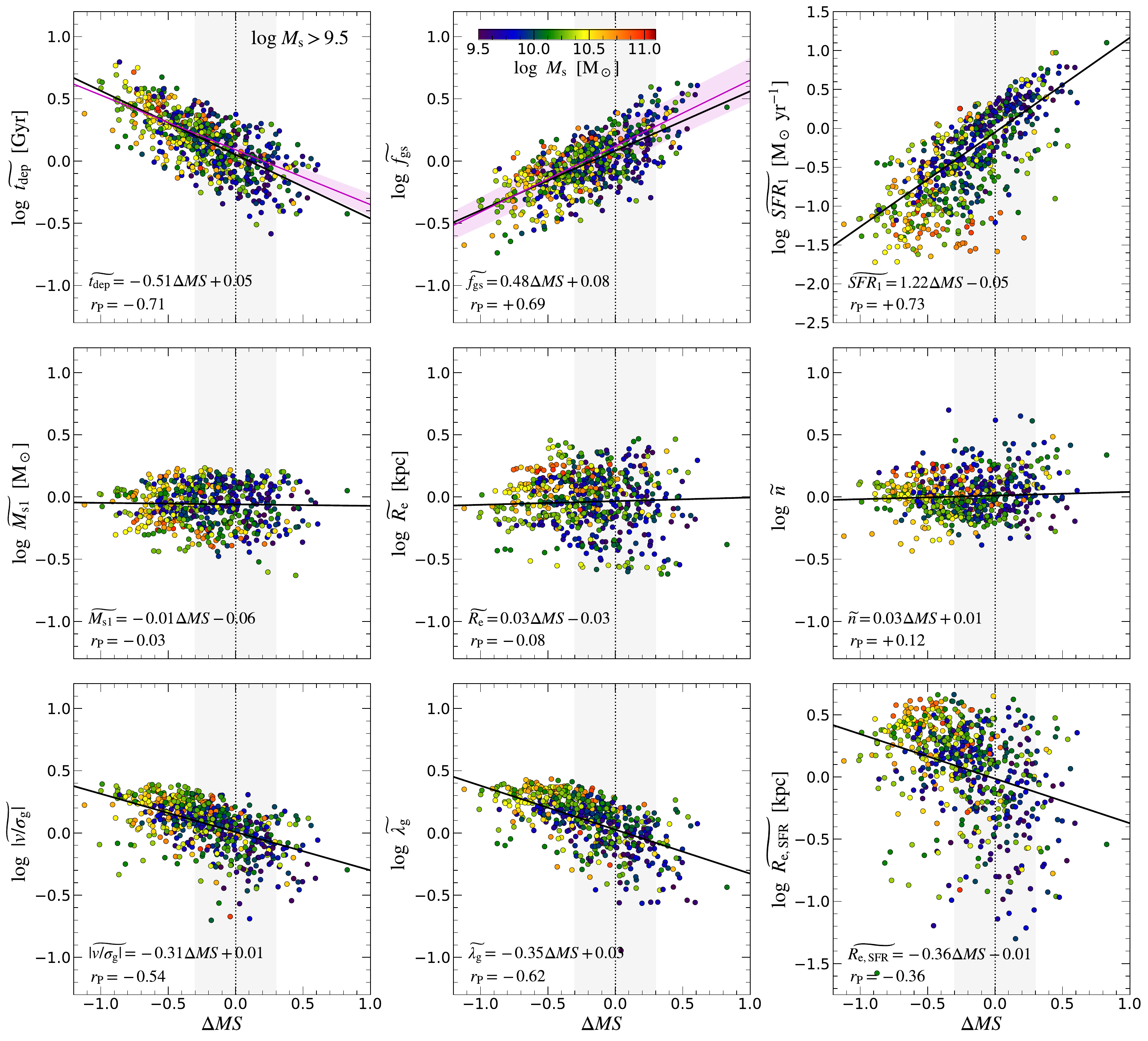}
\caption{
Gradients across the MS for galaxies {above} the critical BN mass. The systematic redshift and mass dependencies at the MS ridge are eliminated. {The stellar mass, $\Ms$, is indicated by colour. {\bf Top:} properties involving gas, the depletion time ($\tdep$) and gas fraction ($\fgs$) for the whole galaxy, and SFR within $1\kpc$ ($SFR_1$). The magenta lines and shaded regions in $\tdep$ (top-left) and $\fgs$ (top-middle) indicate the best-fitting and the uncertainties of observations by \citet{tacconi18}. {\bf Middle:} properties involving stars. The stellar mass within $1\kpc$ ($\M1s$), the 3D half mass radius of stars ($\Re$), and the \sersic index ($n$). {\bf Bottom:} gas rotational velocity over the velocity dispersion ($\vosg$), the spin parameter of the gas ($\lambdag$) and the 3D effective radius of the SFR.} {We find that significant gradients across the MS in quantities which involve gas and SFR. Conversely, the stellar properties show neither significant gradients nor tight correlations.} 
}
\label{fig:gradients_q-dMS}
\end{figure*}

\smallskip
In \citet{tacchella16_ms}, we explain how this evolution pattern confines the SFGs into a flat, narrow sequence of width $\pm 0.3\,$dex in $\DMS$ {\citep[as observed,][]{noeske07, whitaker14}}, how it makes the MS bend down above the critical mass of $\Ms\ssim10^{10}\msun$ \citep{schreiber16}, and how it is responsible for gradients of gas-related properties of galaxies across the MS ridge in the post-BN, massive end. {In particular, it was shown in \citet{tacchella16_ms} and discussed here bellow (see \se{gradients}, \fig{gradients_q-dMS}), that both the depletion time and the gas fraction in the \vela simulations reproduce a consistent dependence on $\DMS$ as seen in observations by \citet{genzel15} and \citet{tacconi18}.}

\smallskip
The confinement is explained by the bound oscillations about the MS ridge of galaxies below the critical mass. {These oscillations are illustrated in \fig{L-shape} as well as \figApp{L_V01_V27}, which shows the evolution track of low-mass galaxies in the plane of sSFR and $\Sig1$.} These galaxies show several oscillations in sSFR, representing repeating episodes of compaction and quenching attempts before the final BN, which is followed by long-term quenching. The downturn and upturn at the top and the bottom of the MS, respectively, are understood in terms of the balance between the gas inflow rate into the core ($in$) and the gas consumption rate by star formation and outflows from the core ($out$) \citep[][Fig.~9]{tacchella16_ms}. Wet compaction is expected when $in\sgt out$ \citep{db14} while quenching by gas depletion is expected when $in\slt out$. At the {\BNs} after compaction, the central SFR and outflows are enhanced, and the lack of further gas supply from the shrunken disc tips the balance toward a quenching process, causing a downturn. The downturn is assisted by `morphological' quenching, where the growing bulge suppresses the violent disc instability and the associated inflow \citep{martig09,cacciato12}.  

\smallskip
An upturn by a new compaction event can occur once the replenishment of the disc by fresh gas is faster than the gas depletion, $t_{\rm rep}\slt t_{\rm dep}$. The depletion time is observed to be growing rather slowly with time, {$t_{\rm dep}\ssim(1+z)^{-0.5} M_{12}^{-0.19} \Gyr$} \citep{genzel15,tacconi18}, while from theory and simulations, the accretion time grows much faster, {$t_{\rm rep}\ssim 25 (1+z)^{-5/2} M_{12}^{0.14} \Gyr$} \citep{dekel13}, implying that upturn is expected {to be more common} at $z\!\geq\!3$.
 
\smallskip
{The final long-term downturn at the major BN phase, which carries the galaxy off the MS (possibly even forever), occurs when $t_{\rm rep}\sgt t_{\rm dep}$. This is caused by the overall} suppression of gas supply into the galaxy once the CGM becomes hot, which typically occurs once the halo exceeds $\simi10^{11.5}\msun$ and at $z\slt 3$ \citep{db06}. {This threshold mass is associated with the bending down of the MS above a critical stellar mass, which is likely to be triggered by compaction-driven central gas depletion and maintained by the suppression of gas supply due to a hot CGM.}

\smallskip
{
\Fig{dMS-Ms} addresses the evolution of certain galaxy properties with respect to the universal MS ridge. All the snapshots of all the galaxies in our sample are shown in the plane of $\DMS$ versus $\Ms$.} The population reproduces the overall MS, which resembles the observed MS. Below the BN critical mass, the MS is roughly flat, constant as a function of mass. This is because, at a given redshift, the sSFR is roughly mass independent, following the negligible mass dependence of the specific cosmological accretion rate into the halo \citep{dekel13}, and the fact that $(1+z)^{5/2}$ captures the overall decline in time of the sSFR in SFGs. Most of the galaxies are confined to a narrow MS, while a few of the lowest-mass galaxies in our sample (that evolve through the BN phase late) quench below the MS for a short period and rejuvenate back into it. Above the BN mass, the MS naturally bends down in the post-BN phase, reflecting the progressing quenching process at the massive end.

\smallskip
The colour in \fig{dMS-Ms} denotes the desired property in each panel. The top panels show quantities within the central $1\kpc$, while the bottom panels refer to the whole galaxy.
The gas mass within $1\kpc$ (top-middle) is the prominent tracer of compaction, the BN stage, and the following central depletion and quenching. A peak of central gas mass (blue) is reached at $\Ms\ssim10^{10}\msun$ in the upper part of the MS, demonstrating that this is where the BNs are expected to reside. A similar peak in the same location is seen in the SFR within the inner $1\kpc$ (top-right), which naturally traces the central gas mass.
On the other hand, the stellar mass within $1\kpc$ (top-left), which is a proxy for the observable $\Sig1$ commonly used to identify the BN, shows only {a horizontal gradient as a function of $\Ms$} {\citep[see also][]{ceverino15_e}}, with the stellar density growing rapidly during the compaction phase and settling to a constant (blue) after \BNs, where $\Ms\!\geq\!10^{10}\msun$.

\smallskip
Observable global properties that are related to the gas mass $\Mg$ are the depletion time and the gas-to-star fraction defined by
\be
\tdep = \Mg/{\rm SFR}, \quad
\fgs = \Mg/\Ms .
\ee
The colours in the bottom panels of \fig{dMS-Ms} indicate vertical gradients across the MS for both $\tdep$ and $\fgs$. There is no horizontal gradient along the MS in $\tdep$, but $\fgs$ also varies systematically with $\Ms$, reflecting the overall decline of gas fraction with cosmological time because our sample follows the evolution of a given set of galaxies.

\subsection{Gradients Across the Main Sequence}
\label{sec:gradients}

\smallskip
\Fig{gradients_q-dMS} shows the vertical gradients of different galaxy properties ($\tilde{Q}$) across the MS for the galaxies above the critical BN mass. In these gradients, the systematic redshift and mass dependencies at the MS ridge were scaled out in the following way, similar to what we did in \citet{tacchella16_ms}, which is close to the method applied by \citet{genzel15} to their observed sample. We assume the model,
{
\be
\log Q = A + B f(z)  + C \DMS + D g(\Ms),
\label{eq:scaling}
\ee
where $f(z)\seq B_0/B + \log(1+z)$, $g(\Ms)\seq D_0/D+\log(\Ms/10^{10.5})$. We then determine the best-fitting values for the parameters $A$, $B_0$, $B$, $C$, $D_0$, and $D$. The specific procedure adopted here is as follows. First, we select the subsample of galaxy snapshots that lie near the MS ridge $|\DMS| \slt  0.15$, temporarily assume $D\seq 0$ and obtain the best-fitting $B$ and $B_0$. Second, using the same subsample near the MS ridge, we eliminate the systematic $z$ dependence by referring to the scaled quantity $\log Q - B f(z)$ and obtain the best-fitting $D$. Finally, we return to the full sample where $\DMS$ spans its full range of values, scale out the $z$ and $\Ms$ dependencies by referring to the scaled quantity 
\be
\tilde{Q} = \log Q - B f(z) - D g(\Ms) = A + C \DMS ,
\ee
and obtain the desired best-fitting C.

\smallskip
One could determine the best-fitting parameters in several alternative ways, e.g., by first eliminating the mass dependence and then the redshift dependence, by a more sophisticated procedure of successive fits \citep{tacconi18}, or by a straightforward simultaneous fit of all the parameters. Given the biases that the sample selection may introduce and the possible deviations from the model assumed in \equ{scaling}, in principle, the results may differ from method to method and sample to sample. {Testing with mock catalogues} reveals that the derived value of $C$ is very robust to the sampling and the best-fitting method, while $B$, and to some extent also $D$, can be sensitive to both. Therefore, we show in \fig{gradients_q-dMS} only the gradients across the MS.

\smallskip
The gradients of some properties may be different pre-BN and post-BN. Here, we limit the discussion to the galaxy snapshots above the BN critical mass. In this regime, a decline of $\DMS$ reflects evolution across the MS, from the BN stage at the top of the MS, through the MS ridge, and finally descending below the MS toward the quenched regime. 

\smallskip
In \fig{gradients_q-dMS}, the gas-dependent quantities show significant gradients with rather tight correlations (Pearson correlation coefficient {$\rp\ssim 0.7$}), which are associated with the central gas depletion process. The slope of the gradients for the global $\tdep$, global $\fgs$ and the SFR within $1\kpc$ are $-0.51\pm 0.07$, $+0.48\pm 0.04$ and $1.22\pm 0.06$ respectively. The first two, similar to \citet{tacchella16_ms}, are in excellent agreement with the observational results \citep{genzel15,tacconi18}. The strong super-linear gradient for the SFR within $1\kpc$ across the MS is an observable prediction.
}

\smallskip
{On the other hand, the observable quantities that refer to the stellar component (second row) do not show tight correlations nor significant gradients across the MS ($\rp\ssim 0.1$),} as the stellar core is established during the BN phase and it remains roughly the same in the post-BN phase during the gas depletion process. In particular, the simulations predict no significant gradients for the stellar $\Sig1$, the half-mass radius $\Re$, or the \sersic index $n$. However, the young stars (not explicitly shown here) show correlations similar to the quantities that involve gas mass and density.

\smallskip
{During the pre-BN phase (not shown in here), below the critical mass $\Ms\slt 10^{9.5}\Msun$, the gradients are not necessarily similar to the post-BN phase. We find similar gradients for the global $\tdep$, global $\fgs$ and the SFR within $1\kpc$, with slopes of: $-0.47\pm 0.03$, $+0.47\pm 0.03$, $1.10\pm0.08$ respectively, and Pearson correlation coefficient of $\rp \sgt 0.64$. Other properties shown in \fig{gradients_q-dMS} show either mild or no significant gradient in the pre-BN phase. The best-fitting slopes of these gradients are listed in \tabApp{gradients_dMS_app} for both the pre-BN and the post-BN phase. }

\subsection{Comparison to Observations}
\label{sec:MS_obs}

As predicted in \fig{dMS-Ms} and reported in \citep{tacchella16_ms}, the blue nuggets are observed preferentially in the upper parts of the Main Sequence of SFGs \citep{vandokkum15,elbaz18}.

As already highlighted in \citet{tacchella16_ms}, the gradients of quantities involving gas across the MS, as seen in \fig{gradients_q-dMS}, are observed. \citet{tacconi18} and \citet{genzel15} presented the scaling relations between galaxy-integrated molecular gas masses, stellar masses and star formation rates as a function of redshift between $z\seq 0$ and $z\seq 4$. They determine molecular gas masses in different ways from CO line fluxes (from PHIBSS, xCOLD GASS and other surveys), far-infrared dust spectral energy distributions (from Herschel), and $\simi$1mm dust photometry (from ALMA), in 758 individual detections and 670 stacks of SFGs, covering the stellar mass range $\log(\Ms/{\rm M_\odot})\seq 9.0-11.8$, and star formation rates relative to that on the MS, $\DMS$, from $-1.3$dex to $+2.2$dex.
\citet{tacconi18} find that the molecular gas depletion time-scales as $\tdep \sprop \DMS^{-0.44}$, and the ratio of molecular-to-stellar mass scales as {$\fgs \sprop \DMS^{+0.53}$.} {These scaling relations agree well with our simulation predictions $\tdep \sprop \DMS^{-0.51 \pm 0.07}$ and $\fgs \sprop \DMS^{+0.48 \pm 0.04}$.}

\smallskip
Similar results have been quoted by other observations. Based on ALMA observations of the long wavelength dust continuum in the COSMOS field, \citet{scoville17} find that the ISM mass is $\propto\! \DMS^{0.32}$ and $\tdep \sprop \DMS^{-0.7}$, which is in qualitative agreement with our simulations. Likewise, \citet{sargent14} {and \citet{puglisi21}} find that $\tdep$ for molecular gas declines with SFR in starburst galaxies, though the gas fraction is lower in starbursts, presumably because it has already been consumed. \citet{silverman_daddi15} find short $\tdep$ well above the MS at $z\ssim1.6$. \citet{huang_kauffmann14} find that the primary correlation of $\tdep$ is with sSFR, and quote $\tdep \sprop {\rm sSFR}^{-0.37}$ to ${\rm sSFR}^{-0.50}$ based on the COLD-GAS survey at low redshifts.

\section{Radius-Mass}
\label{sec:radius}

\begin{figure*} 
\includegraphics[width=0.95\textwidth]{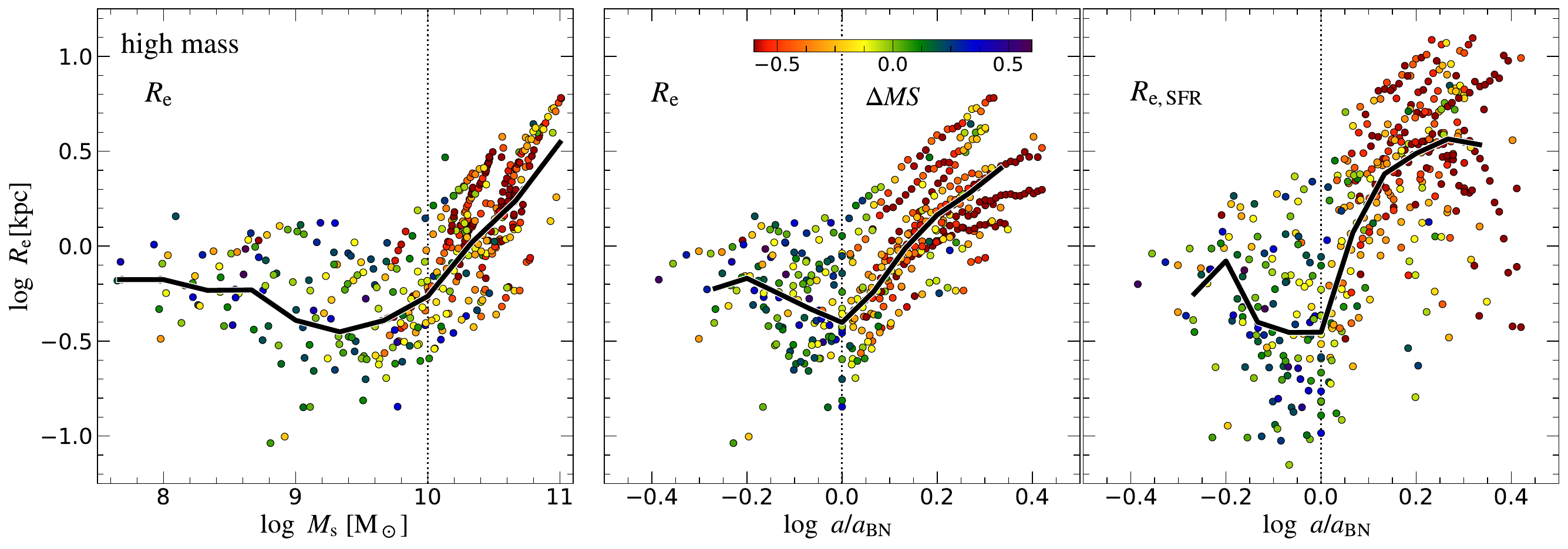}
\caption{
{Size evolution of the high-mass subsample (defined at $z\seq 2$, see similar \figApp{Re-Ms_LM} for the low-mass subsample). The distance from the MS ridge, $\DMS$, is indicated by colour. {\bf Two left panels:} Evolution of effective radius, as a function of $a/\abn$ (left)  and of $\Ms$ (middle). {Pre-BN, the size evolution with mass tends to be roughly horizontal (with individual tracks slowly rising/decreasing). After a brief reduction in $\Re$ during compaction, the post-BN effective radius of the high-mass subsample rises rather steeply as they fall below the MS and start to quench. In the low-mass subsample \figApp{Re-Ms_LM}, their inclination for repeated episodes of small compactions tends to smooth out some of the above-mentioned features}. {\bf Third-panel :} Evolution of the effective radius of the SFR, $\Resf$ as a function of $a/\abn$.{$\Resf$ shows similar behaviour as $\Re$, albeit with a larger scatter.}}
}
\label{fig:Re-Ms}
\end{figure*}

\begin{figure} 
\includegraphics[width=0.95\columnwidth]{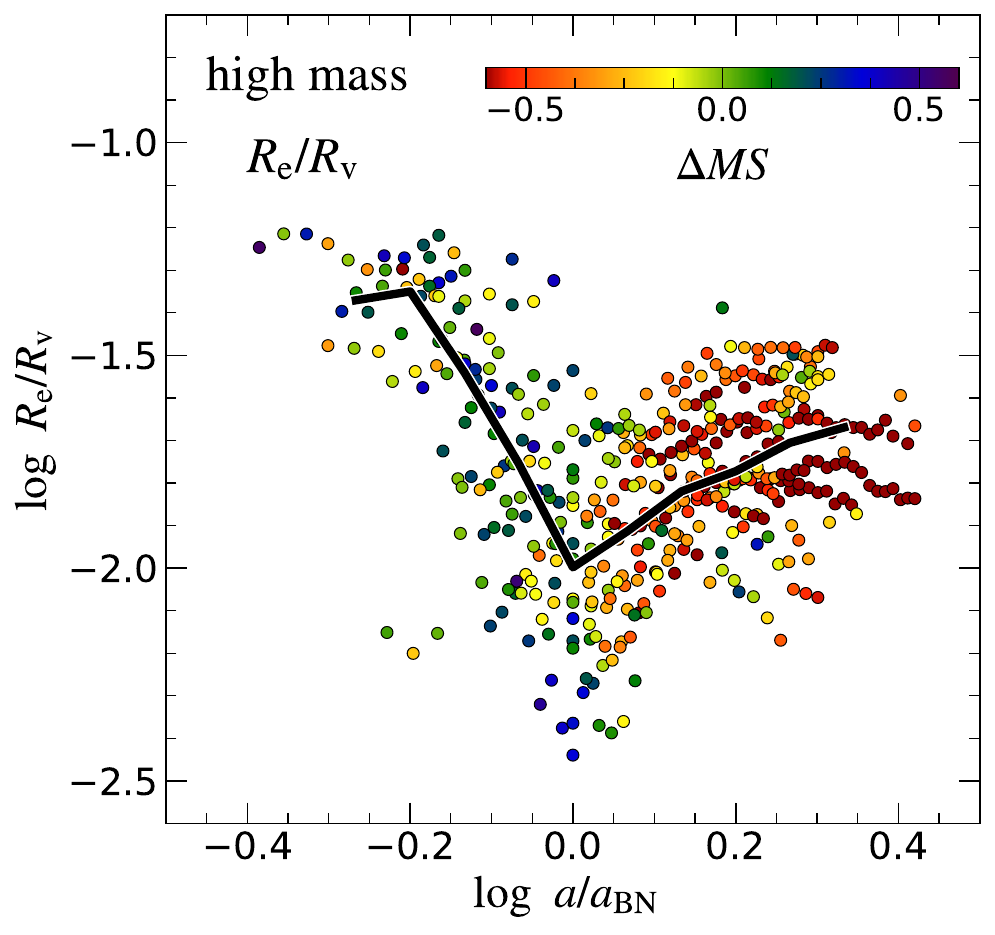}
\caption{
{Evolution of the effective radius scaled by the virial radius, $\Re/\Rv$, as a function of $a/\abn$  (see \figApp{ReORv-abn_LM} for the low-mass subsample). The distance from the MS ridge is indicated by colour.}
{The tendency for $\Re$ to grow with time, as deduced from the growth of $\Rv$ and the universal spin parameter with angular-momentum conservation, is partly compensated by repeated episodes of compaction, in which $\Re$ is pushed down}
}
\label{fig:ReORv-abn}
\end{figure}

The evolution in the effective radius of the stars and the SFR (or young stars) clearly reveals the BN phase and two different characteristic behaviours in the pre-BN and post-BN zones or below and above the BN characteristic mass. This is also reflected in transitions in the surface density profiles $\Sigma(r)$ of gas and stars shown in \fig{profiles_re} \citep{tacchella16_prof}.

\smallskip
\Fig{Re-Ms} show the effective radius for the high-mass subsample, $\Re$ for stars, versus $a/\abn$ and $\Ms$, as well as for SFR $\Resf$ versus $a/\abn$. Pre-compaction, for galaxies below the critical BN mass, the overall evolution of the SFGs tends to be roughly horizontal, namely about a roughly constant $\Re$ as the mass grows, with the individual tracks either slowly rising or slowly decreasing. The scatter is naturally larger for the instantaneous SFR than for the integrated stellar mass. 

\smallskip
Post-BN, the $\Re$ of the quenching galaxies is rising rather steeply with stellar mass, as the latter is growing slowly during this quenching phase. {Similar behaviour of transition to a steep size growth during the quenching phase is also shown in \citet[Fig.~20]{rodriguez17} using semi-empirical modelling.} The compaction itself is marked by a shrinkage in radius over a short period, reaching a minimum at the BN, after which the steep rise commences. The shrinkage is somewhat smeared out when plotted against $\Ms$, showing at the critical BN mass a transition from a flat and spread-out distribution to a steeply rising branch, with only a trace of shrinkage at the BN mass itself.

\smallskip
{The right panel in \Fig{Re-Ms} shows the time evolution of the half radii for SFR, $\Resf$. Pre-BN, $\Resf$ is similar to $\Re$; their ratio ($\Resf/\Re$) is about unity, describing a diffuse star-forming galaxy. A short-term shrinkage can be seen at the BN phase, typically by 0.3dex, as the SFR becomes largely confined to the compact BN. Post-BN, $\Resf$ grows rapidly, and the ratio of $\Resf/\Re$ becomes higher, typically 0.4dex above unity, reflecting the appearance of a gas-rich star-forming extended ring.}

\smallskip
{In \Fig{ReORv-abn}, we show the ratio of the effective radius and the virial radius, $\Re/\Rv$. The tendency for $\Re$ to grow with time, deduced from the growth of $\Rv$ and a universal spin parameter ($\lambda$) with angular-momentum conservation, is partly compensated by repeated compaction episodes, associated with the oscillations about the MS, in which $\Re$ is pushed down. 
{Using an earlier version of the VELA simulations, \citet{ceverino15_e} showed that the assumption of a fixed spin parameter with AM conservation does not reproduce the evolution of the effective radius as derived from simulated galaxies. A good match was achieved only under the assumption of $\lambda(z)$ decreasing with time.}
In \citet{jiang19_spin}, it was demonstrated that the halo spin is not a good predictor for the galaxy size on a one-to-one basis due to episodes of compaction or mergers; however, a model with the halo concentration parameter proved to reproduce a tighter relationship between the halo and galaxy size.}

\smallskip
{In \fig{gradients_q-dMS} we show the post-BN gradients across the MS of both $\Re$ (second row, middle panel) and for $\Resf$ (third row, right panel). After removing the systematic dependencies on $z$ and $\Ms$ at the MS, one can see practically no gradient for $\Re$, while $\Resf$ does show a gradient with a slope of $-0.36\pm 0.09$ ($\rp\ssim 0.36$). The effective radius of SFR shows smaller values above the MS when the galaxy is compact and intensely star-forming at the BN phase. Below the MS, at a later post-BN stage, $\Resf$ becomes larger due to the formation of an extended star-forming disc/ring around the gas-poor stellar nugget.
The lack of MS gradient in $\Re$ is expected since the central stellar component remains roughly the same in the post-BN phase. However, $\Resf$ is related to the young stars, which show trends more closely related to quantities involving gas mass.}

\subsection{Comparison to Observations}
\label{sec:radius_obs}

The behaviour in the $\Re-\Ms$ diagram in \fig{Re-Ms} (left panel) resembles the observed population of this diagram {\citep[e.g.,][]{shen03, franx08, mosleh_franx11, cibinel13, barro13, vanderwel14_MR, tacchella15_sins, vandokkum15, mosleh20, suess21}}. We note that it is difficult to constrain the evolutionary paths of individual galaxies in the $\Re-\Ms$ diagram from the way galaxies populate this plane because of progenitor effects \citep[e.g.,][]{vandokkum96, lilly16}. \citet{vanderwel14_MR} combines redshifts, stellar mass estimates, and rest-frame colours from the 3D-HST survey with structural parameter measurements from CANDELS imaging to determine the galaxy size-mass distribution over the redshift range $0\slt z\slt 3$. They find at all redshifts a shallow slope of $\Re \sprop \Ms^{0.22}$ for SFGs, and a steep slope $\Re \sprop \Ms^{0.75}$ for quiescent galaxies.

\smallskip 
\Fig{Re-Ms} (left) roughly matches the cartoon in Figure 28 of \citet{vandokkum15}. Using observational samples with matching number densities, \citet{vandokkum15} (also \citealt{vandokkum13, patel13}) suggest that even though individual galaxies may have had complex histories with periods of compaction and mergers, the ensemble-averaged evolution of star-forming galaxies in the size-mass plane is well described by a simple inside-out growth track of $\Re \sprop \Ms^{0.3}$. This rather shallow track is confirmed by the sSFR profile within galaxies, which are roughly flat or slightly increasing at $\Ms\slt 10^{11}\Msun$ \citep{tacchella15_sci, nelson16_Ha, liu17}. The flat size growth in the simulations before compaction is rather consistent  \citep[see also][]{tacchella16_prof}. After quenching at high masses, the size grows steeply with mass, both in the simulations and observations. It is still debated how much size (and stellar mass) growth takes place in individual systems due to minor mergers {\citep[see, e.g.,][]{suess19}}, and it probably depends on the stellar mass and environment of a galaxy \citep{carollo13, fagioli16, tacchella17_zens}. In our simulations, we find that high-mass galaxies are going through a significant size growth on a steep track in the $\Re-\Ms$ diagram mainly due to star formation in the outskirts (rising sSFR profiles toward the outskirts).

\smallskip
{\citet{somerville18} used a combination of observed CANDELS galaxies at $z\!<\!3$ with the stellar-to-halo mass ratio from abundance matching, to find the statistical relationship between the effective stellar radius and the virial radius, $\Re/\Rv$. It was shown that at high redshifts of $z\ssim2$, the ratio has a median value of $\Re/\Rv\ssim 0.02$, roughly constant with mass except for the most massive bins around $\simi10^{11}\msun$. 
To compare these results, we select simulated galaxies at similar mass and redshift ranges, choosing post-BN galaxies ($\Ms\sgt 10^{9.5}\msun$) at $z\seq1\sdash3$. As seen in \fig{ReORv-abn}, the high-mass subsample shows a trend of rising $\Re/\Rv$ with $a/\abn$ (or $\Ms$) and a somewhat lower ratio. However, when the entire sample is considered, now including \figApp{ReORv-abn_LM}, we find a rather constant $\Re/\Rv$ as a function of mass with a median value of $\simi 0.02$, mostly consistent with \citet{somerville18}. We do not recover a decline in the ratio in the most massive bins, possibly due to the small number of simulated galaxies with such masses.}

\section{Global Shape}
\label{sec:shape}

\begin{figure*} 
\includegraphics[width=0.95\textwidth]{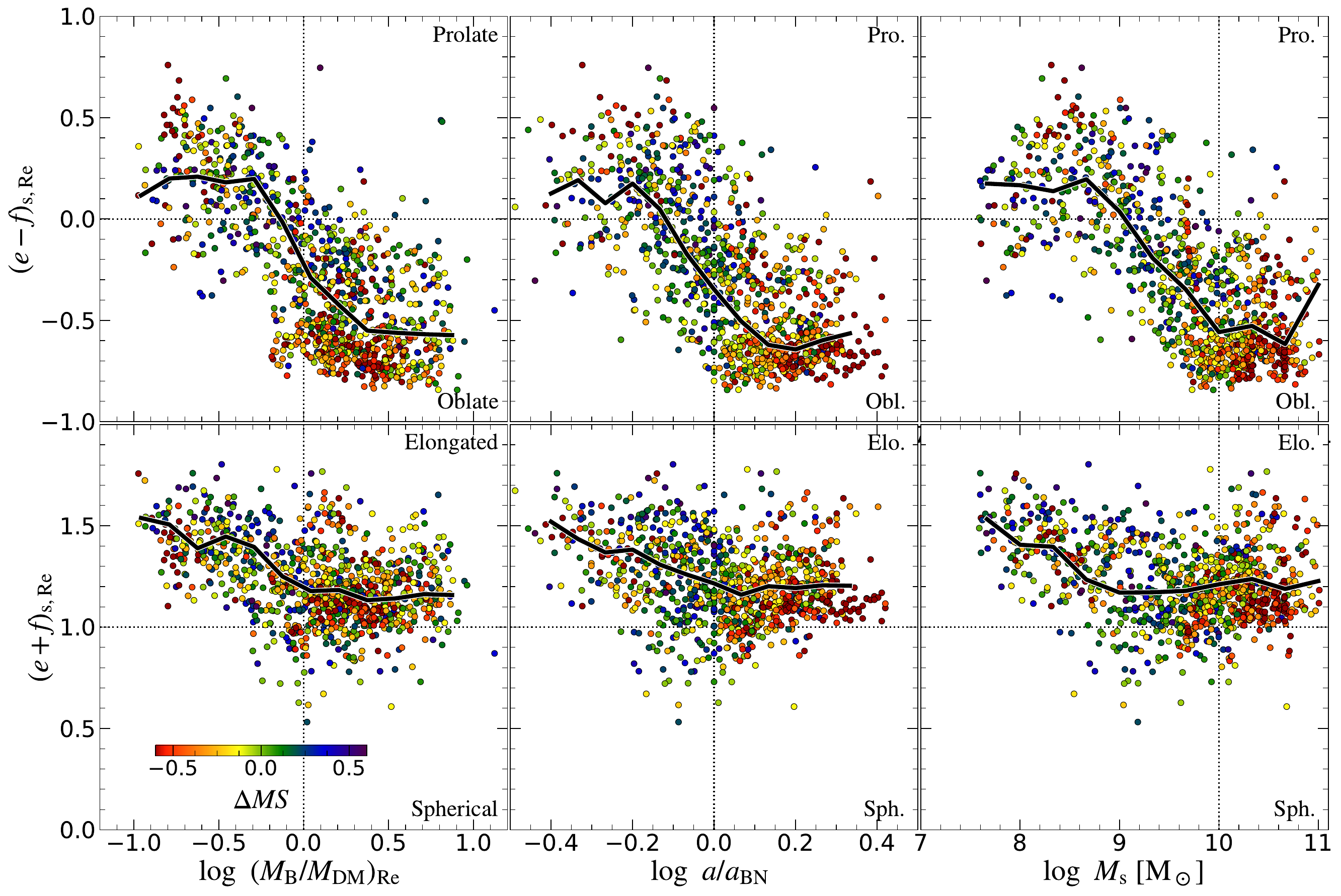}
\caption{
Transition in shape. We characterize the shape of the stellar distribution within $\Re$ by the following two parameters: (a) $e-f$ refers to deviations from triaxiality, positive and negative for prolate and oblate configurations, respectively. (b) $e+f$, referring to variation along the triaxial line $e-f\seq 0$, from elongated at large $e+f$ to spherical at low $e+f$. The shape is plotted against the baryon-to-DM ratio within $\Re$, $a/\abn$, and $\Ms$. The colour marks the deviation from the universal MS ridge.
{Pre-BN, the stellar systems are elongated, triaxial and prolate. Post-BN, the systems become less elongated and oblate, tending towards a disc-like structure.}
}
\label{fig:shape}
\end{figure*}

\begin{figure*} 
\includegraphics[width=0.8\textwidth]{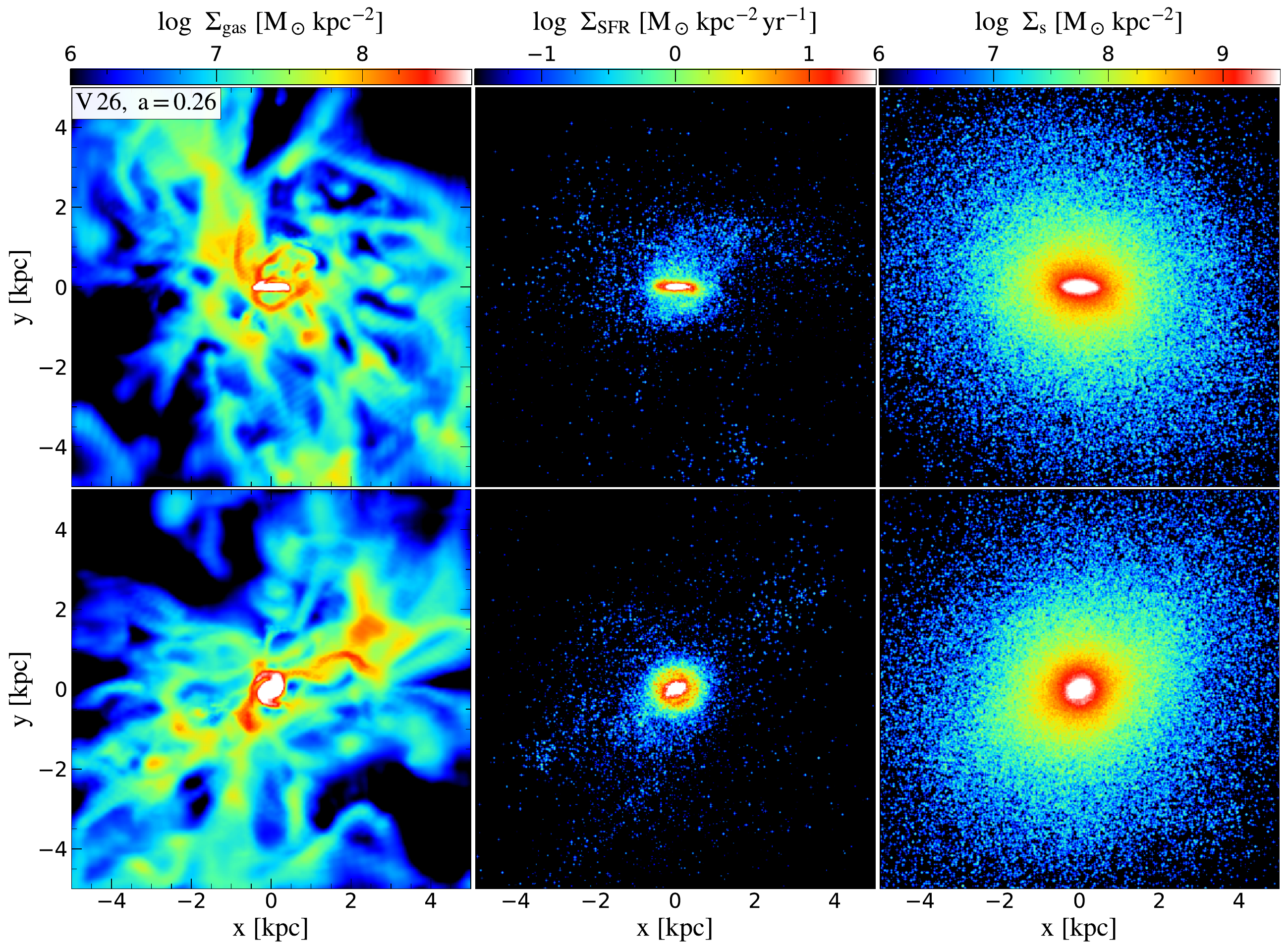}
\caption{
The shape of a typical BN is shown in the projected density (from left to right) of the gas, SFR and stars, edge-on (top) and face-on (bottom). The sub-kpc BN is a disc in gas and SFR (or young stars) and a flattened spheroid in stars.}
\label{fig:shape_bn}
\end{figure*}

\subsection{Transition in Shape}

The BN phase is associated with a transition in the global shape of the stellar system, from elongated-prolate before the BN event to oblate and disc-like structure post-BN. This has been analysed in the same simulations in \citet{ceverino15_shape} and \citet{tomassetti16}, explaining the observational trends with redshift and mass \citep{vanderwel14_shape}, based on the transition from DM to baryon dominance in the core caused by the compaction, discussed in \se{DM-B}.

\smallskip
The DM halo is expected to be elongated and prolate as a result of its assembly along a dominant large-scale filament. This is demonstrated by a correlation between the velocity-dispersion anisotropy and the large-scale geometry of structure associated with the merger history \citep[][Figs.~12-15]{tomassetti16}. The stellar system follows the elongation and prolateness as long as the DM halo dominates the potential in the central galaxy. The torques exerted by the elongated halo indeed indicate that the halo is capable of inducing the elongated stellar shape \citep[][Fig.~16]{tomassetti16}.

\smallskip
Once a compact baryonic core forms and dominates the inner potential, it is expected to make the inner system of old stars and DM rounder by deflecting small-pericenter box orbits. Then, the system of new stars follows the new high-AM gas into an oblate rotating system \citep[][Fig.~11]{tomassetti16}.

\smallskip
As in \citet{tomassetti16}, the shape within $\Re$ is quantified by fitting an ellipsoid to the spatial mass distribution with semi-axes $a\geq b \geq c$. The shape is characterized by the parameters of `elongation' and `flattening',
\be
e=[1-(b/a)^2]^{1/2} , \quad
f=[1-(c/b)^2]^{1/2} \, .
\label{eq:shape}
\ee
Fig.~2 of \citet{tomassetti16} illustrates the interpretation of shape based on location in the plane of $e$ and $f$. In particular, a large $e$ with a small $f$ corresponds to a prolate system, while a small $e$ with a large $f$ corresponds to an oblate system. Thus, the difference $e-f$ is a measure of prolateness versus oblateness. Triaxial systems span the diagonal line $e-f\seq 0$, with the sum $e+f$ measuring the degree of elongation from a sphere to very elongated triaxial systems. Fig.~7 of \citet{tomassetti16} shows the main features of the evolution of the shape of stars, gas and DM as a function of the ratio of baryonic to DM mass within $\Re$, which we denote here $\rm B/DM$.

\smallskip
\Fig{shape} addresses the evolution of stellar shape within $\Re$
through the BN phase by showing $e-f$ and $e+f$ as a function of three different quantities, $\rm B/DM$, $a/\abn$ and $\Ms$. Pre-BN, $e\!-\!f\ssim0.25$ and $e\!+\!f\ssim1.4$ with a scatter $\pm 0.2$, namely the systems are elongated, triaxial and prolate. Post-BN, we see $e\!-\!f\ssim-0.6$ and $e\!+\!f\ssim1.2$, namely, the systems are rounder and oblate, tending toward a disc-like structure. The very clear transition in shape occurs near the BN event, associated with the compaction-driven transition from DM to baryon dominance in the inner galaxy (see \fig{fdm_Re}).

\smallskip
{The colour in \fig{shape}, which marks $\DMS$, predicts a mild gradient of shape across the MS for galaxies above the BN mass, where the stellar systems become more oblate at the lower parts of the MS. In the post-BN phase, the most oblate and least elongated regimes in \fig{shape} tend to have a high density of red symbols, namely oblate and less elongated systems tends to be below the MS. However, except for this tendency, the colours do not show a clear correlation with the shape.
Indeed, in \figApp{shape_dMS_app} (see also \tabApp{gradients_dMS_app}), we show that after the elimination of redshift and mass dependence, there are no gradients across the MS in the $f$ parameter for the post-BN galaxies, the $e$ parameter shows a very mild gradient with a best-fitting slope of $+0.15\pm0.03$ ($\rp\ssim 0.3$).}

At the BN phase itself, focusing on the inner $1\kpc$, the typical stellar shape parameters are $e\!-\!f\ssim-0.3$ and $e\!+\!f\ssim1.2$, namely an oblate spheroid. \Fig{shape_bn} shows the projected mass density in one galaxy at its BN stage, referring to gas and young stars in addition to the stellar system, edge-on and face-on, with respect to the angular momentum. While the inner stellar system is a flattened spheroid, the gas and young stars populate a compact, thick disc. This appearance is typical for our simulated galaxies at the BN phase. This is a prediction for high-resolution sub-kpc observations of gas or young stars, such as expected from ALMA.

\subsection{Comparison to Observations}
\label{sec:shape_obs}

Constraining the intrinsic, three-dimensional shape from the observed two-dimensional shape is complicated by projection effects. However, it is possible to determine the distribution of shapes under certain simplifying assumptions.

\smallskip 
In the present-day universe, star-forming galaxies of all masses $10^9-10^{11}\Msun$ are predominantly thin, nearly oblate discs \citep[e.g.,][]{sandage70, lambas92, padilla08}. At higher redshifts, \citet{ravindranath06} demonstrated that the ellipticity distribution of a large sample of $z\seq2-4$ Lyman break galaxies is inconsistent with randomly oriented disc galaxies, lending credence to the interpretation that a class of intrinsically elongated (prolate) objects exists at high redshift. By modelling ellipticity distributions, \citet{law12} and \citet{yuma12} concluded that the intrinsic shapes of $z\sgt 1.5$ star-forming galaxies are strongly triaxial.

\smallskip 
\citet{vanderwel14_shape} determined the intrinsic, three-dimensional shape distribution of star-forming galaxies at $0\slt z\slt  2.5$, as inferred from their observed projected axis ratios from SDSS and CANDELS data. They find that among massive galaxies ($\Ms \sgt 10^{10}\Msun$), discs are the most common geometric shape at all $z\slt 2$. However, lower-mass galaxies at $z\sgt 1$ possess a broad range of geometric shapes: the fraction of elongated (prolate) galaxies increases toward higher redshifts and lower masses. Galaxies with stellar mass $\Ms \approx 10^{10}\Msun$ are a mix of roughly equal numbers of prolate and oblate galaxies at $z\ssim2$. Similarly, \citet{takeuchi15} find that the shape of the star-forming galaxies in the main sequence changes gradually and that the round discs are established at around $z\ssim0.9$. \citet{zhang19} improved the analysis using CANDELS galaxies. By approximating the intrinsic shapes as triaxial ellipsoids and assuming a multivariate normal distribution of galaxy size and two shape parameters, they construct the intrinsic shape and size distributions to obtain the fractions of prolate, oblate, and spheroidal galaxies in each redshift and mass bin. They confirmed the finding that galaxies tend to be {prolate at low $\Ms$ and high redshifts, and oblate at high $\Ms$ and low redshifts.} Overall, these analyses of observed shapes qualitatively agree with the simulation predictions, where the lower-mass, pre-compaction objects tend to be prolate systems before they become oblate after compaction.

\smallskip 
The images of \fig{shape_bn} illustrate that the blue nuggets within the inner $1\kpc$ are puffy discs.  This is a prediction to be tested with high-resolution, sub-kpc observations, e.g., by ALMA or JWST.

\section{Kinematics}
\label{sec:kinematics}

\begin{figure*} 
\includegraphics[width=0.95\textwidth]{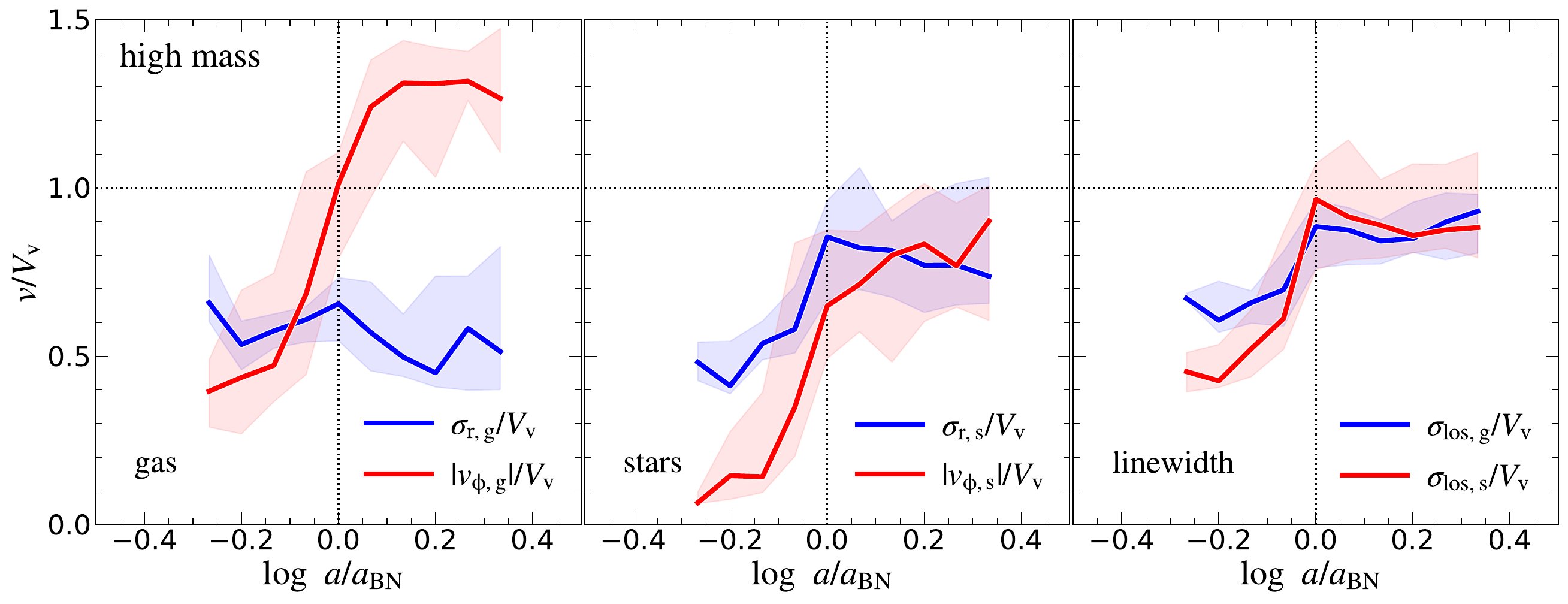}
\caption{
{Kinematic properties versus $a/\abn$ for the high-mass subsample (see same properties as a function of $\Ms$ in \figApp{kin_Ms}).} All velocities are measured in units of $\Vv$ at the BN event (which is about $125\kms$, see \fig{char_M-a_event}). Shown are the rotation velocity and the radial velocity dispersion for the gas (left) and stars (middle), as well as the dispersion along the line of sight (linewidth) for gas and stars. {Pre-BN, the galaxies are dispersion dominated, reflecting the highly perturbed structure of the galaxy at this phase. During compaction, the rotation velocity increases significantly, transitioning into rotationally supported gas systems. Post-BN, the gas is typically in an extended, perturbed ring/disc with $\vrot/\sigma\ssim3$.}
}
\label{fig:kin}
\end{figure*}

\begin{figure*} 
\includegraphics[width=0.97\textwidth]{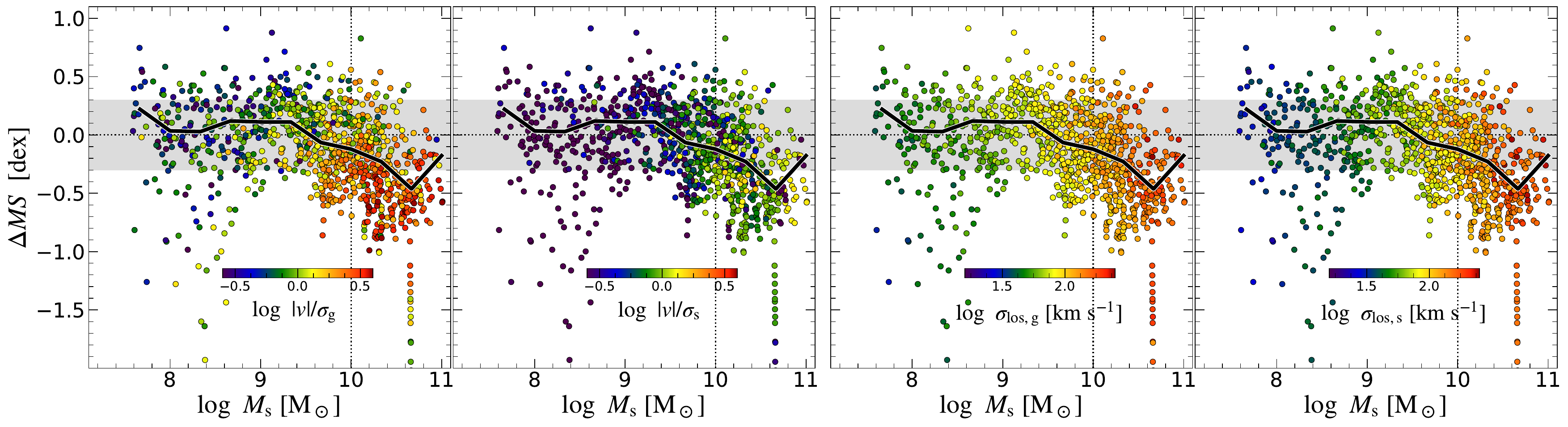}
\caption{
Kinematics with respect to the MS. All snapshots are shown in the plane of $\DMS$ versus $\Ms$. The colour refers to $\Vrot/\sigma$ and $\slos$ for gas and for stars in the galaxy.
}
\label{fig:MS_kin}
\end{figure*}

\begin{figure*} 
\includegraphics[width=0.95\textwidth]{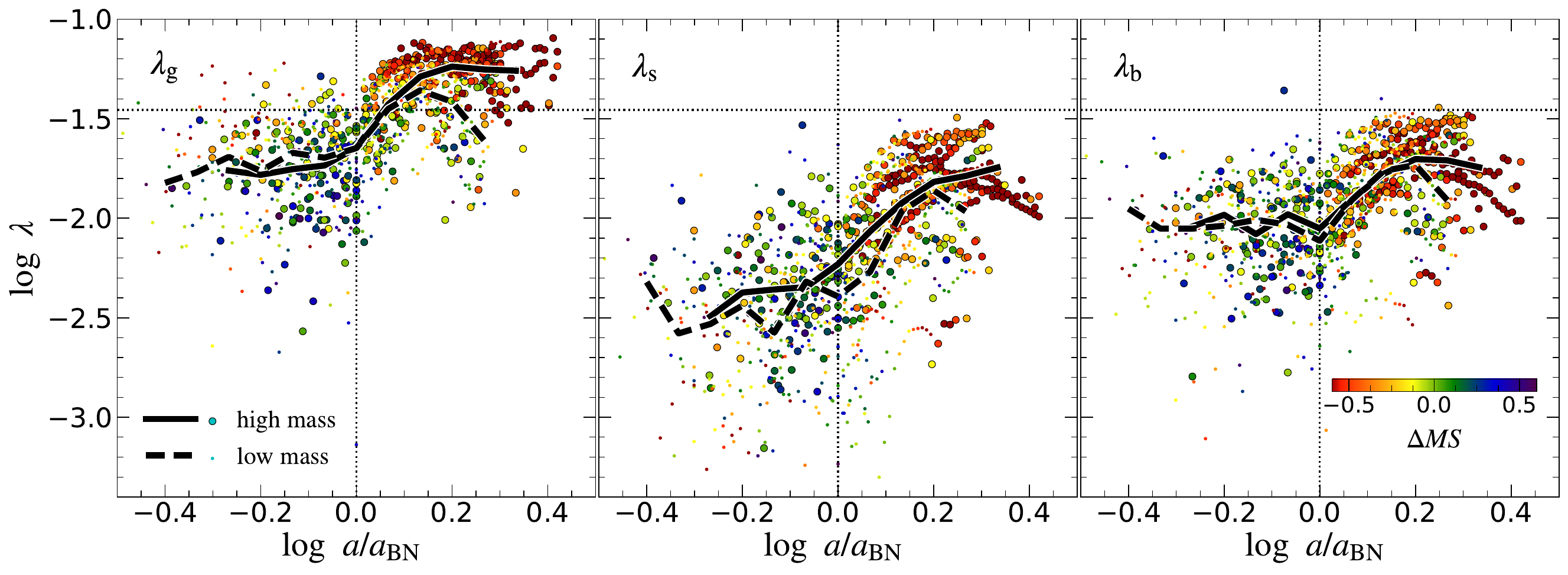}
\caption{
Galaxy spin parameter for gas, stars and baryons, as a function of $a/\abn$. Colour refers to $\DMS$. The medians in bins of $a/\abn$ refer to the high-mass (solid) and low-mass (dashed) subsamples (defined at $z\seq 2$). The horizontal line refers to $\lambda\seq 0.035$, the typical halo spin parameter. {The spins are roughly constant pre-BN. Following the BN phase, the spin steeply rises, saturating to a plateau at $a \ssimeq 1.5 \abn$. The high spin is associated with the formation of a new extended gaseous ring originating from fresh inflow through the streams.}
}
\label{fig:spin}
\end{figure*}

\subsection{Transition in Kinematic Properties}
\label{sec:kin_trans}

{\Fig{kin} shows basic kinematic properties within $0.1\Rv$ as a function of $a/\abn$ for the high-mass subsample (see the complementary \FigApp{kin_Ms} as a function of $\Ms$).} Shown for gas and for stars are the rotation velocity $\Vrot$, and the radial velocity dispersion $\sigr$. Also shown is the line-of-sight velocity dispersion $\slos$, observable as a linewidth, averaged over {64 random directions for the line of sight. The velocities are normalized by the virial velocity $\Vv$ and stacked together.} The same kinematic properties for four example galaxies were shown in \citet[][Figs.~10-11]{zolotov15}.

\smallskip
The pre-compaction galaxies are dispersion dominated, both for the gas and the stars, reflecting the intense feeding by streams, including mergers and the highly perturbed structure of the galaxy. The velocities are $\simi (0.4-0.6) \Vv$ for the gas and $\simi (0.2-0.4)\Vv$ for the stars.
During the compaction, the rotation velocity increases significantly. Post-BN, the gas is mostly in an extended, perturbed ring/disc with $\vrot/\sigma\ssim3$. The stars are spheroid dominated with an overall $\vrot/\sigma\ssim1$, while the {kinematically selected stellar disc component (not shown here) has $\vrot/\sigma\ssim3$, similar to the gas.} This results in a jump in the line-of-sight velocity dispersion $\slos$, which has important contributions from both $\vrot$ and $\sigma$. High line-of-sight velocities are indeed observed in BNs and RNs \citep{barro14_kin,barro15_kin,barro16_kin}.

\smallskip
For galaxies above the critical BN mass, the simulations show for the gas disc a median rotation velocity {$\simi 1.3\Vv$}, and a velocity dispersion of $\simi 0.6\Vv$. For the stars, both are $\simi (0.7-0.8) \Vv$. This results in $\slos \ssim (0.8-1) \Vv$ both for the gas and the stars. Galaxies with masses below the critical mass are expected to have $\slos \ssim 0.6 \Vv$ for the gas and $\simi 0.4 \Vv$ for the stars.

\smallskip
\Fig{MS_kin} shows the kinematic properties of galaxies with respect to their position on the MS in the plane of $\DMS$ versus $\Ms$. Shown by colour are $\Vrot/\sigr$ and $\slos$ for gas and stars within {$0.1\Rv$}. In both quantities, we see a steep rise along the MS (namely as a function of $\Ms$) near the critical BN mass, reflecting the time evolution involving a steep rise during the compaction phase (\fig{kin}). For the gas, above the critical BN mass, we see a strong gradient of $\Vrot/\sigr$ across the MS (as a function of $\DMS$), from about unity at the top of the MS to $3-5$ in the lower part of the MS. This gradient is due to the gradual appearance of an extended gas ring or disc after the BN phase. The gradient is largely smeared out in $\slos$ when averaged over all inclinations, showing mostly a gradual correlation with $\Ms$.
{In \fig{gradients_q-dMS} we show the gradient across the MS for $\vosg$ (third row, left panel). After eliminating systematic redshift and mass dependencies, we find a slope of $-0.31\pm0.03$ with a rather tight correlation, $\rp\ssim 0.54$.} Thus, an observational detection of the gradient across the MS would require an estimate of $\Vrot/\sigr$, namely a decomposition of $\Vrot$ and $\sigr$ from a two-dimensional line-of-sight velocity map. The kinematics of stars do not show a gradient across the MS, as both $\Vrot$ and $\sigr$ grow together during the compaction phase (\fig{kin}).

\smallskip
The evolution through the BN phase is also associated with changes in the spin parameter of the galaxy, reflecting the angular-momentum loss associated with the compaction and the buildup of a new extended gaseous disc/ring. We define the spin parameter of component $i$ {(gas, stars, or baryons within a given volume) by $\lambda_i\seq  (J_i/M_i)/(\sqrt{2}\Vv\Rv)$,} where $J_i$ and $M_i$ are the angular momentum and mass of that component. The volume considered here is for the entire galaxy, a sphere of radius {$0.1\Rv$.}

\smallskip
\Fig{spin} shows the evolution of the gas spin parameters in the entire sample as a function of $\aabn$ {(see the spin parameter as a function of $\Ms$ in \figApp{spin_Ms})}. Shown are all snapshots of all galaxies, with the colour referring to $\DMS$, and the two thick lines are medians in bins of $\aabn$ for the subsamples of high-mass and low-mass (ranked at $z\seq 2$). The spins are roughly constant pre-BN. The angular-momentum loss associated with the central compaction is marginally indicated on the scale of the whole galaxy. Following the BN phase, the spin steeply rises, saturating to a plateau at $a \ssimeq 1.5 \abn$. This high spin is associated with the formation of a new extended gaseous ring from a freshly accreted inflow.
The violent disc instability in this ring is partly suppressed by `morphological' quenching due to the massive bulge \citep{martig09}, preventing a torque-driven shrinkage of the ring and keeping it extended { (see \se{ring})}.
{A post-BN plateau in the spin parameter} is typical for the high-mass subsample, where the BN identified is indeed the last major BN event {which is followed by a prolonged quenching process} (yellow and red colours in the figure) with no further significant compaction events. {In the low-mass subsample, some galaxies may go through one or more compaction events after the identified BN, which triggers central SFR rejuvenation and lower the galaxy spin.} 

\smallskip
{The gradient across the MS of $\lambdag$ is shown in \fig{gradients_q-dMS} (third row, middle panel). Low values of $\lambdag\ssim1$ are seen above the MS ridge. When the central region is depleted from gas, and the galaxy is driven below the MS toward the green valley, we see higher values of $\simi2$ reflecting the buildup of an extended gas ring. We find a best-fitting slope of $-0.35\pm0.04$ and a fairly tight correlation, $\rp\ssim0.62$.}

\subsection{Jeans Equilibrium and Dynamical Mass}
\label{sec:jeans}

\begin{figure*} 
\includegraphics[width=0.455\textwidth]{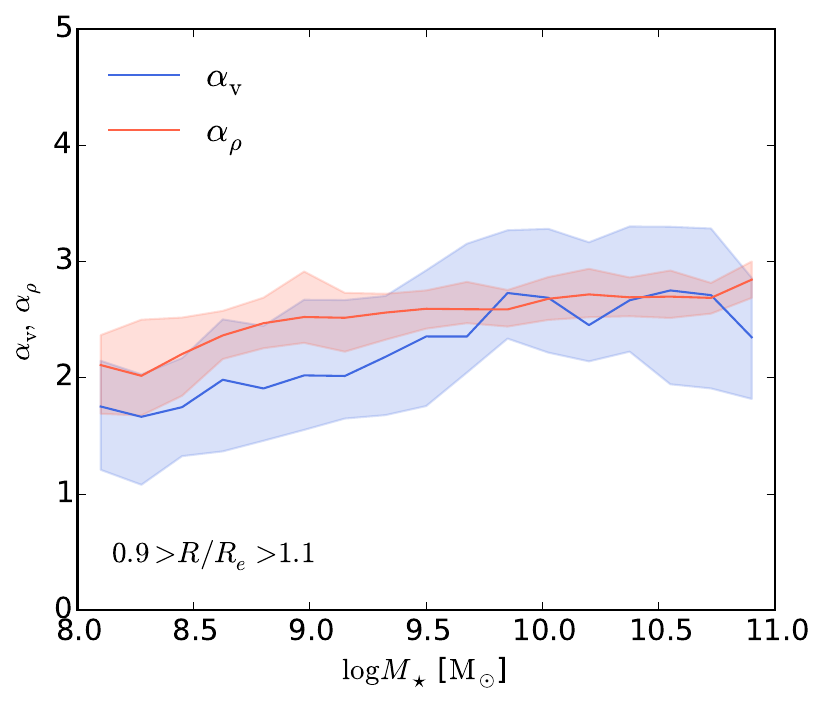}
\includegraphics[width=0.445\textwidth]{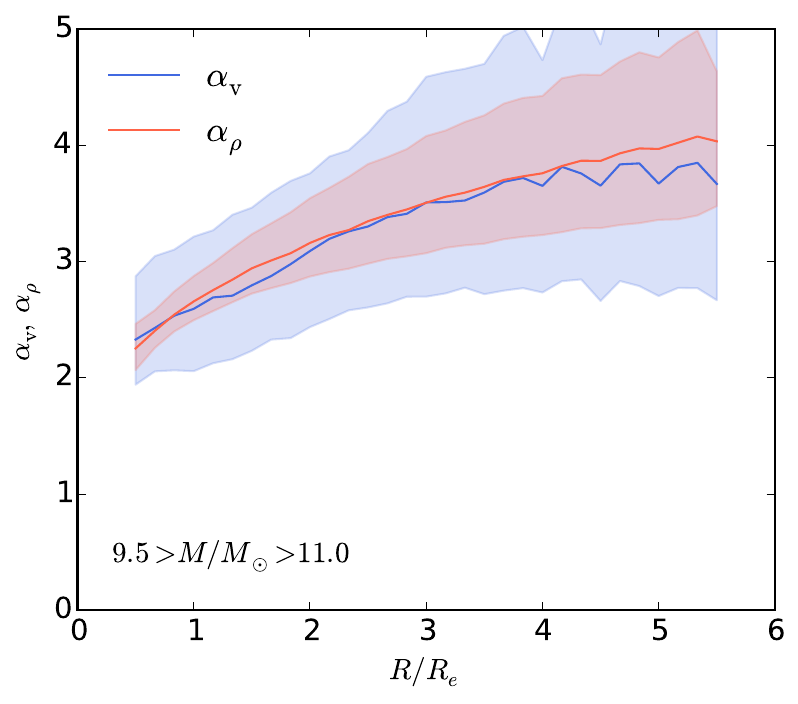}
\caption{
Validity of the simplified Jeans equation for stars. Shown are the value of $\alphav$ and $\alpharho$, derived from \equ{jeans} and \equ{alpha}, respectively. For Jeans equilibrium with isotropy and $\sigr\seq\const$, one expects $\alphav\seq \alpharho$. {\bf Left:} at $\Re$ as a function of stellar mass. The simplified Jeans equation is valid in the massive galaxies post-BN, with $\alphav\seq 2.6\pm 0.5$. {\bf Right:} for post-BN massive galaxies as a function of radius. The simplified Jeans equation is valid out to $5\Re$, with $\alpha$ increasing with radius.
}
\label{fig:jeans}
\end{figure*}

Observational estimates of the total (dynamical) mass within a given radius (say $\Re$) are based on observed kinematic properties and the assumed validity of Jeans (or hydrostatic) equilibrium for stars (or gas) along the radial direction in the major plane. This serves, for example, for evaluating the fractions of DM and baryons within this radius, which we already know is expected to change drastically during the compaction phase. Unfortunately, the kinematic observational estimates commonly yield a dynamical mass which is larger than the baryonic mass, indicating that either the system is out of equilibrium or the way the Jeans (hydrostatic) equilibrium is applied is erroneous. In particular, the ability to measure the dynamical mass, namely the validity of Jeans (or hydrostatic) equilibrium, may vary with $a/\abn$ and with $r/\Re$.

\smallskip
The Jean equation in the plane is 
\be
\vcirc^2 = \vrot^2 + \alpha \sigr^2 ,
\label{eq:jeans}
\ee
where $\vcirc$ is the circular velocity derived from the potential ($\vcirc^2 \seq  G M(r)/r$ for a spherical mass distribution), $\vrot$ is the rotation velocity in the given plane, and $\sigr$ is the radial velocity dispersion, both approximated as constants in the relevant range of radii. The value of $\alpha$ depends on the density profile,
\be
\alpha = -\frac{\dd \ln \rho}{\dd \ln r} .
\label{eq:alpha}
\ee
For an isothermal sphere $\alpha\seq 2$. For a sphere with a projected \sersic profile of index $n$, $\alpha$ at $\Re$ ranges from $\simeqi 2$ for $n\seq 1$ to $\alpha \ssimeq 2.8$ for $n\seq 4$, and it is increasing with the radius within the galaxy. For a thin self-gravitating exponential disc with an exponential radius $\rd$ ($\Re \seq  1.67 \rd$) one predicts a steep rise with radius, $\alpha \seq  2(r/\rd)$ \citep{burkert10}.

\smallskip
{Using both analytic considerations and simulations, \citet{kretschmer21} provided a prescription to evaluate the dynamical mass from kinematic measurements of stars or gas.} The validity of the simplified Jeans equation for stars is evaluated in \fig{jeans}, which shows the values of $\alphav$ and $\alpharho$ as derived from \equ{jeans} and \equ{alpha}, respectively. {When plotted at $\Re$ against stellar mass {(left)}, one can see that the Jeans equation is approximately valid for masses above the critical BN mass; namely post-BN, where the median values of $\alpha$ deviate by less than 10 per cent, and the scatter in $\alphav$ is $\pm 20$ per cent. However, for masses below the BN mass, the medians of the two $\alpha$s deviate by $\simi 25$ per cent, and the scatter in $\alphav$ is $\pm 25$ per cent, indicating somewhat larger deviations from Jeans equilibrium in the highly perturbed pre-BN galaxies. When the $\alpha$ values are plotted for the high-mass galaxies against radius {(right)}, one sees that the medians agree to within 5 per cent {out to $\simi 5\Re$}.
} 

\smallskip
The value of $\alpha$ varies with mass and especially with radius. At $\Re$, for galaxies $\geq 10^{10}\msun$ where Jeans equilibrium is valid, we measure $\alphav\seq 2.6\pm 0.5$. This is consistent with the values obtained for a spheroidal system with \sersic indices in a wide range of about $n \ssim 2$ and is smaller than what is expected from a thin disc. For the galaxies below the BN mass, at $\Ms \ssim 10^9\msun$, where Jeans equilibrium is a cruder approximation, $\alphav \seq  2.0 \pm 0.5$. For the massive galaxies post-BN, $\alpha$ is increasing with radius, from $\alphav\seq 2.5\pm 0.5$ at $r\seq \Re$, through $\alphav\seq 3.0\pm 0.7$ at $r\seq 2\Re$ to $\alphav\seq 3.5\pm 1.0$ at $r\seq 4\Re$. This is significantly shallower than the linear rise with radius predicted for a thin self-gravitating disc.

\smallskip
{For the gas, when plotted at the gas half mass radius ($R_{\rm e, gas}$), Fig.~5 in \citet[][]{kretschmer21} showed that hydrostatic equilibrium is achieved in massive galaxies with $\Ms \sgt 10^{9.5} \msun$ only if corrections for non-constant velocity dispersion and deviation from spherical potential are included. The large errors in the gas \citep[see Fig.~2 and 5][]{kretschmer21} may indicate deviations from equilibrium. When $\alphav$ and $\alpharho$ are plotted against radius in Fig.~2d in \citet{kretschmer21}, we see that the agreement between the medians indicates a hydrostatic equilibrium in post-compaction galaxies out to $3.5R_{\rm e, gas}$.}

\subsection{Comparison to Observations}
\label{sec:kinematics_obs}

Information on galaxy kinematics at $z\ssim2$ has become available through integral-field spectroscopy of optical emission lines such as H$\alpha$, tracing the ionized gas \citep[see][for a detailed review]{glazebrook13}. A key point that has emerged from these studies is that substantial fractions of high-redshift galaxies have regular kinematics despite irregular photometric morphologies; this is likely due to the presence of a large number of highly gas-rich discs. Particularly, it appears that at least {50 per cent} of $10^{10}-10^{11}\msun$ SFGs on the $z\ssim2$ MS are rotationally supported structures with typical rotation velocities of $100-300\kms$, and with high values of velocity dispersion $\sigma\ssim 50-100\kms$ \citep{forster06, genzel06}. Using the ratio of circular rotation velocity to dispersion $v_{\rm rot}/\sigma$ as a tracer of the kinematics, larger galaxies ($\Ms\sgt 5\times10^{10}\msun$) have typically $v_{\rm rot}/\sigma\seq 1-10$ at $z\ssim2$ \citep{forster09,genzel11,gnerucci11,jones10,law09,swinbank12,wisnioski15_KMOS,swinbank17,forster18}. These observations are in agreement with our simulated galaxies above the critical mass, which have $v_{\rm rot}/\sigma\ssim 3$ due to the post-BN extended gas ring or disc.

\smallskip 
At lower masses, at $10^{9}-10^{10}\msun$, the abundance of dispersion-dominated SFGs increases significantly. This is again consistent with our simulated, pre-compaction galaxies being dispersion dominated for both the gas and the stars. However, caution is advised because a significant contribution to the velocity dispersion deduced from observations may be due to instrumental broadening and beam smearing \citep{newman13}, so observations at higher resolution (e.g. with adaptive optics) are needed to shed more light on the true nature of these dispersion-dominated systems.

\smallskip
{Recent observation of massive $10^{10}\msun$ galaxies at very high redshifts, $z\sgt 4$, reveal the existence of dynamically cold discs \citep{smit18, neeleman20, lelli21, rizzo20, rizzo21, Herrera-Camus22}. We find that the gas in post-BN galaxies, namely galaxies above the critical mass, tend to be rotationally dominated with typical values similar to observed massive galaxies at $z\ssim2$ of $\vos\ssim3$, consistent with some observations \citep{smit18, neeleman20, Herrera-Camus22}.
Several studies find indications for surprisingly large values of $\vos\ssim 7-30$ \citep{lelli21, rizzo20, rizzo21,fraternali21}, well above the expected values from the evolution of observed SFGs between $z\ssim0-4$. While the differences between observational studies could be partly caused by a difference in the nature of these sources, the different implemented modelling approaches may also affect the results. In \citet{Kretschmer22}, it was shown using {MIGA} simulation \citep{kretschmer20} that $\vos$ is highly dependent on the tracer used in the analysis. Namely, while moderate values of $\vos\ssim3$ were found when using HI gas, very high values of $\simi8$ were obtained for molecular CO or $\rm H_2$, which tend to reside closer to the mid-plane.}

\section{Metallicity}
\label{sec:metallicity}

\begin{figure} 
\includegraphics[width=0.95\columnwidth]{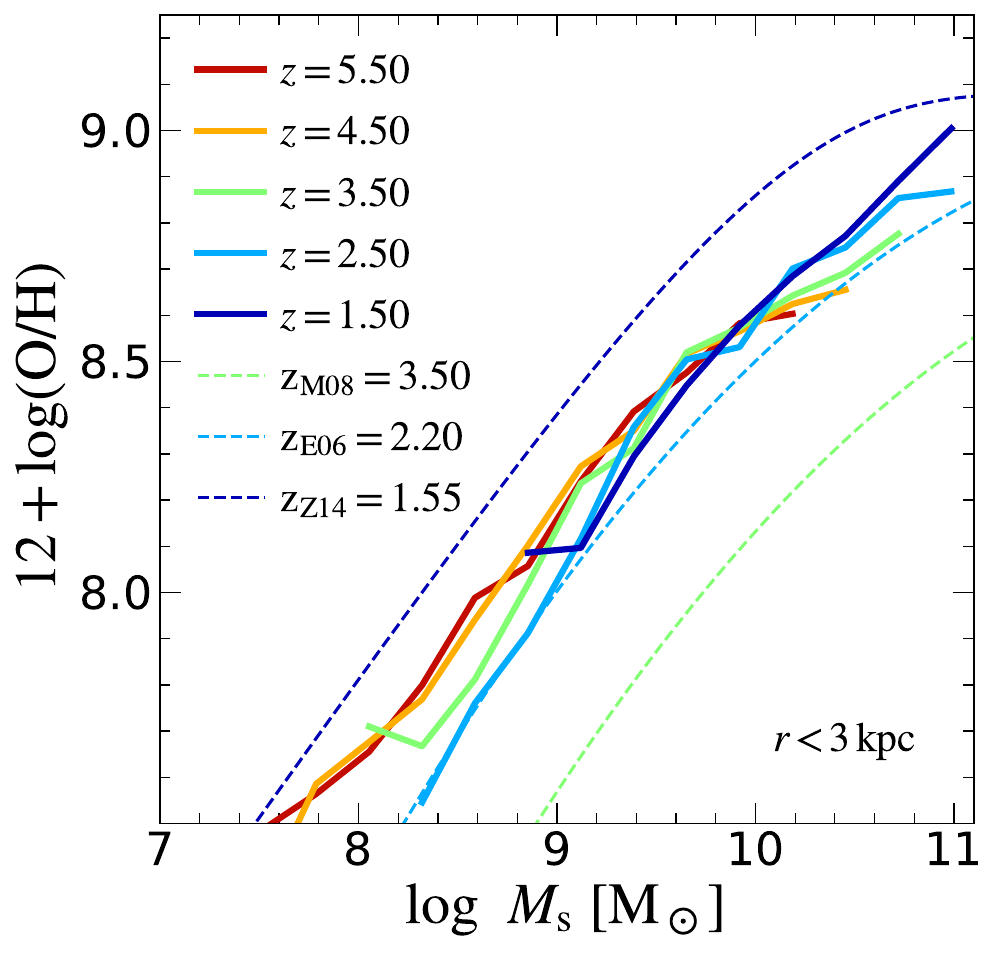}
\caption{
Gas metallicity versus stellar mass in different redshift bins. The metallicity is measured within the inner $3\kpc$. Dashed lines show fits of observational data at redshifts: $z\seq 1.55$ from \citet{zahid14a}, $z\seq 2.2$ from \citet[fits are taken from \citealt{maiolino08}]{erb06}, $z\seq 3.5$ from \citet{maiolino08}. {At a given redshift, we find a continuous rise of $Z$ with $\Ms$ for masses below the critical BN mass and a shallower rise for masses above the BN mass. The simulations do not show evolution with redshift in the zero point for a given mass, and the shoulder remains near the same mass at all redshifts, defined by the critical BN mass.}
}
\label{fig:Z-M}
\end{figure}

\begin{figure} 
\includegraphics[width=0.95\columnwidth]{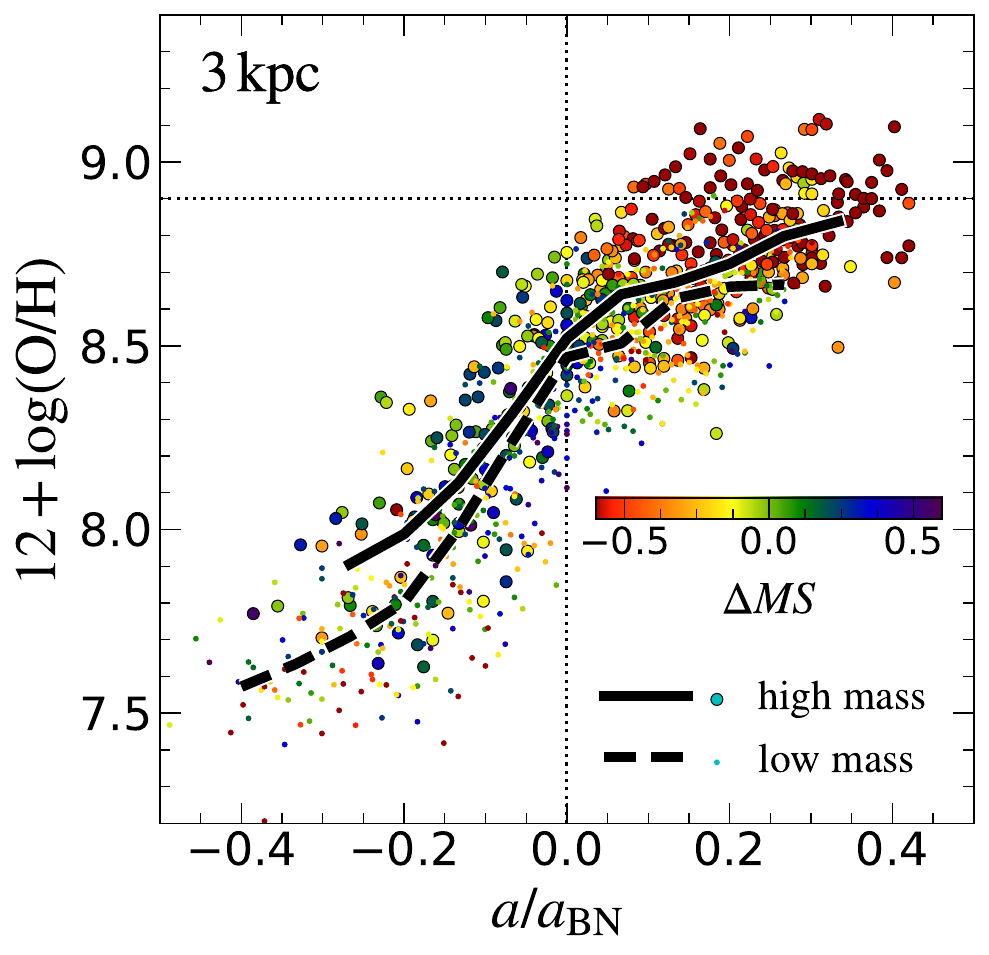}
\caption{
Gas metallicity as a function of $a/\abn$. The metallicity is measured within the inner $3\kpc$. The colour indicates the distance from the MS, $\DMS$, the medians refer to the high-mass and low-mass subsamples (defined at $z\seq 2$), and the horizontal dotted line marks solar metallicity ($0.02\,\rm Z_\odot$ or $\rm 12+\log(O/H)\ssim8.9$). {The metallicity sharply rises during the compaction phase and transitions into a shallower slope post-BN. In \figApp{Z-abn_app} we show the metallicity measured within $1\kpc$ and $0.1\Rv$.}
}
\label{fig:Z-abn}
\end{figure}

\subsection{Transition in the Metallicity-Mass relation}

{In \fig{Z-M}, we show the distribution of our simulated galaxies in metallicity-mass relation $Z-\Ms$, in bins of redshift (distinguished by colour). To crudely mimic observations, the gas metallicity is measured within $3\kpc$ of the galaxy centre. As suggested in \citet{maiolino08}, the volume selection is motivated by the typical aperture size of $\simi 6\kpc$ in several surveys \citep{savaglio05, erb06, maiolino08, zahid14a} observed within a wide range of redshifts ($z\ssim 0.7-3.5$). One can see at a given redshift a continuous rise of $Z$ with $\Ms$ for masses below the critical BN mass, with a linear or slightly sub-linear slope, and a shallower rise towards a plateau for masses above the BN mass. We do not find evolution in the zero point for a given mass, and the shoulder remains near the same mass at all redshifts, defined by the critical BN mass. The steep rise pre-BN reflects a continuous enrichment by star formation, while the post-BN flattening reflects saturation as the SFR is gradually suppressed.}

\smallskip
{\Fig{Z-abn} shows the evolution of gas metallicity as a function of $a/\abn$. The colour indicates $\DMS$. The two thick lines show the medians in bins of $a/\abn$ for the subsample of high-mass and low-mass galaxies (defined at $z\seq 2$). The metallicity is rising in the pre-BN phase. 
Post-BN, we see a transition to a shallower slope (measured within $3\kpc$). The gas metallicity within $1\kpc$ or $0.1\Rv$ (see \figApp{Z-abn_app}) shows a plateau after compaction. The metallicity in the inner $1\kpc$ tends to plateau due to the characteristic central gas depletion and low central SFR in the post-BN phase. At the galaxy scale of $0.1\Rv$, an extended disc/ring forms around the quenching nugget. During this phase, the expected increase in metallicity due to star formation in the disc may be compensated or overpowered by low-metallicity gas accreted from the cosmic web.
Although, on the one hand, there is an increase in the formation of metals as a result of star formation in the disc, the gas in the disc is diluted by low-metallicity accreted gas from the streams.}

\smallskip 
{The colour in \fig{Z-abn} indicates $\DMS$. We do not find a significant gradient across the MS after we remove the dependence of $z$ and $\Ms$. This is true when measured within $3\kpc$ as well as within $0.1\Rv$ (as seen in the two right panels of \figApp{Z_dMS_app}). However, the metallicity within the inner $1\kpc$ of the post-BN galaxies shows a significant gradient (see the left panel in \figApp{Z_dMS_app}). Moreover, when we compare the post-BN and the pre-BN galaxies, we find a sharp transition in the gradient across the MS occurring at the BN. While the pre-BN galaxies show no gradient across the MS and no correlation (listed in \tabApp{gradients_dMS_app}), the post-BN galaxies show a significant gradient with a slope of $-0.48\pm 0.02$ and a rather tight correlation with $\rp\ssim 0.66$.}

\subsection{Comparison to Observations}
\label{sec:metallicity_obs}

The characteristic shape of the distribution in the $Z-\Ms$ diagram at each redshift, as seen in \fig{Z-M}, is consistent with observations \citep{savaglio05, erb06, maiolino08, mannucci09, mannucci10,troncoso_maiolino14,zahid14a,zahid14b,wuyts14,sanders_shapley15,onodera_carollo16}. {We do not find evolution in the zero point for a given mass as seen in certain observations \citep{zahid14a,zahid14b,wuyts14}, it is closer to} evolution of galaxies along a universal track at all times. This may be due to the large uncertainties in the observations (e.g., metallicity calibration), which are performed differently at different redshifts. It may alternatively reflect an inaccuracy in the simulations, perhaps associated with the too-early star formation and the too-low gas fraction. The estimation of metallicity is especially uncertain because it is a ratio of two uncertain quantities, the mass in metals and the gas mass. Nevertheless, the universal shape in the $Z-\Ms$ diagram and the {characteristic turnover} at the critical BN mass is robust and confirmed in observations.

\section{Profiles} 
\label{sec:profiles}

\subsection{Transition in profiles}

\begin{figure*} 
\includegraphics[width=0.95\textwidth]{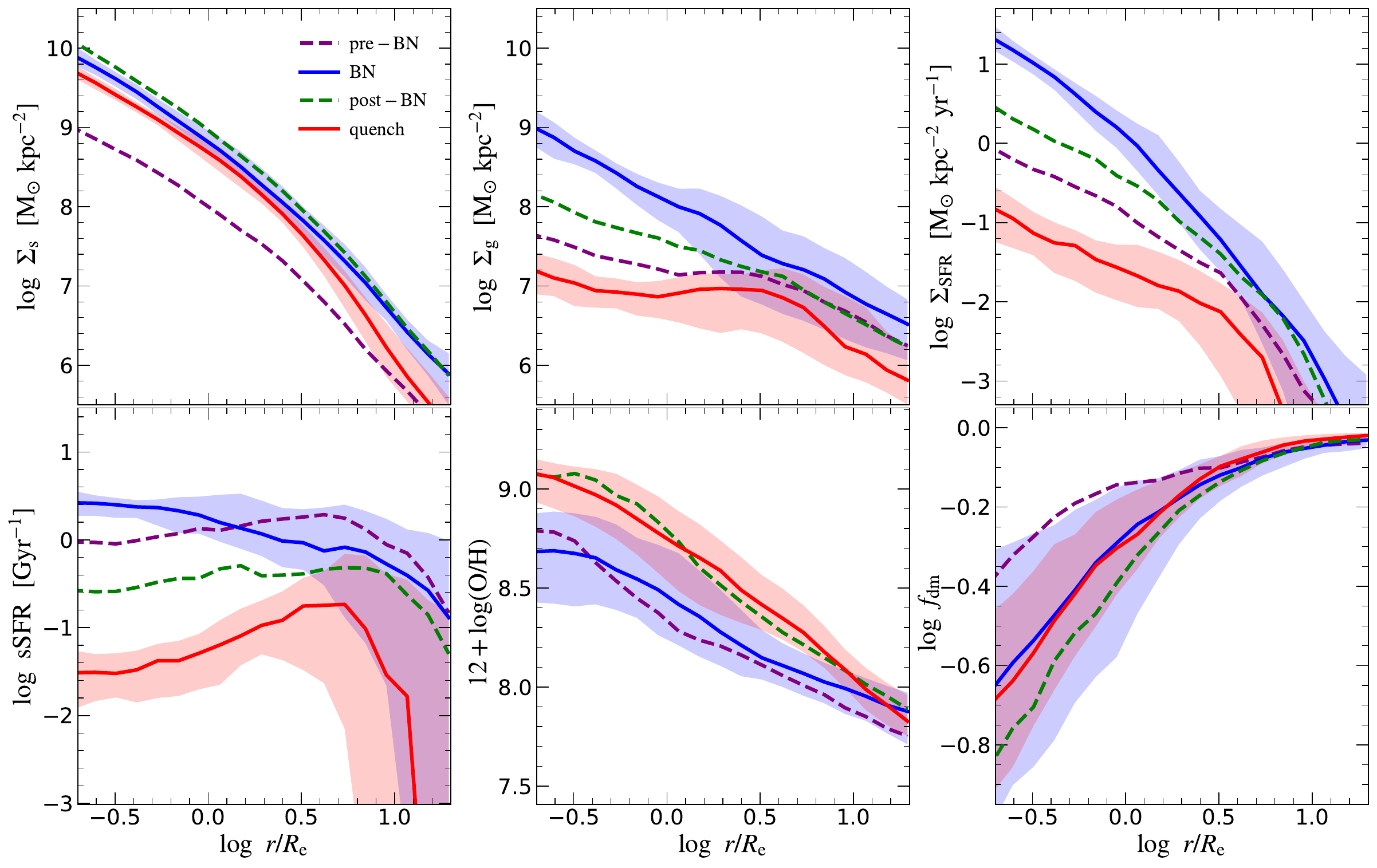}
\caption{
{Profiles of different quantities stacked at four galaxy evolutionary phases as a function of the radius with respect to $\Re$. {\bf Top, from left to right:} shown are the profiles of the stellar surface density ($\Sigs$), gas surface density ($\Sigg$) and SFR surface density ($\Sigsf$). {\bf Bottom:} The specific SFR ($sSFR$), gas metallicity ($\Zg$) and the dark matter fraction ($\fdm$). All quantities are measured within concentric spherical shells. The four phases are (1) pre-BN (dashed purple) measured at the onset of compaction. (2) BN (blue). (3) post-BN (dashed green). (4) quenching phase (red). The lines indicate the median of the profiles. The shaded areas show the $1\sigma$ scatter about the median of the BN and the quenching phase.} {Each individual profile of the quantities $\Sigs$, $\Sigg$, $\Sigsf$ and $sSFR$ is scaled to have the same median $\Sigma(r\slt \Re)$ of all the galaxies in each phase. We do not scale the y-axis in the gas metallicity and dark matter fraction before stacking (see raw profile medians without scaling in \figApp{profiles_r}).}
}
\label{fig:profiles_re}
\end{figure*}

\smallskip
The evolution through the phases of compaction, BN and quenching is reflected in the evolution of the profiles of the relevant quantities as a function of radius in the galaxy. This has been studied using the same simulations in \citet{tacchella16_prof}. We summarize these results here and add the profiles of gas metallicity and {DM fraction}. 

\smallskip
{\Fig{profiles_re} shows the medians of individual galaxy profiles stacked at times which refer to the four phases of evolution: pre-BN, BN, early post-BN, and the late post-BN quenching phase. The BN event is defined as described in \se{identify}, referring to the global peak of gas mass (and SFR) within the inner $1\kpc$. The pre-BN phase is defined at the onset of compaction, OC (see \se{identify}), where the central gas density starts to rise towards the peak of compaction, which typically occurs $\simi 300\Myr$ before the BN event. The early post-BN phase is taken to be the first snapshot where the galaxy is at the lower edge of the MS with $\DMS\slt -0.2$, which generally corresponds to the time when the inner $1\kpc$ starts to quench, typically $\simi 300\Myr$ after the BN event. Finally, the late post-BN quenching phase occurs when the galaxy drops below the lower edge of the MS, at its lowest $\DMS$, typically one of the latest snapshots of $z \ssim 1-2$.} {For all quantities except for the two mass fractions ($\Zg$ and $\fdm$), before stacking, we scale each individual galaxy profile to have the same median effective density ($\Sigma(r\slt \Re)$) of all the galaxies in each phase. In \figApp{profiles_r}, we show raw profile medians, scaling neither the x nor the y-axis.}

\smallskip
The key to the wet-compaction-driven BN phenomenon is the inner $\simi 1\kpc$ cusp that has developed in the gas density profile at the BN phase, {while the inner profile of both the pre-BN and post-BN shows a rather flat core. The central gas surface density (top, middle panel) at BN exceeds $10^{8}\msun\kpc^{-2}$ within the effective radius, while all the other phases are significantly lower. The pre-BN and late post-BN are lower by an order of magnitude or more in the innermost regions.}
%
{The SFR profile (top, right panel) closely follows the gas profile in accordance with the local SFR recipe in the simulations.} At the BN phase, the SFR profile is cuspy within the inner few kpc, and the central SFR density is higher by an order of magnitude than before the BN and late after the BN.

\smallskip
The sSFR profile tells us the relative activity of star formation at different radii. Pre-BN, the sSFR is weakly rising with radius, while at the BN, the central sSFR is higher by a factor of a few such that the profile is gradually declining. This reflects the vanishing of the outer gas disc into a compact gaseous and star-forming BN. It can be interpreted as an early `outside-in' quenching process associated with the compaction. {During the quenching phase,} the central sSFR becomes lower than at the BN phase by two orders of magnitude, while at {$2-6\Re$ it is only a factor of $\simi3-10$} below its value at the BN phase. The sSFR profile rises with radius, more and more so with time. This reflects the gas depletion in the core and the formation of an extended, gaseous, star-forming ring. It can be interpreted as `inside-out' quenching.

\smallskip
The self-similar growth of the stellar-mass density profile during the {pre-BN, BN, and post-BN} reflects the weak dependence of sSFR on radius during these phases. {We find little evolution in the central density during the three later phases (BN, post-BN and quenching phases), reflecting saturation of growth in the centre as the BN completes its main period of star formation and becomes a passive RN.} {In \figApp{profiles_r} we see that the median stellar density profile at the quenching phase is generally somewhat larger, as expected for a later stage in the evolution. However, when we scale the x-axis with respect to $\Re$, as seen in \fig{profiles_re}, the density at the quenching phase lies slightly below the BN and post-BN profiles. This reflects the central stellar growth saturation after compaction, combined with a typically larger $\Re$, by a factor of $\simi 2.5$, compared to the BN or post-BN phase (see \figApp{profiles_r} for the median $\Re$ in each phase).}

\subsection{Gas Phase Metallicity profiles}

{The gas metallicity profiles (\fig{profiles_re} bottom, middle panel) do not show significant evolution between the pre-BN and BN phases. Both display a relatively low sub-solar metallicity at all radii. The BN phase is characterized by the contraction of gas from the outskirts of the galaxy and intense star formation closely followed by SN feedback from the recently formed massive stars. As a result, competing processes may affect the metallicity during this phase: On one hand, intense star formation during compaction will cause a rise in metal production from explosions of massive stars. On the other hand, feedback from these SN explosions may expel the recently enriched gas. In addition, the ongoing contraction of gas may also help sustain low metallicity by diluting the centre as gas from the outskirts of the galaxy is funnelled inside.
Post-BN, we see enrichment by a factor of $\simi2$ above the BN phase in the inner regions $<\!\Re$, 
exceeding solar metallicity, while the metallicity in the outskirts remains relatively low. As a result, the profiles show a steeper decline with radius at the post-BN phases. 
The two post-BN phases have very similar metallicity profiles when we scale the radius by $\Re$ (\fig{profiles_re}), with a hint of a steeper slope around $1\sdash 3\Re$ in the post-BN compared to the quenching phase. For a given radius (\figApp{profiles_r}) at the quenching phase, we see metallicity of $\simi0.3\sdash 0.4$dex above the BN phase in the outskirts ($2\sdash 6\kpc$), while the metallicity in the post-BN phase is only $<\!0.1$dex above it.
The post-BN phase is taken to be the time in which the galaxy goes down to the lower edge of the MS, namely about the time it is processing toward the `green valley'. After this stage, we typically find that the centre becomes increasingly passive and later often surrounded by a gaseous, low-metallicity, and star-forming ring. 
The metal production increase due to star-formation in the post-compaction ring competes with dilution by accretion of high angular-momentum fresh gas from cold streams (see \se{ring}). Although the metallicity for a given radius during the time between the post-BN and quenching phase increased in the galaxy outskirts, one can still see that in both post-compaction phases, the metallicity is declining with a steeper  slope than that of the BN, similar to the behaviour seen in \fig{profiles_re}.}

\subsection{Dark matter Fraction and Declining Rotation Curves}
\label{sec:rot_curves}

{The profiles of dark matter fraction, $\fdm(r)$, are shown in \fig{profiles_re} (bottom right) and \figApp{profiles_r}. $\fdm(r)$ is measured within thin concentric spherical shells around the galactic centre. The profiles generally show an increase with radius at the four phases, all reaching values close to $1$ at large radii where the dark matter dominates over all other components.
Pre-BN, the central dark matter fraction shows significantly higher values ($\simi 0.7$ at $\Re$) than the three following stages ($<\!0.5$ for $r\ssim\Re$). 
The profiles at the BN, post-BN, and quenching phases show similar fractions when the radius is scaled by $\Re$. This reflects what we have already shown in \fig{fdm_Re}, namely the reduction of the dark matter fraction when measured within $r\slt \Re$ from high values $>\!0.5$ before compaction to a median of $\simi 0.3$ at the BN. Only mild evolution is seen after compaction, which can be partly attributed to the post-compaction increase in $\Re$.}

\begin{figure*} 
\includegraphics[width=0.95\textwidth]{"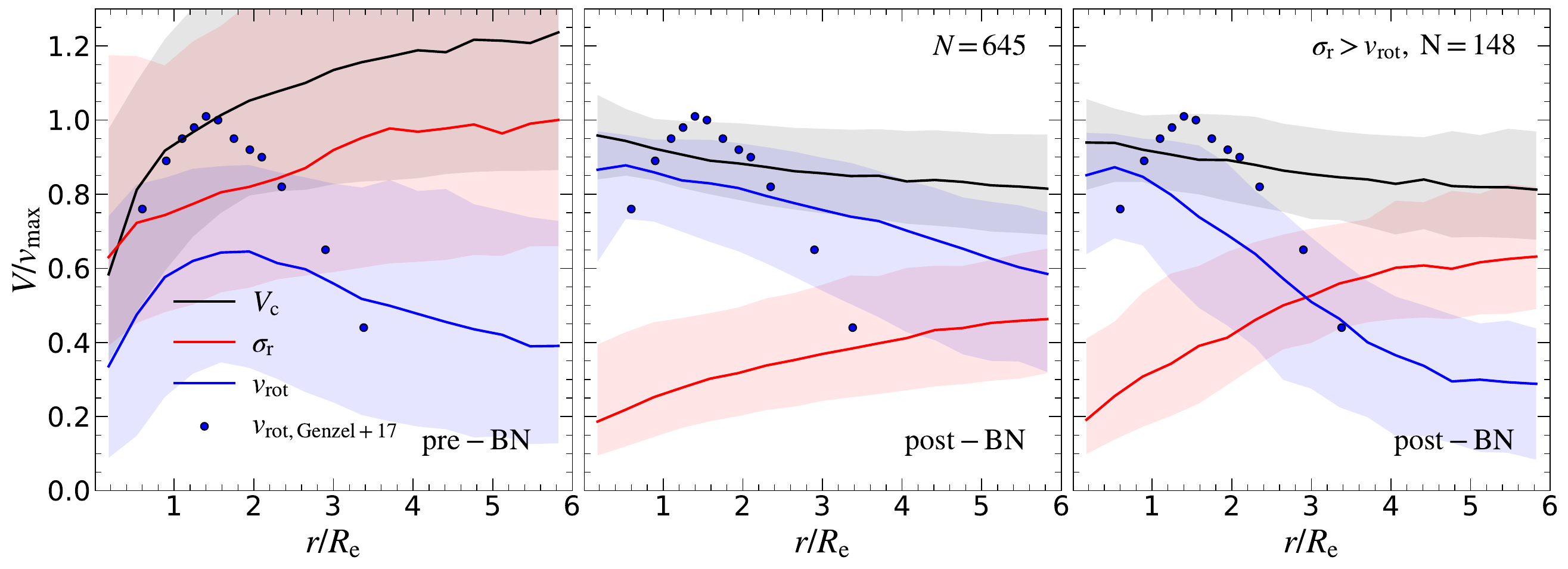"}
\caption{
{Rotation curves: $\Vcirc(r)$, $\Vrot(r)$ and $\sigr(r)$, normalized by the maximum $\Vrot$ for each snapshot. {\bf Left:} Pre-Compaction, stacked are all snapshots (454) with $\Ms\slt 10^{9.5}\msun$. {\bf Middle:} Post-compaction, based on the snapshots (645) with $\Ms\sgt 10^{9.5}\msun$. {\bf Right:} Post-compaction subsample containing the $\simi 20$ per cent of the snapshots (148) selected by $\sigr\sgt \Vrot$ at $4\slt r/\Re\slt 6$. The blue symbols mark $\Vrot$ from observed $z\ssim2$ massive discs \citep{genzel17}. The selected post-BN subsample resembles the observations, showing declining rotation curves and dispersion dominance at large radii.}
}
\label{fig:rotation_curves}
\end{figure*}

\smallskip
\Fig{rotation_curves} shows stacked rotation curves, $\Vcirc(r)$, $\Vrot(r)$ and $\sigr(r)$, normalized by the maximum $\Vrot$ for each snapshot.
{The snapshots are divided into two subsamples, pre and post-BN, with stellar masses below and above $10^{9.5}\msun$, respectively. Also shown is a subsample of 20 per cent of the post-BN sample, selected to have $\sigr\sgt \Vrot$ at $4\slt r/\Re\slt 6$.} The post-BN rotation curves are declining beyond $\Re$, and $\sigr$ is rising, summing up to an almost flat $\Vcirc$. The subsample selected to have $\sigr\sgt\Vrot$ at large radii shows steeply declining rotation curves, compatible with the ones observed by \citep{genzel17}, shown in the figure for comparison as blue symbols. 

\subsection{Extended Ring/Disc} \label{sec:ring}

If fresh gas is supplied after compaction and {onset of inside-out} quenching, a new extended disc/ring may form around the depleting nugget and develop a new VDI phase with some star formation, as seen in \fig{pics_V07}. Zooming out from such pictures clearly shows that the ring is forming due to new high-angular-momentum gas that spirals in through streams of cold gas, penetrating the DM halo from the cosmic web into its centre. {The fact that the post-compaction ring originates from fresh gas can be seen by the metallicity profiles in \fig{profiles_re}. Although some increase in metal enrichment is expected at the outskirts of the galaxy as a result of star-formation in the post-compaction ring, dilution by fresh gas sustains a relatively low metallicity at large radii ($>\! 3\Re$) as seen in \fig{profiles_re}, where both post-compaction phases show a steeper decrease with radius than the slope at the two earlier phases.}

\smallskip
{Using both an analytical model and simulations, \citet[][]{dekel20_ring} studied the origin and formation mechanism of extended gas rings in massive galaxies at high redshifts $z\slt 4$. By analytically evaluating the torques which are exerted by tightly wound spiral structure, it was shown that ring shrinkage depends on the cold-to-total mass ratio ($\delta_{\rm d}$). A long-lived ring is expected to form when the ring transport time, $\propto\!\delta_{\rm d}^{-3}$, becomes longer than the timescales of ring replenishment by accretion and depletion by SFR interior to the ring. These conditions become valid when $\delta_{\rm d}\slt 0.3$. When compared to simulations, it was shown that long-lived rings tend to form once a compaction-driven increase in the central mass results in a lower value of $\delta_{\rm d}$.}

\subsection{Comparison to Observations}
\label{sec:profiles_obs}

The predictions for the evolution of the surface-density profiles for stellar mass, gas mass and SFR, as in \fig{profiles_re}, {have been compared to observations in \citet{tacchella16_prof}.}
The robustness of gaseous rings about the quenched red nuggets is a prediction to be tested. Galaxies with star-forming H$\alpha$ rings about massive bulges have been observed at $z\ssim2$ \citep{genzel14_rings}, but their abundance has not been verified yet as they may be of low surface brightness and, therefore, hard to detect.
\citet{nelson16_Ha, nelson21} created deep stacked H$\alpha$ images of 3D-HST data to map out the SFR distribution in $z\ssim1$ SFGs, finding that the H$\alpha$ emission is more extended than the stellar continuum emission {\citep{matharu22}}, consistent with the inside-out assembly of galactic discs. This effect grows stronger with mass, indicating that the more massive systems have an extended star-forming component in their outskirts.
{\citet{abdurro'uf23} studied galaxies at $0.3\slt z\slt6$ in two clusters and a blank field by combining data from HST and JWST to perform a spatially resolved spectral energy distribution modelling. The high spatial resolution, combined with lensing magnification in the cluster fields, allowed a sub-kpc scale resolution in some galaxies. They find flat sSFR radial profiles and similar $\Re$ and $\Resf$ radii at the highest redshift bin, a significant fraction of central starbursts in SF galaxies at $1.5\sleq 2.5$ and $\Resf<\Re$. Finally, at the low redshift bin, they find a higher fraction of galaxies with suppressed central SF while they continue to form stars in the disc. This is consistent with the inside-out assembly \citep[see also][for nearby galaxies]{abdurro'uf22}.}
\citet{tacchella15_sci} measured the H$\alpha$ emission distribution in individual systems at $z\ssim2.2$, comparing the stellar mass and SFR surface density profiles of individual galaxies. The main finding is that lower mass systems have roughly flat sSFR profiles, while galaxies at $\Ms\sgt10^{11}\msun$ have reduced sSFR profiles, which indicates those galaxies have their star-formation activity taking place in an extended ring, consistent with the simulated post-compaction galaxies. Similar results have been reported by \citet{tacchella18_dust} after taking special care of the dust attenuation distribution within the galaxies.
{\citet{mancini19} studied the star formation history (SFH) in 10 observed `green-valley' galaxies at $0.45\slt z\slt1$. By performing a multiwavelength bulge-to-disc decomposition, the SFH for the two components were derived separately. It was shown that these galaxies, at the bending of the MS, tend to have old bulges and significantly younger discs. It was argued that the bending of the MS is mostly due to galaxies which formed their bulge at very high redshifts, with ages approaching the age of the Universe \citep[see also][]{taccella22}, and later experienced a rejuvenation of SF in the outer regions of the disc \citep{dimauro22}. This is again consistent with the simulated post-compaction galaxies shown here.}

\smallskip
While we find a good agreement between the reported shapes of the observed and simulated profiles, the SFR surface density profiles in the simulations have a too-small normalization in comparison to observations at a given mass and redshift. This might indicate that the galaxies in the simulations are consuming the gas too early, resulting in too small SFR at low redshifts ($z\ssim1$).

\smallskip
As highlighted above, during the BN phase, the star formation takes place primarily in the central $\simi1\kpc$. We expect that this star formation is heavily obscured by dust. Recent observations with ALMA {\citep{barro16_alma, tadaki16, tadaki17, tadaki20}} detected massive, dusty SFGs that are experiencing a dusty nuclear starburst and are doubling their mass quicker in their centres than in their outskirts, consistent with the rising sSFR as seen in the simulated BN galaxies, \fig{profiles_re}.

\smallskip 
{The simulation prediction of central baryon dominance after the compaction phase, namely above the critical mass, as seen in \fig{fdm_Re}, is confirmed observationally \citet{price16,genzel17,genzel20}.} We find in \fig{rotation_curves} that a fraction of the simulated galaxies at $z\ssim2$ have declining rotation curves beyond $\Re$, while observationally this seems to be the typical behaviour \citet{lang17}. {The modelling involved in the dynamical analysis of observed rotation curves includes an assumption of radially constant velocity dispersion, in the absence of any clear contradicting evidence, it is utilized as the simplest assumption. In \fig{rotation_curves}, we see that the simulations indicate a radially rising velocity dispersion in massive post-compaction galaxies; such behaviour may leave room for somewhat higher dark matter fractions within $\Re$.}

\section{SN Feedback and BH Growth}
\label{sec:SN_BH}

\subsection{Transition in BH Growth at a Critical Mass}

We propose that the compaction event has a crucial role in the growth of the black hole in the galaxy centre, and in triggering AGN feedback that helps the quenching of massive galaxies. The characteristic mass of BNs, at $\Ms\ssim 10^{10}\msun$ or $\Mv \ssim 10^{12}\msun$, indeed appears in many other observed properties of galaxies, such as the peak mass of galaxy formation efficiency $\Ms/\Mv$, defining the scale of bimodality in galaxy properties \citep[e.g.][]{db06}. In particular, strong AGN activity tends to show up in observed galaxies above this critical mass, indicating rapid BH growth in this regime. 

\smallskip
At least two relevant physical processes give rise to a characteristic mass in the same ballpark. First{, the lower limit is set by} the critical scale for supernova feedback, corresponding to a potential well with a virial velocity of $V \ssim 100 \kms$ at all redshifts, below which the SN energy deposited in the ISM and CGM is capable of significantly heating a large fraction of the gas and ejecting massive outflows, while above which SN-feedback becomes ineffective \citep{ds86}. Second{, the upper limit is set by} the critical mass for virial shock heating, $\Mcrit \ssim 10^{12}\msun$ at all redshifts, below which most of the gas entering the halo streams cold into the galaxy, while above which a virial shock is stable and can support a hot CGM \citep{bd03,keres05,db06}{, which in turn may lower the efficiency or even suppress cold gas supply to the galaxy}. Cold streams can penetrate the hot haloes even slightly above the critical mass at high redshifts, but cold gas supply to the galaxy is significantly suppressed in such haloes at $z \leq 2$ \citep{db06,ocvirk08,keres09,dekel09}. To some extent, these two scales are related as they both arise from a balance between cooling and dynamical times, but they arise in two different contexts. A third scale, of unrelated origin, is the scale for non-linear clustering, the Press-Schechter mass, which is rapidly growing with time but is in the ballpark of {$\simi 10^{11-12}\msun$} at $z\ssim 1$, thus possibly relevant to the characteristic scale of galaxy formation at moderate redshifts $z\ssim 1-2$ {\citep[e. g., see Fig.~2 of][for a summary of the three processes as a function of $\Mv$ and redshift]{db06}}.

\smallskip
{Several cosmological simulations which incorporate both SN feedback and an accreting BH yield similar robust results concerning the cross-talk between these two processes. These include simulations in various codes (such as AMR, SPH and quasi-Lagrangian) with different sub-grid models, and more to the point, using different implementations of BH growth and SN feedback \citep{dubois15,angles17,bower17,dubois20_NH,habouzit17,habouzit19}. All are showing consistent results} that the BH growth is suppressed below the critical halo mass of $\simi 10^{11.5}\msun$ ($\Ms \slt  10^{10}\msun$), and rapid BH growth starts quite abruptly and prevails at larger masses. This can be seen, for example, in \citet[][Fig.~2]{dubois15} and \citet[][Figs.~6-7]{bower17}.

\smallskip
This asymptotic behaviour below and above the critical mass can be naturally understood in terms of the characteristic masses associated with SN feedback and virial shock heating. Well below the critical mass, SN feedback is efficient in removing gas from the galaxy potential well, helped by {the buoyancy of SN-driven gas} through the cooler CGM, thus suppressing accretion onto the BH and starving it. Well above the critical mass, the SN-driven winds are bound to the galaxy, both by the deep halo potential and by the hot ($\Tv$) CGM that prevents buoyancy of the cooler SN-driven gas, so any outflows recycle and cannot prevent accretion onto the black hole {\citep[The role of buoyancy is nicely argued in][]{bower17}}. Note that the rapid BH growth naturally activates AGN feedback, which in turn (a) can suppress star formation and, therefore, SN feedback, and (b) can keep the CGM hot after the gas was heated by the {virial shock.}

\smallskip
The wet compaction event, which tends to occur at a similar critical mass, triggers the transition between these asymptotic regimes. The sudden condensation of gas into the inner $1\kpc$ drives a stimulated accretion onto the BH, which overcomes any pushback by SN feedback and starts the phase of rapid BH growth. In the simulation described in \citet[][Fig.~8]{dubois15}, the onset of rapid BH growth is indeed associated with a compaction event driven by a minor merger. This role of compaction is being studied in detail in the \nh simulations \citep{lapiner21}. 

\subsection{Comparison to Observations}
\label{sec:SN-BH_obs}

There is observational evidence for a correlation between BNs and AGN activity.
\citet{kocevski17} find in the CANDELS sample at $z\ssim2$ that in the BN regime of the {SFR-$\Sig1$ diagram, namely galaxies with high $\Sig1$ and high SFR,} the fraction of galaxies that host X-ray AGNs is {about {40 per cent}}. While in the pre-compaction SFG regime and the post-compaction quenched regime, it is less than {25 per cent.}
{\citet{juneau13}, using a sample of X-ray obscured AGN at $z\seq 0.3\sdash1$, find that they tend to be hosted in compact galaxies of high sSFR above the MS ridge, namely in the BN regime.}
{\citet{chang17} analysed the morphology of obscured AGNs at $0.5\slt z\slt 1.5$ in the COSMOS survey, finding that such galaxies have a typical stellar mass of $>\!10^{10}\msun$, and tend to be more compact in terms of both $\Re$ and their \sersic index than normal SFGs of a given redshift and stellar mass. Again, consistent with BNs.}
\citet{forster19} identify AGN-driven winds preferentially above a threshold stellar mass of $\simi 10^{10}\msun$, and an anti-correlation with $\Re$, consistent with the predicted compaction-driven black hole growth. {In \citet[fig.~9]{lapiner21} it was shown, using eight massive galaxies in the \nh simulations, that a similar trend is seen in the fraction of AGNs as a function of stellar mass. That is, the fraction of AGNs is very low below the turn-up mass and steeply rising above it. The results are overall consistent with \citet[figs.~6,13]{forster19}, with a slight offset to lower masses or higher AGN fractions, which depends on the selected AGN luminosity threshold.}

\section{Trigger of compaction}
\label{sec:trigger}

\subsection{Mergers}
\label{sec:mergers}

\begin{figure} 
\includegraphics[width=0.95\columnwidth]{"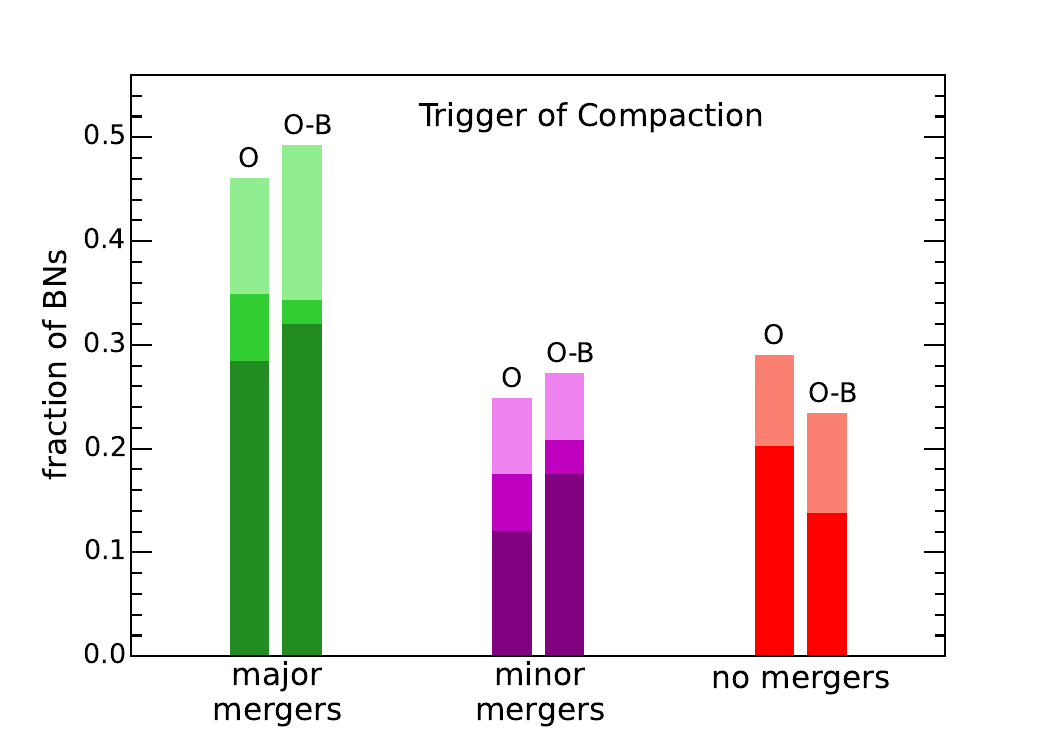"}
\caption{
Fraction of compaction events that are triggered by major mergers ($>\!1:3$), by minor mergers ($1:10$ to $1:3$), and by another mechanism. The identification of a co-occurring merger and compaction is made near the onset of compaction (O) or during the compaction from the onset to the BN (O-B). Dark colours refer to solid identifications of bound mergers, medium colours refer to fly-byes (that mostly end up as mergers), and light colours refer to uncertain mergers. The best-estimate fractions of these events are $0.48, 0.26$ and $0.26$, respectively. 
}
\label{fig:mergers-bn}
\end{figure}

What are the mechanisms that trigger wet compaction? In principle, it could be any mechanism that causes significant dissipative AM loss. Wet major mergers are obvious candidates, as they are known to be capable of pushing gas in \citep{barnes91,mihos96,hopkins06,dc06,covington08,hopkins10,covington11}. However, their expected frequency has to be compared to the abundance of BNs. Indeed, we see in the simulations compaction events that are not associated with major mergers and some not even with mergers of any kind. Therefore, we attempt to estimate the statistics of the triggers of compaction.

\smallskip
The galaxy merger rate could be approximated by the halo merger rate, which could be estimated based on the EPS formalism from Fig.~6 and eq.~24 of \citet{neistein08_m}. We obtain that the rate of major mergers (with a mass ratio larger than $1:3$) for a $\simi10^{12}\msun$ halo in the EdS regime is
\be
\frac{dN}{dt} \ssimeq 0.4\, [(1+z)/3]^{2.5} \Gyr^{-1}
\label{eq:mergers}
\ee
We wish to compare this to the rate of onsets of compaction events. We note in \fig{char_M-a_event} that most galaxies undergo a major compaction event when their halo mass is typically in a mass range of $\Delta \log M\ssim0.4$ about the critical BN mass. In order to relate this mass range to a redshift range, we appeal to the average growth of haloes in the EdS regime \citep[][eq.~9]{dekel13},
\be
M(z) \propto M_0 e^{-\alpha (z-z_0)} , 
\label{eq:M_growth}
\ee
where the mass at $z_0$ is $M_0$ and $\alpha \ssimeq 0.8$ in the relevant mass range. This gives $\Delta z \ssimeq 3 \Delta \log M \ssimeq 1.2$. In the EdS regime, where $(1+z)\ssimeq (t/t_1)^{-2/3}$ with $t_1\ssimeq 17.5\Gyr$, this corresponds at $z\seq 2$ to $\Delta t \ssimeq 1.4\Gyr$. A comparison to \equ{mergers} implies that about half the galaxies in the given mass range undergo major mergers, so the latter can be responsible for about half the onsets of compaction. Minor mergers ($1:10$ to $1:3$) are expected to have a similar frequency.

\smallskip
We study the origin of compactions in our simulations in Ginzburg et al. (in preparation) and summarize the main result here. \Fig{mergers-bn} summarizes the correlation between the BN events and merger events {using the entire sample}. The merger events are identified by rapid variations of stellar mass within the inner $(0.1-0.2)\Rv$ over the output timestep of $\simi100\Myr$. The events are divided into BNs associated with (a) major mergers (mass ratio $>\!1:3$), (b) minor mergers ($1:10$ to $1:3$), and (c) BNs that are not associated with any merger event. We also identified a couple of BNs associated with mini-minor mergers ($1:20$ to $1:10$) and counted them as `no mergers'. A small fraction of the mergers (medium darkness in the figure) is identified as flyby events, most of which are first-passages before the final coalescence. The association with a compaction event is evaluated by the co-occurrence (to within $\pm 100\Myr$) of the merger event either with the onset of compaction (O) or with any time during the compaction process, between the onset and the BN (O-B). 

\smallskip
We thus estimate that major mergers trigger about 48 per cent of the compaction events, about 26 per cent are associated with minor mergers, and another 26 per cent are not associated with stellar mergers and thus must be triggered by a different mechanism. We find no significant difference between the triggers of all the compaction events and those of the major BNs that lead to significant quenching. We also find that about one-third of the major mergers do not lead to compaction events. These results are indeed in the ballpark of the toy estimates above. They indicate that, as expected, major mergers serve as a major trigger of compaction events and that, together with minor mergers, they are responsible for about 70 per cent of the compaction events. However, about 30 per cent of the compaction events are not triggered by mergers as identified by their stellar component.

\subsection{Other Potential Triggers of Compaction}

The other mechanisms that may trigger AM loss and lead to compaction events are being studied elsewhere using these simulations. A partial list is as follows.
 
\smallskip
\no
{\bf (b) Counter-rotating streams.}
The streams could drive compaction by generating low AM patches once they are counter-rotating with respect to the disc and other streams. Our simulations indicate that typically one stream out of three is counter-rotating \citep[][Fig.~14]{danovich15}.
 
\smallskip
\no
{\bf (c) Recycling.}
We see a new fountain pattern in our simulations, where feedback-driven winds from the disc outskirts carry out high AM and return to the galaxy with low AM, generating low AM patches and possibly triggering global compaction (DeGraf et al., in prep.). A similar mechanism has been proposed by \citet{elmegreen14}.

\smallskip
\no
{\bf (d) Triaxial cores.}
The torques induced by the pre-BN triaxial haloes discussed in \se{shape} could induce AM loss that could lead to compaction, while the disappearance of such torques in the post-BN rounder halo would end the occurrence of compaction events.
 
\smallskip
\no
{\bf (d) Satellite compression.}
In a study of SDSS satellite galaxies \citep{woo17}, we found that they are compact when quenched, indicating that satellite quenching may also involve compression. This could be induced in the outer host halo by ram-pressure compression when self-gravity overcomes the ram-pressure push. Alternatively, it could be induced by tidal compression in the inner halo where the host halo density profile is core-like \citep{dekel03}.

\smallskip
\no
{\bf (e) VDI-driven disc contraction.}
In \citet{db14}, based on Toomre instability, we analysed the wet shrinkage due to torques in a galaxy undergoing violent disc instability (VDI), energetically powered by the inflow to the bottom of the potential well. We found that VDI in our cosmological simulations is externally stimulated \citep{inoue16}, e.g., by minor or major mergers, tidal interactions, or counter-rotating streams. The VDI is thus commonly associated with the same mechanisms that can trigger AM loss and wet compaction. The toy model based on VDI thus captures many of the properties of wet compaction even when not solely driven by VDI.

\section{Abundance of BNs}
\label{sec:abundance}

\subsection{Evolution of Abundance}

\begin{figure} 
\centering
\includegraphics[width=0.95\columnwidth]{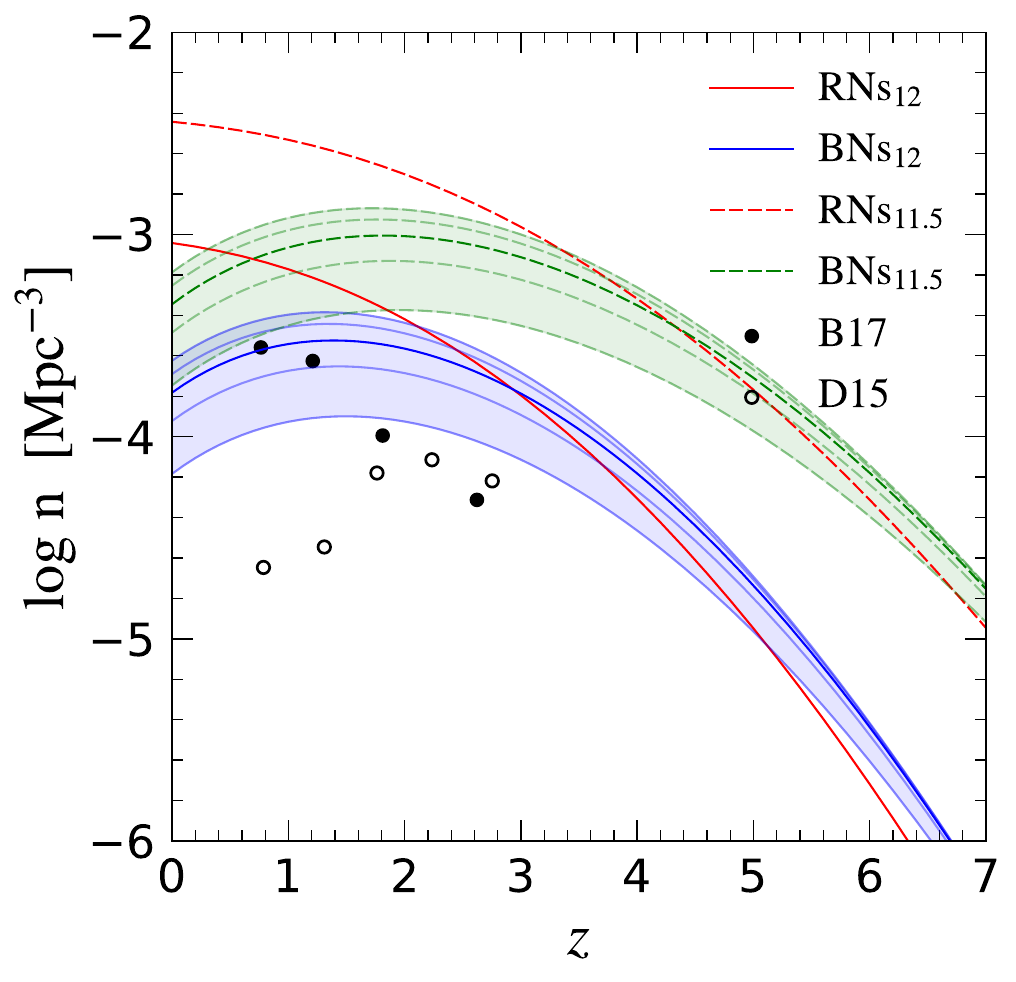}
\caption{
Evolution of the abundance of BNs according to a toy model assuming a constant threshold halo mass of $\Mv\seq10^{12}\msun$, plus a compaction trigger by major mergers, {and assuming several constant duration of BNs with $\Delta t_{\rm BN}\seq0.1, 0.2, 0.3, 0.4, 0.5\thub$ (blue, bottom to top respectively). Also shown is the evolution of BNs abundance assuming a lower threshold halo mass for BNs of $\Mv\seq10^{11.5}\msun$ (green). Our fiducial case (dark blue line) assumes a threshold halo mass of $\Mv\seq10^{12}\msun$ (median of $\Ms\ssim 10^{10.3}\msun$ at $z\ssim2$) and $\Delta t_{\rm BN}\seq0.3\thub$. Recent observational estimates of compact star-forming systems with $\Ms\sgt 10^{10.3}\msun$ are shown for comparison in solid and open black symbols, from \citet[Fig.~8]{barro17_uni} and \citet[Fig.~19]{vandokkum15} respectively. The model (assuming the fiducial case) overestimates the observed abundance at $z\ssim2\sdash3$. However, this should serve as an upper limit, partly because of a larger threshold mass for BNs in the observed abundances and a high threshold of $\Sig1$. In addition, the fixed duration of compaction is expected to be lower for observed compact BNs due to a stellar-based high threshold for $\Sig1$, while our measurements for the duration rely on the evolution of central gas. The abundance of post-BN passive central RN (fiducial case in solid red, dashed red indicates a mass threshold of $10^{11.5}\msun$)} {is expected to rise in time continuously}.
}
\label{fig:n-z}
\end{figure}

The predicted abundance of BNs as a function of redshift is not straightforward to deduce from the simulations, as our sample consists of a given group of galaxies that grow in time and is not a fair sample at any redshift. Therefore, the abundances plotted in Fig.~15 of \citet{zolotov15} are not directly comparable with observations. However, based on the characteristic halo mass $\Mcrit$ for most BNs, and the typical lifetime of a galaxy in the BN phase $\Delta t_{\rm BN}$, both derived from the simulations, one can make a toy-model estimate of the abundance of BNs at any time in the EdS regime ($z\sgt1$). Using the average growth of haloes in \equ{M_growth}, at any given $z$, we translate $\Delta t_{\rm BN}$ to a mass interval $\Delta M$. We then use the Sheth-Tormen version of the Press-Schechter approximation \citep[][appendix]{sheth02,db06} to compute the comoving number density of haloes with masses in the interval $(\Mcrit,\Mcrit+\Delta M)$. Finally, the abundance of post-BN galaxies is estimated as the integral of all past BNs after they are turned off as star-forming BNs.

\smallskip
Assuming that half the compactions are caused by major mergers (as estimated in our simulations, \se{trigger}), one may adopt a delay in the onset of BN by half the mean time between major mergers. Based on \equ{mergers}, the latter is
\be
\frac{\Delta t}{t} \simeq  \frac{{0.56}}{(1+z)/4} \, .
\ee
As above, using \equ{M_growth}, this can be translated to a mass increment to be added to $\Mcrit$.

\smallskip
\Fig{n-z} shows the result of this toy-model estimate, assuming a constant threshold halo mass, a delay of half the mean time between major mergers, and a constant duration for {BNs}. {We show the model in five typical fixed duration of compaction, $\Delta t_{\rm BN} \seq 0.1, 0.2, 0.3, 0.4, 0.5\thub$ (bottom to top respectively), and two threshold halo masses $\Mv\seq 10^{12}\msun$ (blue) and $\Mv\seq 10^{11.5}\msun$ (green).}
{Our fiducial model (dark blue line) assumes a threshold halo mass for BNs of $\simi 10^{12}\msun$, and a fixed duration for BNs of $\Delta t_{\rm BN}\seq 0.3\thub$ \citep[][Fig.~16]{zolotov15}. While the simulations indicate a somewhat lower critical halo mass, we assume a threshold of $10^{12}\msun$ given the possible dependence on feedback and the tendency of the simulations to overestimate the stellar-to-halo mass relation (see \se{limitations}). From abundance matching, this threshold halo mass is expected to correspond to a median stellar mass of $\simi 10^{10.3}\msun$ at $z\ssim 2$ with a scatter of $\pm 0.3-0.4$dex. Assuming the fiducial case, we see that the abundance of BNs is predicted to rise with time from high redshifts till $z\ssim3$, reaching a broad peak somewhat above $n\ssim10^{-4}\Mpc^{-3}$ near $z\ssim1\sdash 2$. Moreover, we see that BNs are expected to be observable also at higher redshifts, e.g., with $n\ssim10^{-5}\Mpc^{-3}$ at $z\ssim5$, but only with $n\slt 10^{-6}\Mpc^{-3}$ at $z\sgt7$. 
The fiducial case overestimates the observed abundance at $z\ssim2\sdash 3$; however, we expect the model to serve as an upper limit to the observed abundance partly due to the following reasons. 
First, we expect a shorter duration for observed BNs when a high threshold for $\Sig1$ is adopted for observed BNs. We measure the BN duration ($\Delta t_{\rm BN}$) based on the evolution of the central gas through the whole compaction event (compact with high SFR), starting from the onset of compaction and up to the post-BN phase before the galaxy is leaving the Main Sequence. When a central stellar-based threshold of $\Sig1$ is introduced, the duration is expected to start after a massive central stellar body is formed, roughly starting the duration from or even after the peak of compaction (BNg).
Second, taking into account, the scatter in the stellar-to-halo mass relation for a $10^{12}\msun$ halo at $z\ssim2$, the observed sample is selected above a threshold stellar mass which is larger than the critical BN mass.
Finally, the prediction would also serve as an upper limit if only a fraction of the galaxies in the mass range following the critical BN mass actually become BNs.
We also show a lower halo mass threshold of $10^{11.5}\msun$ (green). However, this threshold is significantly lower than the mass threshold of the observed abundances.}

\smallskip
The abundance of post-BN passive galaxies is expected to {rise in time continuously}, crudely matching the BN abundance at $z\ssim2-3$, as observed.

\subsection{Comparison to Observations}
\label{sec:abundance_obs}

The evolution of BN abundance has been estimated observationally earlier \citep[e.g.][]{barro13,vandokkum15}, using a constant and rather high threshold in central stellar surface density, and yielding comoving densities that peak at $\simi10^{-4}\Mpc^{-3}$ near $z\seq 2-3$. The most reliable measurement so far is provided in Fig.~9 of \citet{barro17_uni}, based on a threshold in $\Sig1$ that varies with redshift following the observed threshold for compact quiescent RNs. This observed number density of BNs peaks at $n\ssim 3\times10^{-4}\Mpc^{-3}$ near $z\ssim1$, in general agreement with the prediction. The selection threshold is $\Ms\sgt10^{10.3}\msun$, which may explain why the prediction overestimates the observed abundance at $z\ssim2$.

\section{Discussion}
\label{sec:disc}

\subsection{BNs in other simulations}
\label{sec:other_sims}

The occurrence of wet compaction events, and therefore the abundance of BNs and their properties, may depend on the strength of feedback assumed in the simulations. The feedback may suppress the level of dissipative compaction or affect the central depletion and the overall quenching process. In particular, a different feedback strength may affect the critical mass of BNs, perhaps through the evolution of gas fraction. Since there is certain degeneracy between the SFR and feedback that allows a match of galaxy properties with observations, different cosmological simulations use different SFR and feedback recipes and may thus result in different BN properties. It is, therefore, important to verify the validity of our results concerning the BN properties using different simulations with different feedback recipes.

\subsubsection{BNs in the \nihao simulations}
Preliminary tests using the \nihao simulations indicate the occurrence of compaction events at a similar critical mass. {The \nihao cosmological simulations \citep{wang15} use the N-body SPH code {\sc gasoline} \citep{Wadsley04, keller14}. While the \vela and \nihao simulations differ in many respects, one of the key distinctions we would like to explore is the effect of different stellar feedback on the phases of compaction. In \nihao, the efficiency of stellar feedback in the pre-SN stage is boosted from 10 to 13 per cent of the total stellar flux, aiming for a better match with {the stellar to halo mass ratio based on} abundance matching by \citet{behroozi13_b}. In addition, during the SN stage, energy is fed into the gas through a blast-wave SN feedback \citep{stinson06}, with delayed cooling for $\simi 30\Myr$ for particles within the blast region. As a result, the \nihao simulations incorporate stronger stellar feedback than the \vela simulations (see \seApp{sims_app}).} {Due to the nature of compact objects, one must ensure an adequately resolved inner region for the study of compaction events. The choice of particle mass and force softening in \nihao ensures a properly resolved mass profile in the inner $\simi 1$ per cent of the virial radius. Therefore, the \nihao simulations properly resolve the central region of $\simi 1\kpc$ at high redshifts and halo mass about the critical mass of $\simi 10^{12}\msun$, making them suitable for the study of compaction events at high redshifts.}
{Preliminary results show that} the main difference from the current \vela simulations is in {a more temperate} post-BN central depletion, which leads to a slower central quenching and thus a longer lifetime for the BNs. The detailed analysis of BNs in the \nihao simulations is deferred to a separate paper. 

\subsubsection{BNs in the \nh simulations}
{
Using the \nh cosmological simulations, \citet{lapiner21} studied compaction events in eight massive galaxies, selected to have exceeded a stellar mass of $10^{10}\msun$ at some point in their evolution. The \nh simulations exhibit compaction events with similar features and at a similar critical mass as seen in the current work. 
The \nh is a zoom-in cosmological simulation \citep{dubois20_NH}, run with the AMR {\sc RAMSES} code \citep{teyssier02}. Briefly, the \nh simulation is a zoom-in on a spherical patch, with a radius of $10\cMpc$, embedded in the large-scale box of {\sc Horizon-AGN} simulations \citep{dubois14_hagn, dubois16_hagn}. The maximum spatial resolution in this patch is $\simi 40 \pc$. The sub-grid model incorporates both SN and AGN feedback. The relatively strong SN feedback models the phases of momentum and energy conservation \citep{kimm14, kimm15}, considers the effect of SN clustering \citep{kimm17, gentry17}, and takes into account the momentum due to preheating near OB stars. Black holes grow through Bondi-Hoyle accretion \citep{bondi44}, with a spin-dependent radiative efficiency computed on-the-fly and restricted to the Eddington limit. The AGN feedback consists of two modes, a thermal `quasar' and a kinetic `radio' mode, determined by the accretion rate efficiency. \citep[see further details in][]{dubois20_NH}.

\smallskip
\citet{lapiner21} identified major compaction events in six galaxies. The two galaxies which showed only mild compaction events were below the critical mass until relatively low redshifts, hardly exceeding it until the final analysed redshift.
The two main differences from the current \vela simulations are as follows:
(1) In some cases, the post-BN central gas depletion in \nh is more pronounced than the \vela simulations. \citet{lapiner21} showed that the compaction phase marks a transition from suppressed BH growth pre-BN to rapid growth during and after compaction (followed by a self-regulated growth phase at late times), demonstrating a tight correlation between the time of compaction and the onset of BH growth, and therefore, the associated onset of frequent AGN activity episodes. The cause for this tight correlation was ascribed to a compaction-driven deepening of the potential well, shown to affect both the position of the BH and the stable availability of a gas reservoir for its accretion. Consistent correlations between compact central density and onset of high BH accretion rate were also seen in \citet{byrne22} using post-processed BH accretion in the {\sc fire} simulations. While we find that one of the hallmarks of the BN phase is the subsequent central gas depletion and inside-out quenching, as seen in the current work using simulations without AGN, this attribute may be further accentuated by the inclusion of AGN feedback, which is expected to be active once the galaxy is above the critical BN mass scale.
(2) Another difference is the time of transition from dark matter to a baryon-dominated central region. Although the transition usually occurs before the peak of compaction in both simulations, in \vela, it is typically close to the BN time (see \fig{char_M-a_event}), and in \nh the transition tends to happen somewhat earlier. Further analysis is required to quantify and understand the cause of this difference.
}


\subsubsection{BNs in the \velaF simulations}
{Preliminary results using the \vela generation 6 simulations \citep{ceverino22_vela6} show that the compaction events are still robust in the presence of stronger feedback and display very similar results in terms of the strength of compaction. However, in \velaF, we find a delay in the time of compaction when compared to the {current \vela (\velaC) simulations}. The \velaF simulations include the same galaxies with the same initial conditions as \velaC. One of the main differences is the inclusion of stronger SN feedback, introducing the injection of momentum from the unresolved expansion of SN and stellar wind gas bubbles \citep{ostriker11}. In particular, the injected momentum is boosted by a factor of 3 to account for the effect of clustered SN explosions \citep{gentry17}, which may lower the gas density in the vicinity of a star cluster and increase momentum per supernova in the subsequent explosions. A detailed analysis of BNs in the \velaF simulations and comparison with \velaC is deferred to a separate paper (Lapiner et al. in prep.).}

\subsection{Compaction of high- versus low-mass subsamples}
{

The simulations reveal several differences between the high-mass (HM) and the low-mass (LM) subsamples; these are divided by the median of the total stellar mass in the galaxy at $z\seq2$. When these two subsamples are compared, one should keep in mind that the LM typically experiences the major compaction at later times, closer to the final redshift of the simulations; therefore, the late post-BN phase is often unexplored in these galaxies (e.g., as seen at the high mass end for LM in \figApp{Re-Ms_LM} compared to the HM in \fig{Re-Ms}).

\smallskip
\no The LM generally display the following differences when compared to the HM:
App\rf a. Less prominent major compactions with a lower central gas density and SFR (LM: \figApp{M-a_med_lm}, HM: \fig{M-a_med}). 
\rf b. More frequent episodes of central gas rejuvenation, namely small compaction events in the post-BN phase (\figApp{M-a_med_lm}).
\rf c. More oscillations about the MS and reaching lower post-BN central surface density (\fig{L-shape}).
\rf d. Overall higher central dark matter fraction (\fig{fdm_Re}).
\rf e. Overall larger effective radius with a larger scatter (\figApp{Re-Ms_LM},\ref{fig:Re-Ms}). 
\rf f. Lower $\vosg$ (\figApp{kin_LM},\ref{fig:kin}) during the post-BN phase.
\rf g. Post-BN drop in the gas spin $\lambdag$ (\fig{spin}) and in $\vosg$ (\figApp{kin_LM},\ref{fig:kin}), reflecting the late rejuvenation episodes again.
\rf h. Lower metallicity (\fig{Z-abn}, \ref{fig_app:Z-abn_app}).

\smallskip
However, some of the key features of compaction are still seen in most of the simulated galaxies, including the LM subsample. Overall, the evolution of the LM with respect to the BN shows similar trends as the HM subsample, although the typical values are frequently different, and the scatter is larger. We emphasize that the major compaction to a BN occurs about the same critical mass with negligible redshift dependence, as seen in \fig{char_M-a}, where we use the entire sample. The critical BN mass forces a delay in the compaction time of the LM subsample compared to the HM.

\smallskip
It should be noted that the difference in the compaction time and strength of the HM versus the LM subsample may have other implications. For example, given the tendency of the HM subsample to experience a more prominent major compaction event at earlier times than the LM, the central BH may be affected. Namely, early major compactions are expected to form a stable massive stellar body in the centre and deepen the potential well at early times; this may lead to more efficient growth of the massive central BH \citep{lapiner21} and, in turn, to a rapid quenching by the associated AGN feedback. The simulations presented here do not include BHs and AGN feedback. However, the current simulations reveal that the compaction process triggers the onset of central quenching, with the HM subsample typically displaying clearer long-term central gas depletion. Therefore, the addition of early BH growth and the onset of AGN activity may further assist an earlier and faster quenching \citep[e.g.,][]{park22} of the HM subsample compared to the LM.

\smallskip
In summary, the LM subsample reaches the critical mass for compaction at later times when the Universe is less dense and relatively gas-poor. Therefore, the compaction process, which is governed by gas-rich conditions and drastic AM loss, tends to be less pronounced in these galaxies. A deeper understanding of these differences and their origin deserves a more detailed study beyond the scope of this paper. 
}

\section{Conclusion}
\label{sec:conc}

{Cosmological simulations reveal a robust event of dramatic consequences during the evolution of galaxies at high redshifts -- wet compaction to a blue nugget.} The blue nugget phase marks drastic transitions in the galaxy structural, compositional, {and kinematic properties} as follows (see a summary in \tab{transitions}).

\smallskip
\no \ul{1. Structure and star formation:}
\rf a.
The major BN event, the one preceding a significant quenching process, tends to occur near a critical {`golden'} mass, $\Mbn$, independent of redshift, which in the current simulation is a stellar mass $\Ms \ssim 10^{10}\msun$ corresponding to a halo virial mass $\Mv \ssim 10^{12}\msun$ {(\fig{char_M-a_event})}. Therefore, evolution in time with respect to the BN event can {in practice} be translated to dependence on mass with respect to $\Mbn$.
\rf b.
The compaction to a BN makes the surface density within the inner $1\kpc$ (or $\Re$) grow above a critical value (that is slowly decreasing with time) at the BN and post-BN phases, typically $\Sig1\sgt10^{9.5}\msun\kpc^{-2}$ for the stellar surface density.
\rf c.
The typical galaxy evolution track in the sSFR$-\Sig1$ diagram has a characteristic L shape: a pre-BN contraction at a roughly constant sSFR that abruptly changes to post-BN quenching at a roughly constant $\Sig1$ {(\fig{L-shape})}.
\rf d. 
The typical evolution of the effective radius $\Re$ as a function of $\Ms$ is a shallow increase pre-BN that turns into a steep rise post-BN after a short-term shrinkage near the BN mass {(\fig{Re-Ms})}.
\rf e.
The metallicity is rising roughly linearly with $\Ms$ below $\Mbn$, and it saturates to a plateau above $\Mbn$ {(\fig{Z-abn},\ref{fig_app:Z-abn_app})}.

\smallskip
\no \ul{2. Evolution with respect to the Main Sequence:}

\rf f.
{One can translate the evolution of galaxy properties with respect to the BN phase to the evolution with respect to the universal Main Sequence of SFGs in the universal sSFR$-\Ms$ diagram (where the overall decline of sSFR with redshift is scaled out). Pre-BN, the sSFR oscillates about the MS ridge, climbs to the upper MS at the BN and then descends through the ridge and the bottom of the MS to the quiescent regime {(the `green valley')} during the post-BN quenching process \citep{tacchella16_ms}.}
\rf g.
This is reflected in gradients across the MS of galaxy properties that are associated with gas and SFR, especially at the BN and post-BN masses. For example, the depletion time, gas-to-stellar mass ratio, SFR {and gas kinematics}, after the systematic redshift and mass dependence have been eliminated, relate to the deviation from the MS as $\tdep\sprop \DMS^{-0.5}$, $\fgs\sprop \DMS^{0.5}$, {SFR$_1\sprop \DMS^{1.2}$, $\vosg\sprop\DMS^{-0.3}$ and $\lambdag\sprop\DMS^{-0.35}$} {(\fig{gradients_q-dMS})}.

\smallskip
\no \ul{3. Dark matter fraction and shape:}
\rf h.
Pre-compaction, the central region ($1\kpc$ or $\Re$) is dominated by dark matter, and it becomes baryon-dominated at the BN and after, with a DM fraction $\simi0.2$ within $\Re$ {(\fig{fdm_Re})}.
\rf i.
As a result, the pre-BN stellar system tends to be elongated, triaxial or prolate, and it becomes oblate post-BN {(\fig{shape})}. The BN itself is a sub-kpc compact thick disc in gas and SFR and is less flattened in stars {(\fig{shape_bn})}, {with a median global \sersic index of $n\ssim 3.5$ at the peak of compaction.} 

\smallskip
\no \ul{4. Kinematics:}
\rf j.
Pre-BN, the system is kinematically highly perturbed, with $V/\sigma\ssim1$. Both $V$ and $\sigma$ rise during the compaction, such that post-BN, the gas becomes a rotation-dominated extended disc with $V/\sigma\ssim4$, {while the high-dispersion bulge with $V/\sigma\ssim1$ dominates the stars {(\fig{kin})}.}
\rf k.
Pre-BN, the systems are on average in {a rather} crude Jeans (and hydrostatic) equilibrium for stars (and gas) and closer to equilibrium post-BN. This is crudely valid out to several $\Re$; still, the contribution of dispersion velocity to the {estimated} dynamical mass does not necessarily follow the {familiar} predictions for a self-gravitating thin disc or an isothermal sphere {(\fig{jeans})}.
\rf l.
The galaxy spin, which is low at the BN due to the angular-momentum loss associated with the compaction, is steeply rising immediately after the BN, saturating to a high plateau that reflects a post-BN extended ring {(\fig{spin})}.

\smallskip
\no \ul{5. Profiles:}
\rf m.
The gas surface density profile is rather flat pre-BN; {it develops a steep cusp at the BN and rises} with $r$ from a central hole to an extended ring post-BN.
\rf n.
The stellar density profile grows self-similarly at all radii pre-BN, and its growth gradually saturates post-BN.
\rf o.
The sSFR profile is roughly flat pre-BN. It is declining with $r$ at the BN and rising with $r$ post-BN, reflecting inside-out quenching {(\fig{profiles_re})}.
\rf p.
The metallicity profile, which was weakly declining pre-BN, shows a steep decline with $r$ post-BN, reflecting the fresh origin of the extended ring {(\fig{profiles_re})}.
{\rf q.
The dark matter fraction profile, which was high pre-BN, becomes low at small radii during and after compaction -- reflecting the formation of a dominant massive stellar body in the centre during compaction {(\fig{profiles_re})}.
}

\smallskip
\no \ul{6. Supernova feedback, black hole growth, and quenching:}
\rf r.
Pre-BN, the growth of the central black hole is suppressed by supernova feedback. The compaction allows {favourable conditions for} gas accretion onto the BH, triggering rapid BH growth post-BN in masses above $\Mbn$ {\citep{lapiner21}}.
\rf s.
This allows a transition from SN-dominated feedback pre-BN to active AGN feedback post-BN.
\rf t.
Central quenching starts at the BN, and quenching is maintained post-BN by the shutdown of the gas supply through a hot CGM in haloes above the critical mass for virial-shock heating, which may be kept hot by AGN feedback.
{
\rf u.
The major compaction events which are followed by the onset of quenching tend to occur when the galaxy is in the vicinity of the golden mass of galaxy formation, a stellar mass of about $10^{10}\msun$ within dark matter haloes of $\simi 10^{12} \msun$. This mass scale is confined by the supernova and stellar feedback from below and by a hot CGM, as well as AGN feedback from above. These processes limit the dissipative contraction of gas and allow a significant contraction to occur once near the golden mass. The actual trigger of a compaction event is a mechanism that causes a drastic loss of angular momentum. About half the compaction events in our simulations are triggered by wet mergers. Otherwise, they can be caused by collisions of counter-rotating cold streams, violent disk instability, outflows and recycling, as well as other mechanisms {\citep{dekel19_gold}}.
}

\section*{Acknowledgements}

We acknowledge stimulating discussions with Guillermo Barro, Andi Burkert, Yohan Dubois, Sandy Faber, Reinhard Genzel, David Koo, and Nir Mandelker. This work was partly supported by the grants ISF 124/12, ISF 861/20, I-CORE Program of the PBC/ISF 1829/12, BSF 2014-273, PICS 2015-18, GIF I-1341-303.7/2016, DIP STE1869/2-1 GE625/17-1, and NSF AST-1405962.
The cosmological \vela simulations were performed at the National Energy Research Scientific Computing Center (NERSC) at Lawrence Berkeley National Laboratory and at NASA Advanced Supercomputing (NAS) at NASA Ames Research Center. Development and analysis have been performed in the computing cluster at the Hebrew University.

\section*{data availability}

Data and results underlying this article will be shared on reasonable request to the corresponding author.

%% file: appendix.tex

\appendix
\section{Simulations Details}
\label{sec_app:sims_app}

\subsection{Simulation Method and Sub-Grid Physics}
\label{sec_app:art_app}

The \vela simulations utilize the {\sc adaptive refinement tree} ({\sc art}) code \citep{app_krav97,krav03,app_ceverino09}, which accurately follows the evolution of a gravitating $N$-body system and the Eulerian gas dynamics, with an AMR maximum resolution of $17.5-35\pc$ in physical units at all times. The minimum cell size is set to $17.5\pc$ in physical units at expansion factor $a\seq 0.16$ ($z\seq 5.25$), say. Due to the expansion of the whole mesh, while the refinement level remains fixed, the minimum cell size grows in physical units and becomes $35\pc$ at $a\seq 0.32$ ($z\seq 2.125$). At this time, we add a new level to the comoving mesh, so the minimum cell size becomes $17.5\pc$ again, and so on. 

\smallskip
The dark matter particle mass is $8.3\times 10^4 \msun$, and the stellar particles have a minimum mass of $10^3 \msun$. Each AMR cell is split into eight cells once it contains a mass in stars and dark matter higher than $2.6\times 10^5\msun$, equivalent to three dark matter particles, or a gas mass higher than $1.5\times 10^6 \msun$. This quasi-Lagrangian strategy ends at the highest refinement level, which marks the minimum cell size at each redshift. We often refine based on stars and dark matter particles rather than gas, so within the central halo and the star-forming disc, the highest refinement level is reached for gas densities between $\simi 10^{-2}\sdash100\cmc$ and occasionally for densities as low as $\simi 10^{-3}\cmc$. In the outer circum-galactic medium, near the halo virial radius, the median resolution is $\simi 500 \pc$.

\smallskip
Beyond gravity and hydrodynamics, the code incorporates the physics of gas and metal cooling, UV-background photoionization, stochastic star formation, gas recycling, stellar winds and metal enrichment, and thermal feedback from supernovae \citep{app_ceverino10,app_ceverino12}, plus the implementation of feedback from radiation pressure \citep{app_ceverino14}.

\smallskip
We use the {\sc cloudy} code \citep{ferland98} to calculate the cooling and heating rates for a given gas density, temperature, metallicity, and UV background, assuming a slab of thickness $1\kpc$. We assume a uniform UV background, following the redshift-dependent \citet{haardt96} model, except for gas densities higher than $0.1\cmc$ where we use a substantially suppressed UV background ($5.9\times10^6 \ergs\cms\,{\rm Hz}^{-1}$) in order to mimic the partial self-shielding of dense gas. This allows dense gas to cool down to temperatures of $\simi 300\,\mathrm{K}$. The equation of state is assumed to be that of an ideal mono-atomic gas. Artificial fragmentation on the cell size is prevented by introducing a pressure floor, ensuring that the Jeans scale is resolved by at least $N\seq 7$ cells \citep{app_ceverino10}. The pressure floor is given by
\be
\label{eq:Pfloor}
P_{\rm floor}=\frac{G \rho^2 N^2 \Delta^2}{\pi \gamma}
\ee
{\no}where $\rho$ is the gas density, $\Delta$ is the cell size, and $\gamma\seq 5/3$ is the adiabatic index of the gas.

\smallskip
Star formation is allowed to occur at densities above a threshold of $1\cmc$ and temperatures below $10^4\,\mathrm{K}$. Most stars ($>\!90\,\%$) form at temperatures well below $10^3\,\mathrm{K}$, and more than half of them form at $300\,\mathrm{K}$ in cells where the gas density is higher than $10\cmc$. New stellar particles are generated with a timestep of $dt_{\rm SF}\ssim 5\Myr$. We implement a stochastic model, where the probability of forming a stellar particle in a given timestep is
\be
\label{eq:PSF}
\textbf{P}={\rm min}\left(0.2,\:\sqrt{\frac{\rho_{\rm gas}}{1000\:\cmc}}\right)
\ee
{\no}In the formation of a single stellar particle, its mass is equal to
\be
\label{equ:mstar}
m_{*} = m_{\rm gas}\frac{dt_{\rm SF}}{\tau}\ssim 0.42m_{\rm gas}
\ee
where $m_{\rm gas}$ is the mass of gas in the cell where the particle is being formed, and $\tau\seq 12\Myr$ is a parameter of the simulations which was calibrated to match the empirical Kennicutt-Schmidt law \citep{kennicutt98}. We assume a \citealt{chabrier03} stellar initial mass function. Further details can be found in \citet{app_ceverino14}.

\smallskip
The thermal stellar feedback model releases energy from stellar winds and supernova explosions at a constant heating rate over $40\Myr$ following star formation. The heating rate due to feedback may or may not overcome the cooling rate, depending on the gas conditions in the star-forming regions \citep{app_ds86,app_ceverino09}, as we do not explicitly switch off cooling in these regions. The effect of runaway stars is included by applying a velocity kick of $\simi 10\kms$ to $30\,\%$ of the newly formed stellar particles. The code also includes the later effects of Type Ia supernova and stellar mass loss, and it follows the metal enrichment of the ISM.

\smallskip
Radiation pressure is incorporated through the addition of a non-thermal pressure term to the total gas pressure in regions where ionizing photons from massive stars are produced and may be trapped. This ionizing radiation injects momentum in the cells neighbouring massive star particles younger than $5\Myr$ whose column density exceeds $10^{21}\cms$, isotropically pressurizing the star-forming regions. The expression for the radiation pressure is
\be
\label{eq:RP}
P_{\rm rad} = \frac{\Gamma m_*}{R^2 c}
\ee
{\no}where $R$ is set to half the cell size for the cell hosting a stellar mass $m_*$ and the cell size for its closest neighbours. The value of $\Gamma$ is taken from the stellar population synthesis code \texttt{STARBURST99} \citep{starburst99}. We use a value of $\Gamma\seq 10^{36}\ergs\msun^{-1}$ which corresponds to the time-averaged luminosity per unit mass of the ionizing radiation during the first $5\Myr$ of the evolution of a single stellar population. After $5\Myr$, the number of high-mass stars and ionizing photons declines significantly. Since the significance of radiation pressure also depends on the optical depth of the gas within a cell, we use a hydrogen column density threshold of $N \seq  10^{21}\cms$, above which ionizing radiation is effectively trapped and radiation pressure is added to the total gas pressure. See the `RadPre' model of \citet{app_ceverino14} for further details.

\smallskip
The initial conditions for the simulations are based on dark matter haloes drawn from dissipationless N-body simulations at lower resolution in three comoving cosmological boxes (box sizes of 10, 20, and 40 Mpc/h). We assume the standard $\Lambda$CDM cosmology with the WMAP5 cosmological parameters, namely $\Omega_{\rm m}\seq 0.27$, $\Omega_{\Lambda}\seq 0.73$, $\Omega_{\rm b}\seq 0.045$, $h\seq 0.7$ and $\sigma_8\seq 0.82$ \citep{app_WMAP5}. Each halo was selected to have a given virial mass at $z \seq  1$ and no ongoing major merger at that time. This latter criterion eliminates less than $10\,\%$ of the haloes, which tend to be in dense proto-cluster environments at $z\ssim1$. The target virial masses at $z\seq 1$ were selected in the range $\Mv \seq  2\times10^{11}-2\times10^{12}\Msun$, with a median of $5.6\times10^{11}\Msun$. If left in isolation, the median mass at $z\seq 0$ would be $\simi10^{12}\Msun$. In practice, the actual mass range is broader, with some haloes merging into more massive haloes that host groups at $z\seq 0$.

\subsection{The Galaxy Sample}
\label{sec_app:sample_app}
\smallskip
More than half the sample was evolved to $z\le 1$, and all but three were evolved to $z\le 2$. The simulation outputs were stored and analyzed at fixed intervals in cosmic expansion factor $a\seq (1+z)^{-1}$, $\Delta a\seq 0.01$, which at $z\seq 2$ corresponds to about $100\Myr$. \tabApp{sample_app} lists the final available snapshot for each of the 34 runs in terms of expansion factor, $a_{\rm fin}$, and redshift, $z_{\rm fin}$. We detect the central galaxy in the final available output using the \texttt{AdaptaHOP} group finder \citep{app_tweed09,app_colombi13} on the stellar particles. It is then traced back in time until it contains fewer than 100 stellar particles, typically between $a\seq 0.10\sdash0.13$ ($z\seq6.5\sdash9$).

\begin{table*}
\centering
\begin{tabular}{@{}lcccccccc}
\hline
Galaxy & $\Rv$ & $\Mv$ & $M_{\rm s,\,0.1\Rv}$ & $M_{\rm g,\,0.1\Rv}$ &
${\rm SFR}_{\rm 0.1\Rv}$ & $a_{\rm fin}$ & $z_{\rm fin}$
\\
& $\kpc$ & $10^{12}\Msun$ & $10^{10}\Msun$ & $10^{10}\Msun$ & $\sy$ &  &
   \\
\hline
\hline
01      & 58.25  & 0.16 & 0.22 & 0.20  & 2.65  & 0.50 & 1.00 \\
02      & 54.50  & 0.13 & 0.19 & 0.24  & 1.81  & 0.50 & 1.00 \\
03      & 55.50  & 0.14 & 0.42 & 0.15  & 3.72  & 0.50 & 1.00 \\
04      & 53.50  & 0.12 & 0.09 & 0.13  & 0.48  & 0.50 & 1.00 \\
05      & 44.50  & 0.07 & 0.09 & 0.11  & 0.58  & 0.50 & 1.00 \\ \rowcolor{TabCol}
06      & 88.25  & 0.55 & 2.19 & 0.49  & 20.60 & 0.37 & 1.70 \\ \rowcolor{TabCol}
07      & 104.25 & 0.90 & 6.23 & 1.62  & 25.86 & 0.54 & 0.85 \\
08      & 70.50  & 0.28 & 0.35 & 0.21  & 5.69  & 0.57 & 0.75 \\
09      & 70.50  & 0.27 & 1.07 & 0.43  & 3.93  & 0.40 & 1.50 \\
10      & 55.25  & 0.13 & 0.63 & 0.18  & 3.22  & 0.56 & 0.79 \\
11      & 69.50  & 0.27 & 0.92 & 0.70  & 14.59 & 0.46 & 1.17 \\ \rowcolor{TabCol}
12      & 69.50  & 0.27 & 2.03 & 0.28  & 2.88  & 0.44 & 1.27 \\
13      & 72.50  & 0.31 & 0.78 & 0.90  & 13.88 & 0.51 & 0.96 \\ \rowcolor{TabCol}
14      & 76.50  & 0.36 & 1.33 & 0.70  & 25.40 & 0.42 & 1.38 \\
15      & 53.25  & 0.12 & 0.55 & 0.15  & 1.57  & 0.56 & 0.79 \\ \rowcolor{TabCol}
16 $^*$ & 62.75  & 0.50 & 4.27 & 0.67  & 20.26 & 0.24 & 3.17 \\ \rowcolor{TabCol}
17 $^*$ & 105.75 & 1.13 & 8.97 & 1.55  & 64.82 & 0.31 & 2.23 \\ \rowcolor{TabCol}
19 $^*$ & 91.25  & 0.88 & 4.52 & 0.88  & 40.78 & 0.29 & 2.45 \\ \rowcolor{TabCol}
20      & 87.50  & 0.53 & 3.84 & 0.62  & 7.15  & 0.44 & 1.27 \\ \rowcolor{TabCol}
21      & 92.25  & 0.62 & 4.21 & 0.68  & 9.50  & 0.50 & 1.00 \\ \rowcolor{TabCol}
22      & 85.50  & 0.49 & 4.50 & 0.32  & 12.07 & 0.50 & 1.00 \\
23      & 57.00  & 0.15 & 0.82 & 0.24  & 3.28  & 0.50 & 1.00 \\
24      & 70.25  & 0.28 & 0.92 & 0.42  & 4.31  & 0.48 & 1.08 \\
25      & 65.00  & 0.22 & 0.73 & 0.13  & 2.30  & 0.50 & 1.00 \\ \rowcolor{TabCol}
26      & 76.75  & 0.36 & 1.61 & 0.40  & 9.63  & 0.50 & 1.00 \\
27      & 75.50  & 0.33 & 0.83 & 0.61  & 7.92  & 0.50 & 1.00 \\
28      & 63.50  & 0.20 & 0.24 & 0.32  & 5.70  & 0.50 & 1.00 \\ \rowcolor{TabCol}
29      & 89.25  & 0.52 & 2.56 & 0.49  & 18.49 & 0.50 & 1.00 \\ \rowcolor{TabCol}
30      & 73.25  & 0.31 & 1.67 & 0.52  & 3.80  & 0.34 & 1.94 \\ \rowcolor{TabCol}
32      & 90.50  & 0.59 & 2.71 & 0.56  & 14.89 & 0.33 & 2.03 \\ \rowcolor{TabCol}
33      & 101.25 & 0.83 & 5.04 & 0.63  & 32.68 & 0.39 & 1.56 \\ \rowcolor{TabCol}
34      & 86.50  & 0.52 & 1.66 & 0.62  & 14.66 & 0.35 & 1.86 \\
\hline
\end{tabular}
\caption{
The sample of 32 simulated \vela galaxies. Quoted are the following quantities at $z\seq 2$ (except for the three galaxies marked $^*$, where they are quoted at the final output $z_{\rm fin}\sgt2$): the virial radius, $\Rv$, the total virial mass, $\Mv$, the stellar mass, $\Ms$, the gas mass, $\Mg$ and the star formation rate. The galaxy properties $\Ms$, $\Mg$ and SFR are quoted within $0.1\Rv$. Also listed are the final simulation scale factor, $a_{\rm fin}$ and redshift, $z_{\rm fin}$. {Galaxies marked with (without) background colour are the high mass (low mass) def with mass above (below) the median stellar mass at $z\seq 2$, $M_{\rm s, med}(z\seq 2)\ssim10^{10.06}\msun$}.
}
\label{tab_app:sample_app}
\end{table*}

\begin{table}
\centering
\begin{tabular}{@{}lcccccccc}
\hline
Galaxy & $a_{\rm comp^1: BNg}$ & $a_{\rm comp^2}$ & $a_{\rm comp^3}$\\
\hline
\hline
01      & 0.38  & ----  & ----  \\ 
02      & 0.40  & ----  & ----  \\ 
03      & 0.29  & 0.47  & ----  \\ 
04      & 0.34  & 0.42  & ----  \\ 
05      & 0.44  & ----  & ----  \\  \rowcolor{TabCol}
06      & 0.19  & 0.32  & ----  \\  \rowcolor{TabCol}
07      & 0.24  & 0.20  & ----  \\ 
08      & 0.33  & 0.54  & ----  \\ 
09      & 0.22  & 0.40  & ----  \\ 
10      & 0.27  & 0.43  & ----  \\ 
11      & 0.23  & 0.34  & ----  \\  \rowcolor{TabCol}
12      & 0.20  & ----  & ----  \\ 
13      & 0.29  & ----  & ----  \\  \rowcolor{TabCol}
14      & 0.34  & ----  & ----  \\ 
15      & 0.29  & 0.39  & ----  \\  \rowcolor{TabCol}
16      & 0.18  & 0.15  & ----  \\  \rowcolor{TabCol}
17      & 0.15  & 0.24  & ----  \\  \rowcolor{TabCol}
19      & 0.13  & 0.29  & ----  \\  \rowcolor{TabCol}
20      & 0.22  & 0.18  & ----  \\  \rowcolor{TabCol}
21      & 0.25  & 0.20  & 0.14  \\  \rowcolor{TabCol}
22      & 0.19  & ----  & ----  \\ 
23      & 0.26  & 0.20  & ----  \\ 
24      & 0.25  & 0.44  & ----  \\ 
25      & 0.32  & 0.23  & 0.37  \\  \rowcolor{TabCol}
26      & 0.26  & 0.19  & ----  \\ 
27      & 0.32  & 0.25  & ----  \\ 
28      & 0.40  & ----  & ----  \\  \rowcolor{TabCol}
29      & 0.19  & 0.29  & ----  \\  \rowcolor{TabCol}
30      & 0.19  & ----  & ----  \\  \rowcolor{TabCol}
32      & 0.16  & ----  & ----  \\  \rowcolor{TabCol}
33      & 0.21  & 0.26  & 0.32  \\  \rowcolor{TabCol}
34      & 0.29  & 0.23  & ----  \\ 
\hline
\end{tabular}
\caption{
{
The time of compaction events in the 32 simulated \vela galaxies sample. Quoted are the scale factors at the peak of identified compaction events: the major compaction event, $a_{\rm comp^1: BN}$, and up to two more secondary compactions, $a_{\rm comp^2}$, $a_{\rm comp^3}$. 
}
}
\label{tab_app:comp_time_app}
\end{table}

\subsection{Measuring Physical Quantities}
\label{sec_app:physical}

The virial mass, $\Mv$, is the total mass within a sphere of radius $\Rv$ that encompasses a given overdensity $\Delta(z)$ relative to the cosmological mean mass density, $\Delta(z)\seq (18\pi^2-82\oml(z)-39\oml(z)^2)/\omm(z)$, where $\oml(z)$ and $\omm(z)$ are the cosmological density parameters of mass and cosmological parameter at $z$ \citep{app_bryan98}. The virial properties for the 34 galaxies are listed in \tabApp{sample_app}, together with the stellar mass, $\Ms$, gas mass, $\Mg$, and star-formation rate, SFR, within $0.1\Rv$. These are quoted at $z\seq 2$ except for the five galaxies that were stopped at higher redshift, marked by $*$, for which we quote the properties at $z\seq z_{\rm fin}$.

\smallskip
The stellar mass, $\Ms$, is the instantaneous mass in stars after accounting for past stellar mass loss. The simulation calculates stellar mass loss using an analytic fitting formula, where $10\,\%$, $20\,\%$ and $30\,\%$ of the initial mass of a stellar particle is lost after $30\Myr$, $260\Myr$ and $2\Gyr$ respectively.

\smallskip
The SFR is obtained by ${\rm SFR}\seq \langle M_{\rm *,\,i}(t_{\rm age}\slt t_{\rm max})/t_{\rm max} \rangle_{t_{\rm max}}$, where $M_{\rm *,\,i}(t_{\rm age}\slt t_{\rm max})$ is the mass at birth in stars younger than $t_{\rm max}$. The average $\langle\cdot\rangle_{t_{max}}$ is obtained for $t_{\rm max}$ in the interval $[40,80]\Myr$ in steps of $0.2 \Myr$ to reduce fluctuations due to the $\simi5 \Myr$ discreteness in stellar birth times in the simulation. The $t_{\rm max}$ in this range are long enough to ensure good statistics. This represents the SFR on $\simi 60\Myr$ timescales, which is a crude proxy for H$_\alpha$-based SFR measurements, while UV-based measurements are sensitive to stars younger than $\simi 100\Myr$. We define the specific SFR as ${\rm sSFR}\seq SFR/\Ms$ and the gas fraction as $\fg\seq \Mg/(\Mg+\Ms)$.

\smallskip
{We choose a fixed radius of $1\kpc$ to identify the BN phase and for all measurements of the central quantities associated with compactness. This choice is motivated by empirical findings in both simulations and observations, which indicate a universal behaviour and a tighter relation than comparable selection for the central region. A similar qualitative result is obtained when we use the effective radius, which is typically comparable to $\simi 1\kpc$, especially in the pre-BN phase, although with a somewhat larger scatter (e.g., see the effect of using $\Sige$ as opposed to $\Sig1$ in \fig{L_V07_V12}). Choosing a slightly smaller or larger fixed volume does not change the results; however, the signature of compaction may be washed out if the central gas mass is measured within a larger volume that encompasses the galaxy's outer regions.}

\smallskip
{The \sersic index, n, \citep{sersic63} is measured by fitting a single component \sersic to the stellar surface density profiles derived from a face-on view. Using $\Sigma_{\rm s}(r, n) \seq  \Sigma_{\rm e} \exp [-b(n) ((r/\Re)^{1/n} -1)]$, where $b(n)$ is chosen such that $\Sigma_{\rm e}$ is the surface density at $\Re$, and approximated as $b(n)\ssimeq 1.9992n-0.3271$ \citep{graham05,gerbrandt15}.}

{
\smallskip
The strength of compaction can be roughly estimated by quantifying the rise in the central gas (or stellar) mass before the peak of compaction and the post-BN central gas depletion (or central quenching). We start by defining a window in time around the BN, $t_{\pm}\seq t_{\rm BNg}\pm 0.5\thub$, where $\thub$ is measured at the time of the peak of compaction, $t_{\rm BNg}$. 
The strength of compaction using $\M1s$:
\be
S_{\rm BN, stars} = 
\frac{\Delta\log M_{\rm s1, pre}}
{\max(s_{\rm post, min},\ \Delta\log M_{\rm s1, post})},
\ee
with $\Delta\log M_{\rm s1, pre}\seq \log\M1s(t_{\rm BN_\star}) \sdash \log\M1s(t_{-})$, and $\Delta\log M_{\rm s1, post}\seq \log\M1s(t_{+}) \sdash \log\M1s(t_{\rm BN_\star})$. Namely, a typical strong compaction event will show both a steep rise in $\M1s$ from the pre-BN to the $\M1s$ shoulder at $t_{\rm BN_\star}$ and a small change post-BN in the central stellar mass. In some cases, the post-BN $\M1s$ shows a slight decline with time, possibly caused by an adiabatic expansion in response to gas outflows and stellar mass loss. Therefore, we set a minimal value ($s_{\rm post, min}$) in the denominator to avoid values of $\leq\! 0$. Strong compactions typically show a pre-BN rise of $\Delta\log M_{\rm s1, pre}\sgt1.5$ and post-BN plateau with $\Delta\log M_{\rm s1, post}\slt 0.1$. Here we consider compaction event as strong for galaxies with $S_{\rm BN, stars}$ larger than $\simi 15$. The high-mass subsample shows a median value of $S_{\rm BN, stars}\ssim 40$, while the low-mass subsample typically shows indications for a weaker BN with $S_{\rm BN, stars}\ssim 10$.
 
\smallskip
Similarly, one may use the central gas mass, $M_{\rm g1}$ to determine the strength of compaction:
\be
S_{\rm BN, gas} = 
\Delta\log M_{\rm g1, pre} \cdot \Delta\log M_{\rm g1, post},
\ee
with $\Delta\log M_{\rm g1, pre}\seq \log M_{\rm g1}(t_{\rm BNg}) \sdash \log M_{\rm g1}(t_{-})$, and $\Delta\log M_{\rm g1, post}\seq \log M_{\rm g1}(t_{\rm BNg}) \sdash \log M_{\rm g1}(t_{+})$. Strong compactions typically show a pre-BN rise in $M_{\rm g1}$ of an order of magnitude, $\Delta\log M_{\rm g1, pre}\ssim 1$, and post-BN depletion with $\Delta\log M_{\rm g1, post} \sgt 0.6$. Galaxies with compaction strength of $S_{\rm BN, gas} \sgt 0.5$ are considered strong compactions. Using the central gas mass, $M_{\rm g1}$, again shows stronger compactions for the high-mass subsample compared to the low-mass galaxies, with median values of $\simi 0.8, 0.4$ respectively. 
}

\section{Critical mass}\label{sec_app:cmass_app}

\begin{figure*} 
\centering
\includegraphics[width=0.45\textwidth]{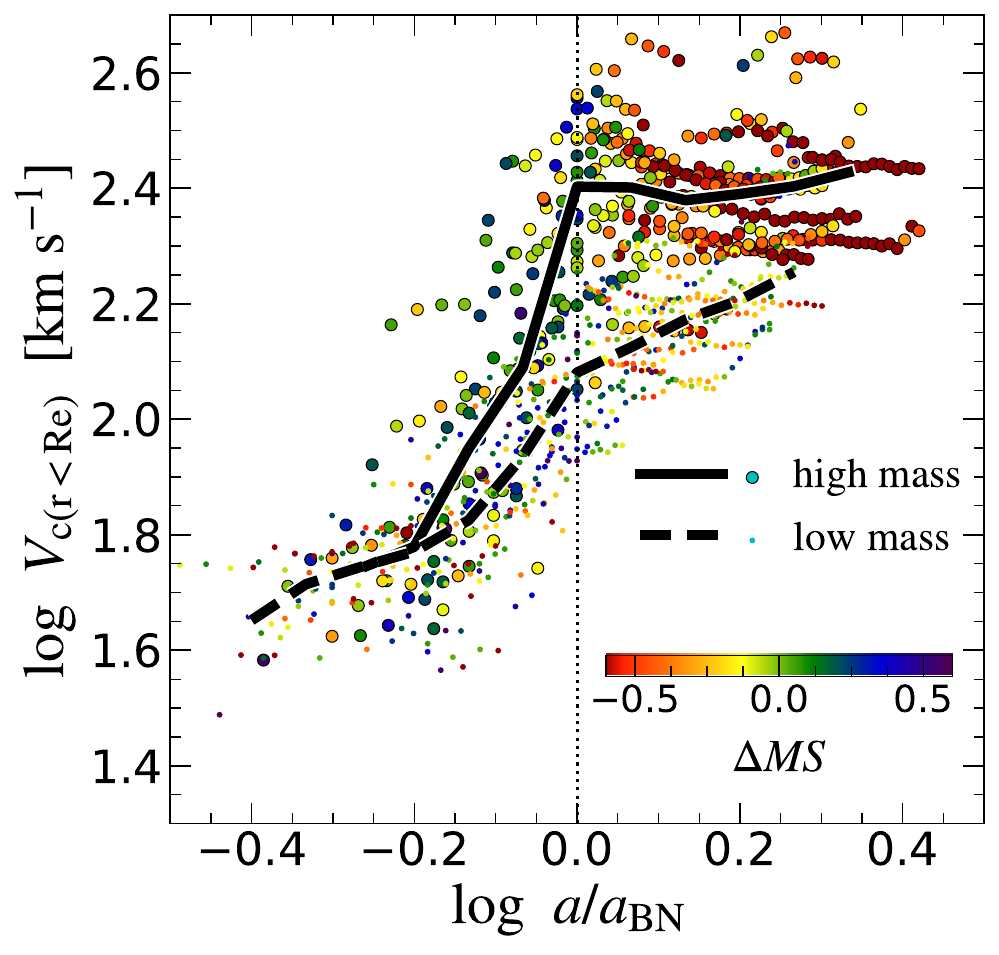}
\caption{{Characteristic BN mass. {Shown is the circular velocity within the effective radius $\Re$ as a function of time (expansion factor), stacked with respect to the BN time ($\abn$). The medians are shown for the high-mass (solid) and low-mass (dashed) subsamples. The colour marks the deviation from the MS. The BN occurs at characteristic values, albeit with a larger scatter when compared to the virial velocity in \fig{char_M-a}. The value of $\Ve$ is significantly lower for the low-mass subsample (a later BN), indicating a weaker compaction event which results in a lower concentration.}} 
}
\label{fig_app:char_Ve-a}
\end{figure*}

{In figure \figApp{char_Ve-a}, we show the circular velocity ($\Ve$) measured at $\Re$. We find a similar transition at the characteristic mass as in $\Vv$. However, $\Ve$ shows a significantly larger scatter than the scatter in the virial velocity, particularly around the compaction time. At the time of the BN, $\Ve$ of the low-mass subsample is significantly lower than the massive subsample by $\simi 0.3$dex. This indicates the weaker compactions experienced by the low-mass galaxies, which results in a lower concentration. When we examine the cause for the scatter in $\Ve$, we find that while the total mass within the effective radius shows a comparable scatter to the one seen for the virial mass, the scatter in $\Re$ is much larger. Around the time of compaction, we find a $1\sigma$ scatter of $\simi 0.37$dex in the virial radius and $\simi 0.6$dex for the effective radius (not shown here). This larger scatter is caused by repeated episodes of compaction, which push $\Re$ down (see \se{radius}). In \fig{Re-Ms}, we showed the evolution of $\Re$ for the high-mass subsample. While the low-mass galaxies show similar qualitative behaviour, their tendency for repeated small compaction events at later times causes larger scatter with less prominent shrinkage in $\Re$ during the BN (see \figApp{Re-Ms_LM}).}

\section{Size-mass}

\begin{figure*} 
\includegraphics[width=0.95\textwidth]{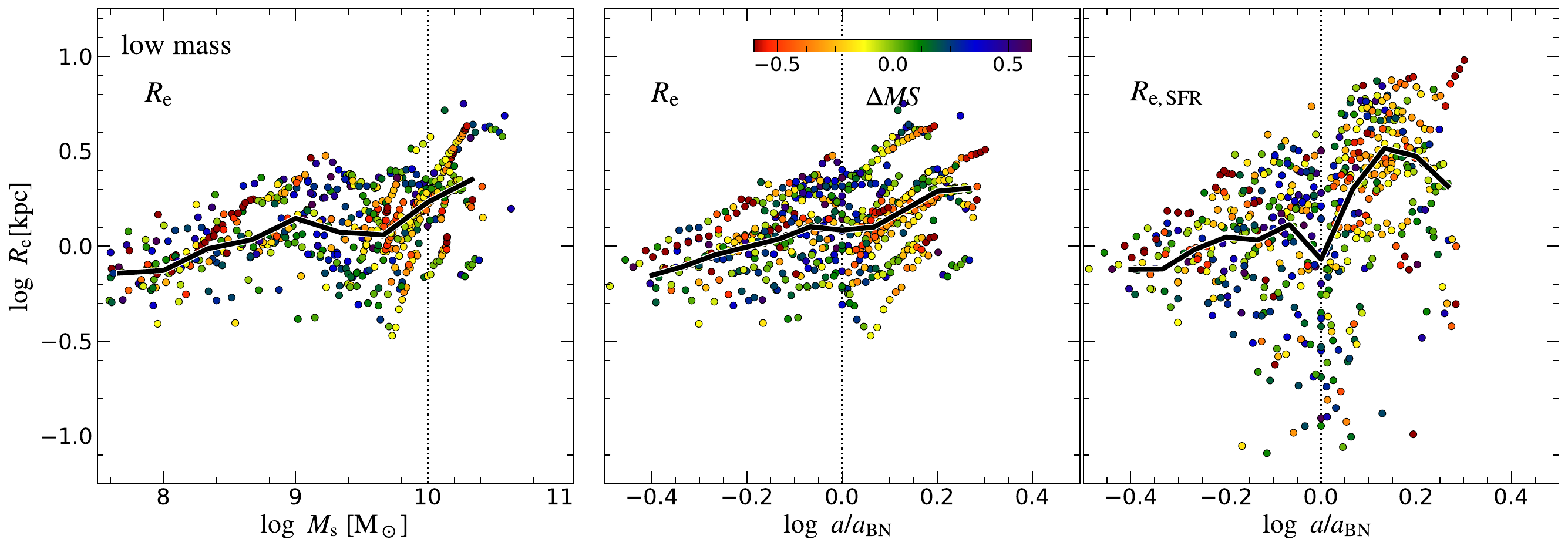}\caption{
{Similar to \fig{Re-Ms}, size evolution for the low-mass subsample (as defined at $z\seq 2$). The distance from the MS ridge, $\DMS$, is indicated by colour. Two left panels: Evolution of effective radius, as a function of $a/\abn$ and $\Ms$. Third panel: Evolution of the effective radius of the SFR, $\Resf$ as a function of $a/\abn$.}
}
\label{fig_app:Re-Ms_LM}
\end{figure*}

\begin{figure*} 
\includegraphics[width=0.85\columnwidth]{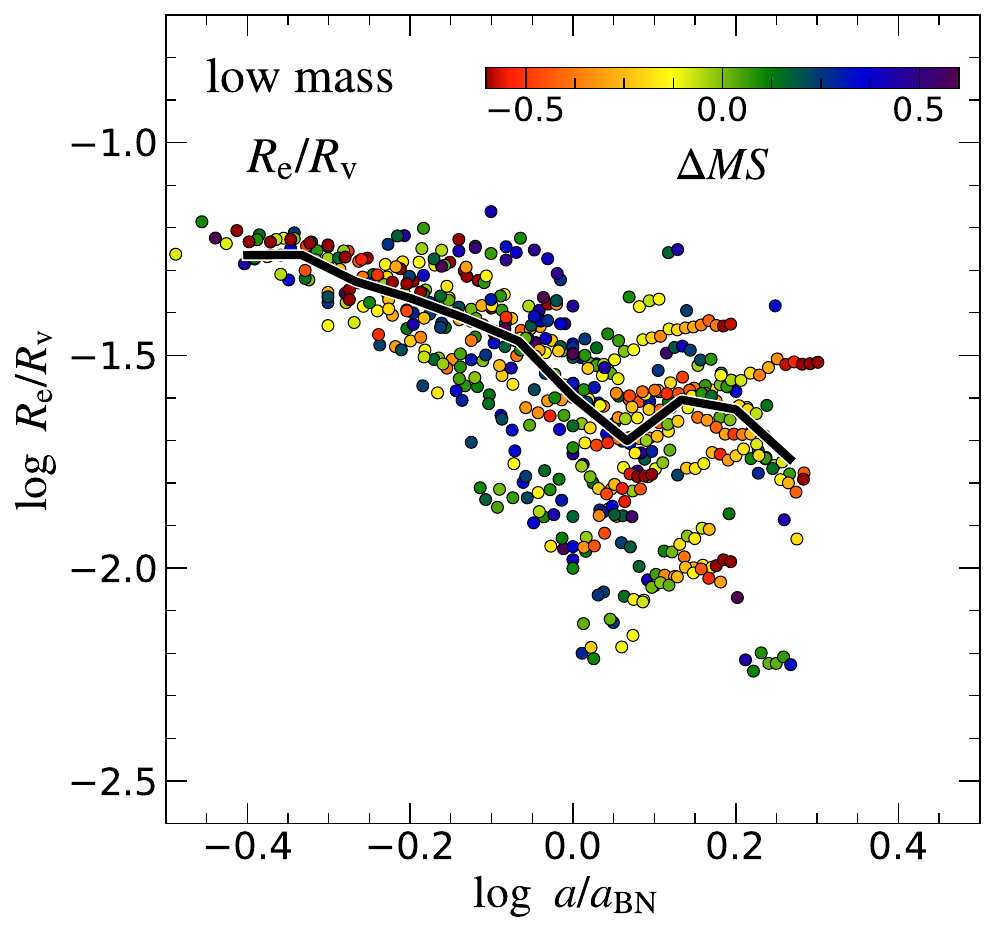}
\caption{
{Similar to \fig{ReORv-abn} for the low-mass subsample. Evolution of the effective stellar radius scaled by the virial radius, $\Re/\Rv$, as a function of $a/\abn$. The distance from the MS ridge is indicated by colour.}
}
\label{fig_app:ReORv-abn_LM}
\end{figure*}

{In \figApp{Re-Ms_LM} we show the same size evolution as seen in \fig{Re-Ms} for the low-mass subsample (defined at $z\seq 2$). The low-mass subsample typically goes through a compaction event at lower redshifts when the universe is less dense, and the gas fraction is lower than that of the high-mass galaxies. As a result, they tend to experience weaker compaction, manifested here in the evolution of $\Re$. Namely, the effective radius does shrink to some extent, yet the effect is not as striking as seen for the high-mass subsample. At the BN the median of $\Re$ and $\Resf$ for the low-mass galaxies is $\simi0.5$dex above that of the high-mass ones.}

\smallskip
{In \figApp{ReORv-abn_LM} we show the evolution of the effective stellar radius scaled by the virial radius ($\Re/\Rv$) as a function of $a/\abn$ for the low-mass subsample (see \fig{ReORv-abn} for the high-mass subsample). Here we see a decline in the ratio with time (a similar decline is also seen without scaling the x-axis to the time of compaction). Although the low-mass subsample tends to have weaker compaction events, the decline in $\Re/\Rv$ shows that even the repeated episodes of relatively weak compaction events, which mildly drive $\Re$ down, still manage to impair the ability to deduce the size of the galaxy based on the classical galaxy-size indicator of $\Re\sprop \lambda_{\rm halo} \Rv$ on a one-to-one basis \citep{app_jiang19_spin}.}

\section{More complementary figures}

\begin{figure*} 
\includegraphics[width=0.95\textwidth]{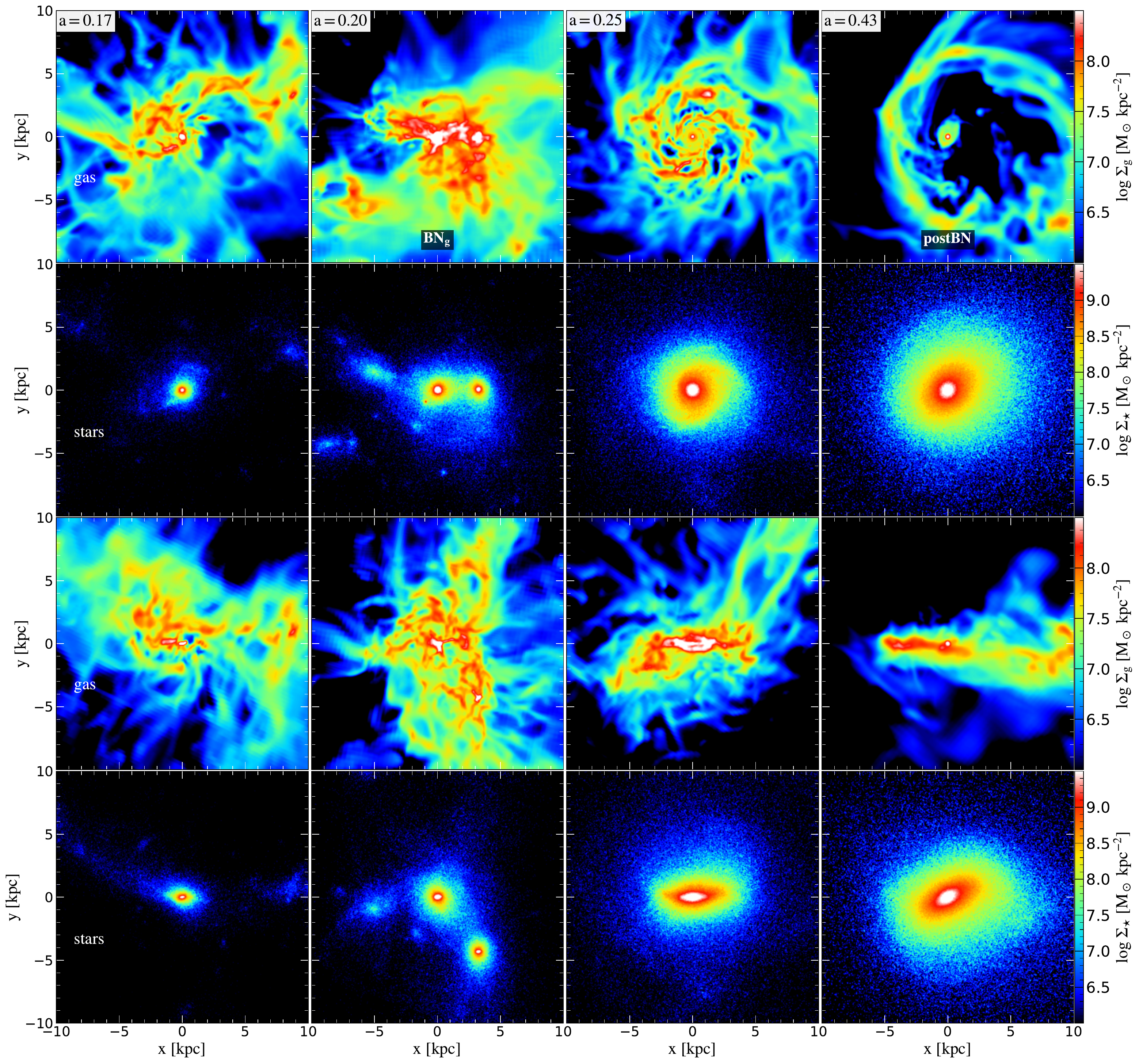}
\caption{
The evolution of V12 through the BN phase. The projected densities in cold gas and stars are shown face-on and edge-on. Gas compaction forms a compact, star-forming BN that leads to a compact stellar core. The latter remains compact while the gas is depleted from the central body, which passively turns into a RN. The newly accreted gas forms an extended star-forming clumpy ring about the RN, and a stellar envelope grows by dry minor mergers.
}
\label{fig_app:pics_V12}
\end{figure*}

\begin{figure*} 
\centering
\includegraphics[width=0.85\textwidth]
{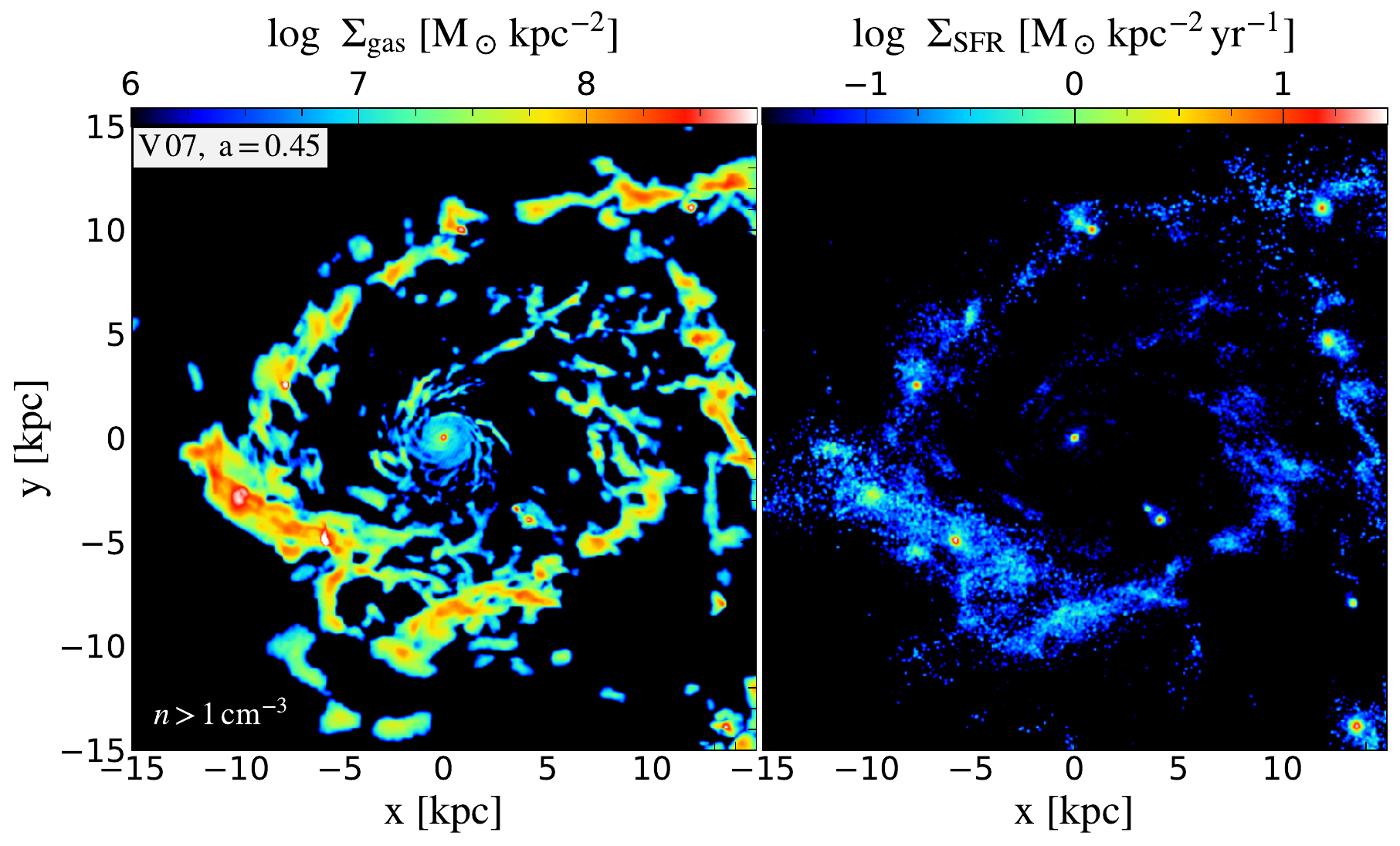}
\caption{
{Projected densities of gas (left) and SFR (right), both shown here in face-on view. On the left panel, we project only cells with a volumetric density of $n\sgt1\cmc$. The local spatial distribution of SFR (or that of young stars) is similar to the distribution of the gas above a certain threshold. Here we use a threshold of $1\cmc$ for a qualitative visual demonstration.}
}
\label{fig_app:pics2D_gas_vs_sfr}
\end{figure*}

\begin{figure*} 
\includegraphics[width=0.8\textwidth]{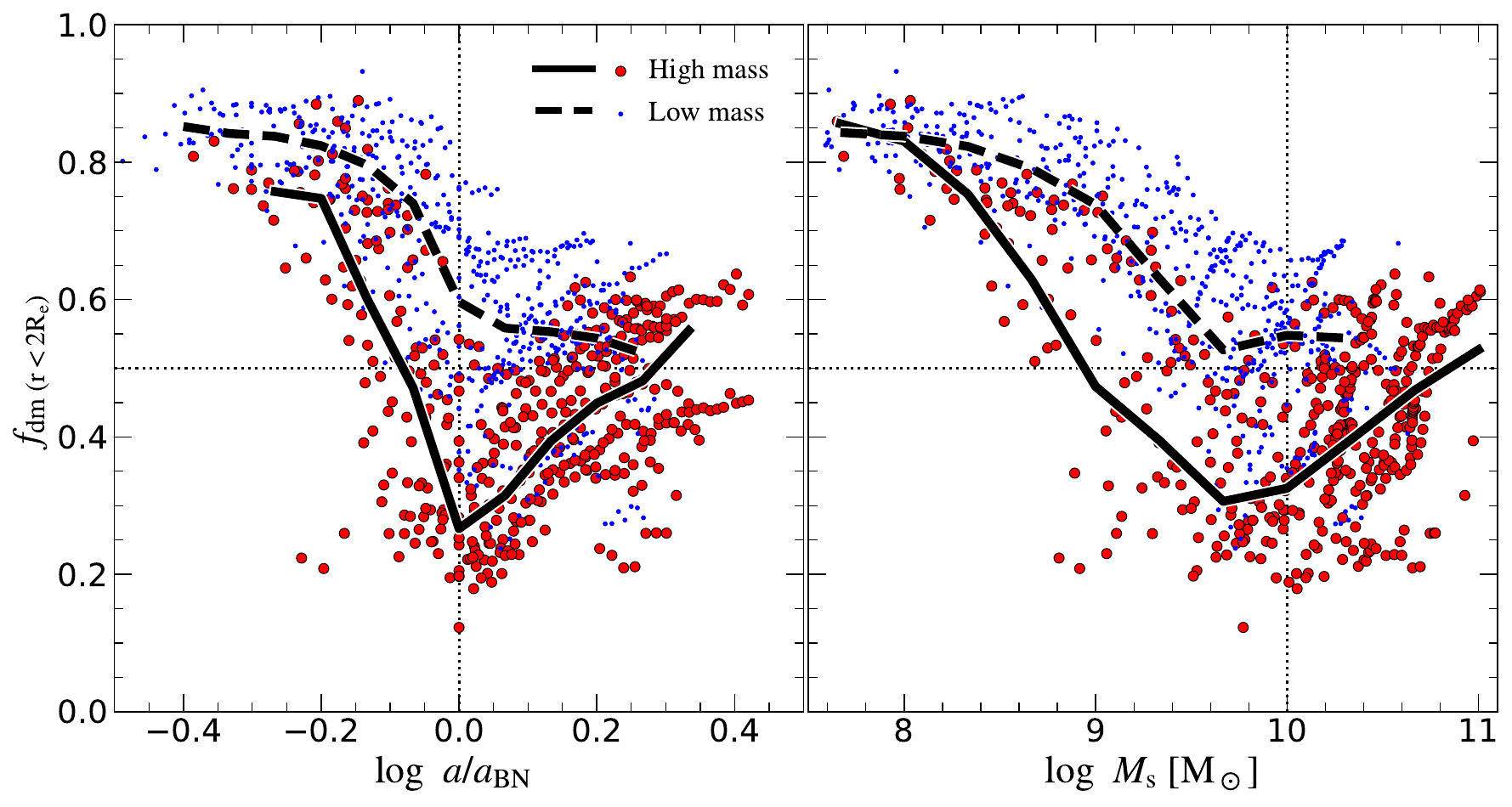}
\caption{
Similar to \fig{fdm_Re}. Here, the dark matter fraction is measured within $\simi 2\Re$.
}
\label{fig_app:fdm_2Re}
\end{figure*}

\begin{figure*} 
\includegraphics[width=0.95\textwidth]{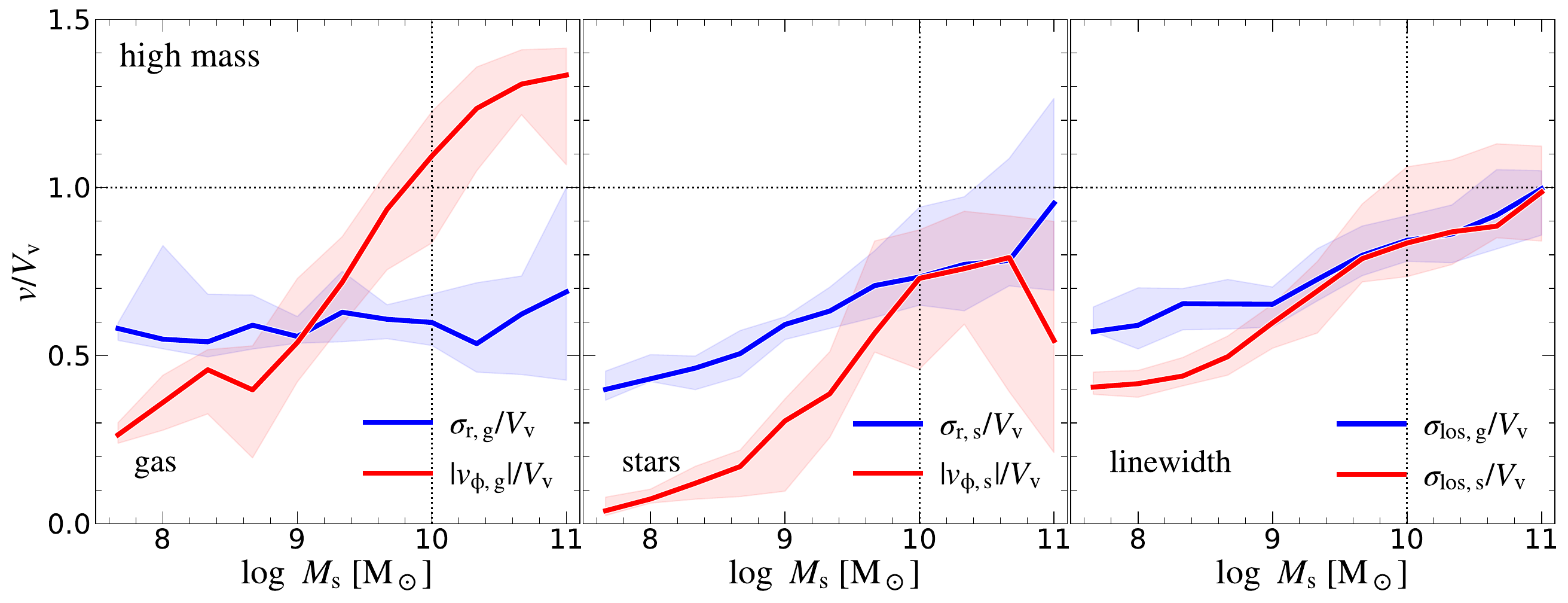}
\caption{
{Same as \fig{kin} but as a function of stellar mass}
}
\label{fig_app:kin_Ms}
\end{figure*}

\begin{figure*} 
\includegraphics[width=0.95\textwidth]{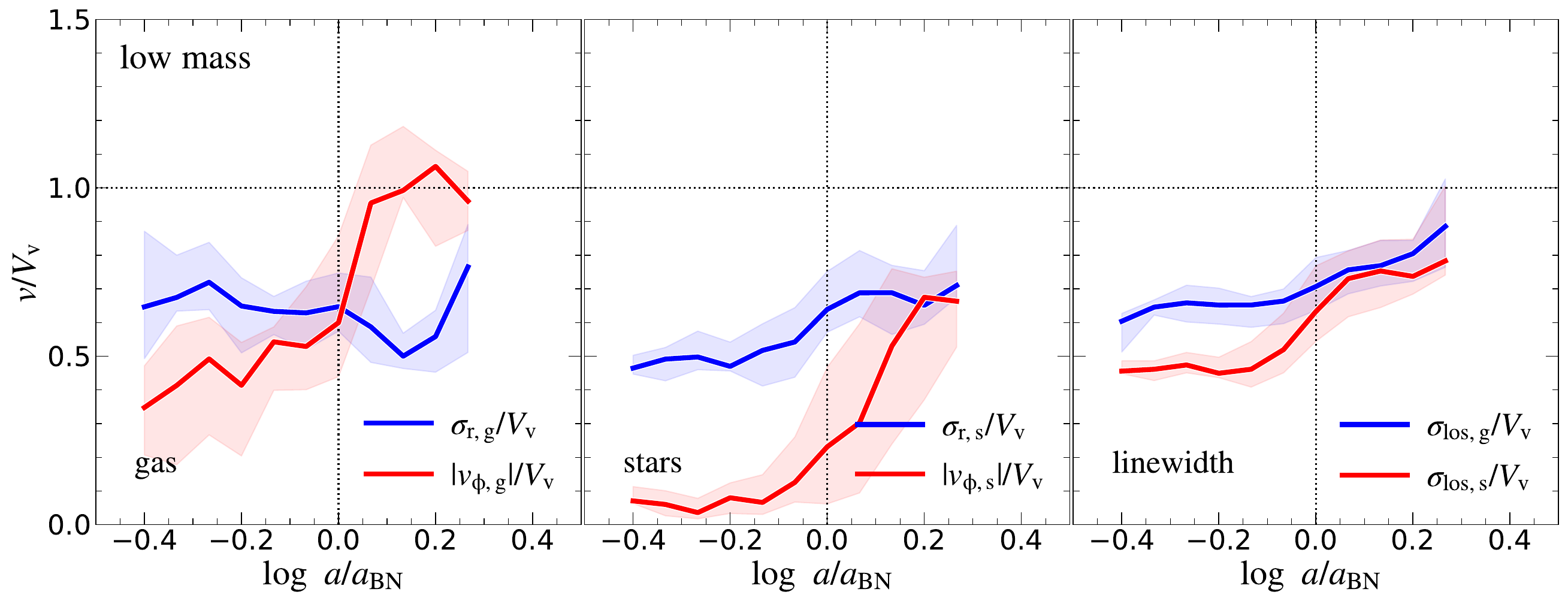}
\caption{
{Same as \fig{kin}, here shown for the low-mass subsample.}
}
\label{fig_app:kin_LM}
\end{figure*}

\begin{figure*} 
\includegraphics[width=0.95\textwidth]{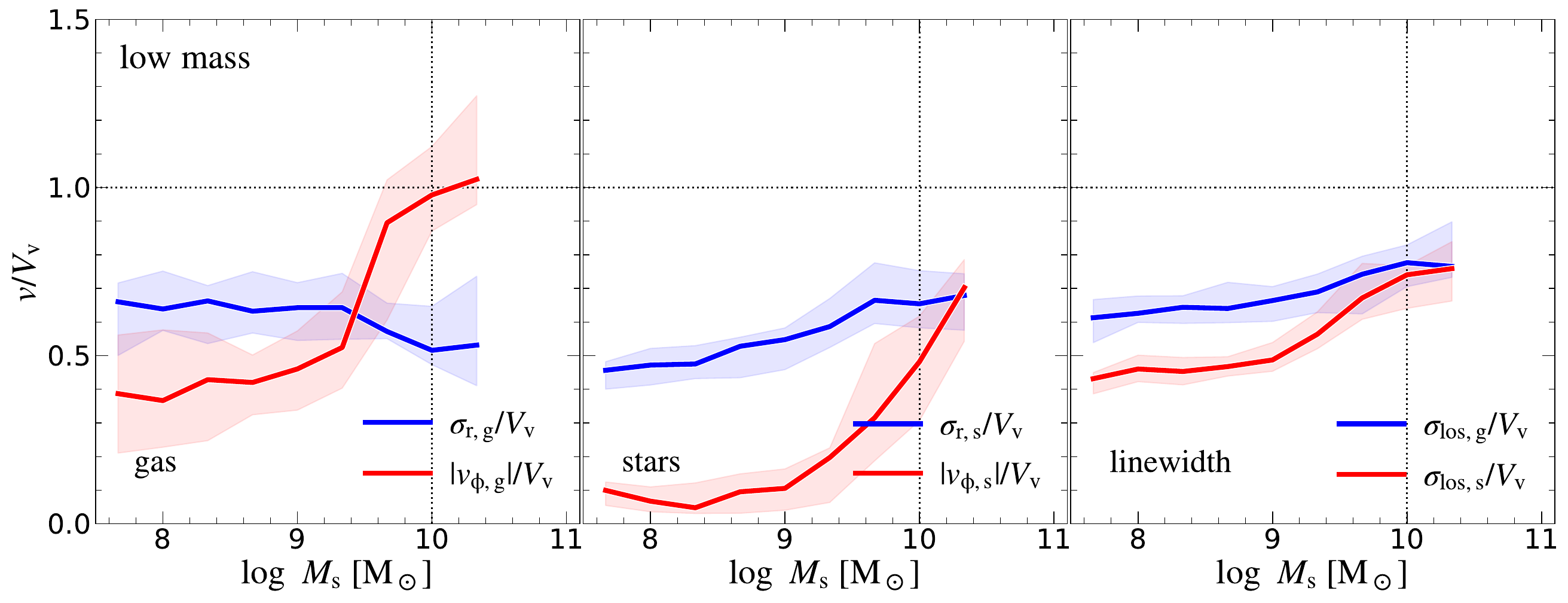}
\caption{
{Same as \figApp{kin_LM}, here shown as a function of the stellar mass.}
}
\label{fig_app:kin_Ms_LM}
\end{figure*}

\begin{figure*} 
\includegraphics[width=0.95\textwidth]{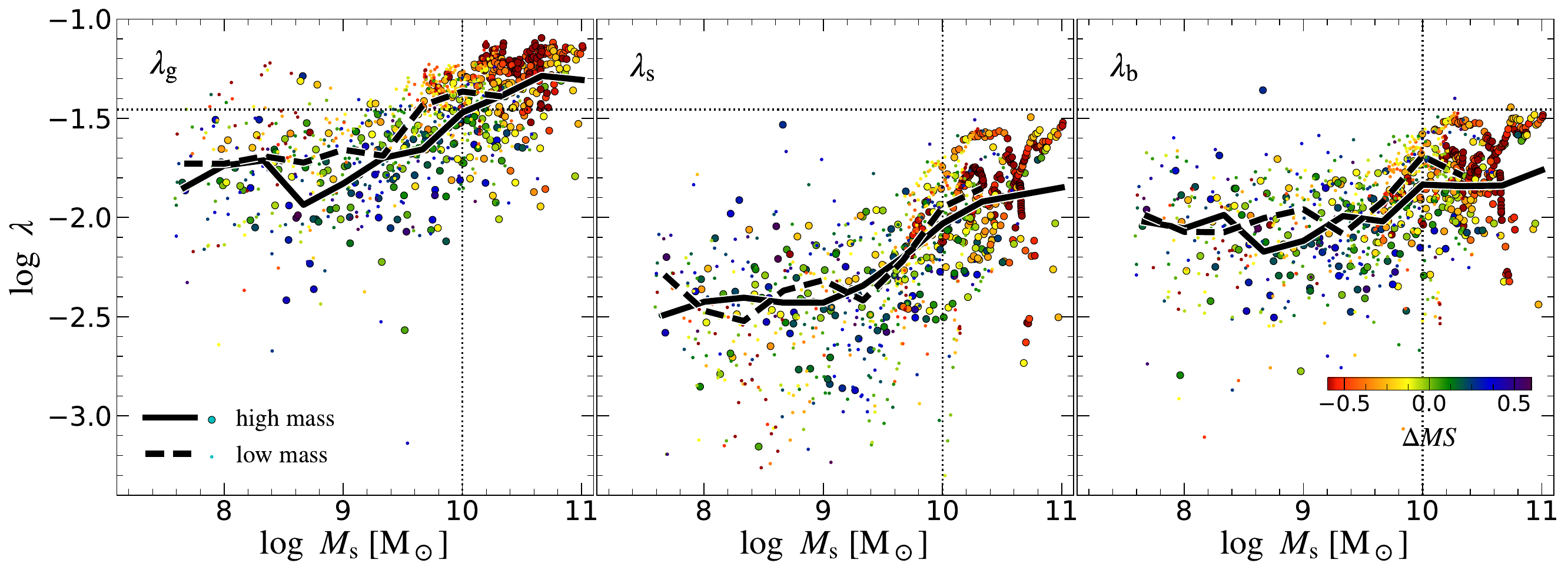}
\caption{
{Same as \fig{spin} but as a function of stellar mass}
}
\label{fig_app:spin_Ms}
\end{figure*}

\begin{figure*} 
\includegraphics[width=0.8\textwidth]{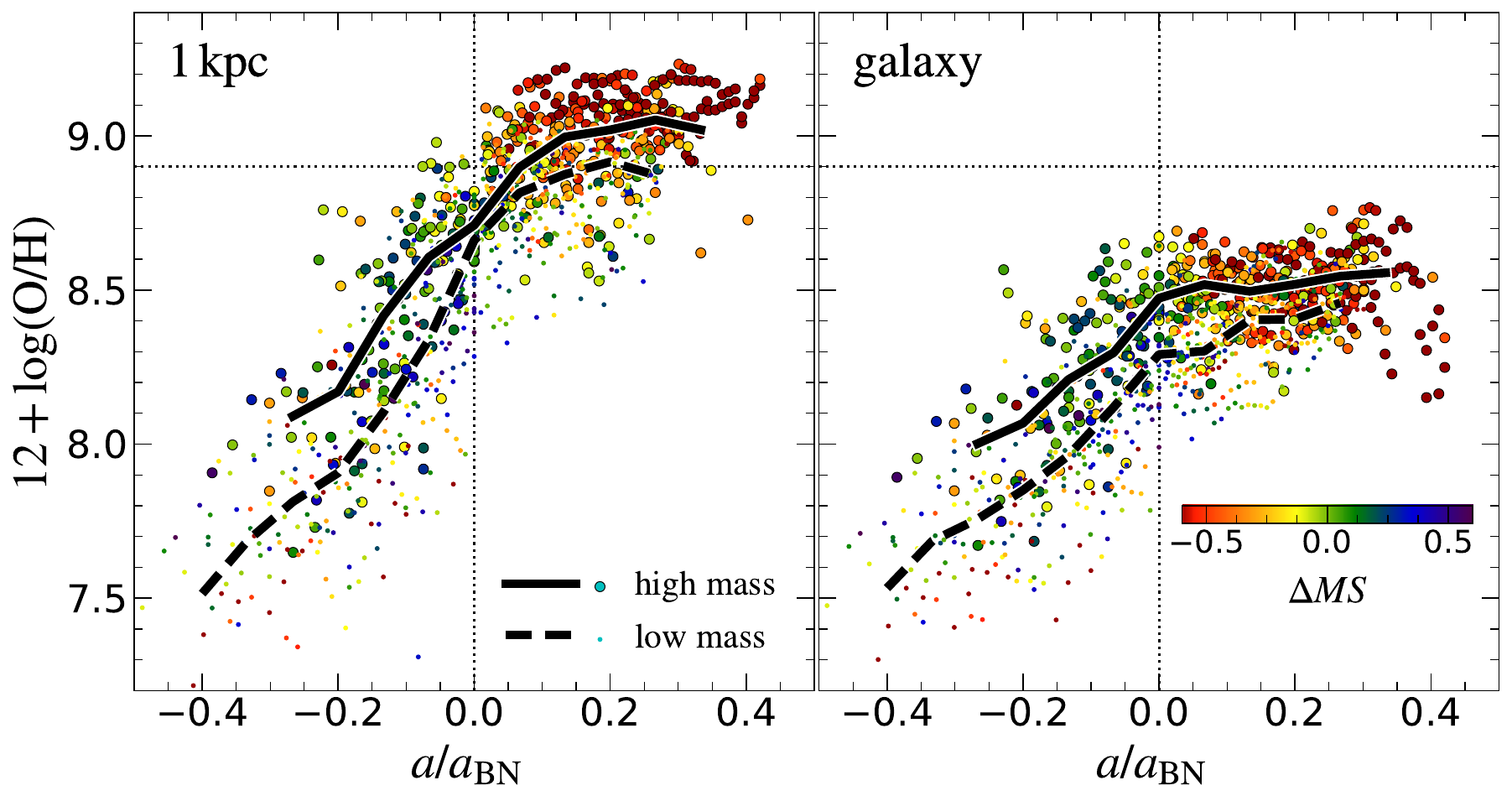}
\caption{
{Similar to \fig{Z-abn}, here the gas metallicity is measured within $1\kpc$ (left), and within $0.1\Rv$ (right) as a function of $a/\abn$. The colour indicates the distance from the MS, the medians refer to the high-mass and low-mass subsamples (defined at $z\seq 2$), and the horizontal dotted line marks solar metallicity. The metallicity sharply rises during the compaction phase and saturates into a plateau post-BN. The relatively low metallicity within $0.1\Rv$ during the post-BN phase reflects a dilution caused by the formation of extended rings of stream-fed fresh gas.}
}
\label{fig_app:Z-abn_app}
\end{figure*}

\begin{figure*} 
\includegraphics[width=0.95\textwidth]{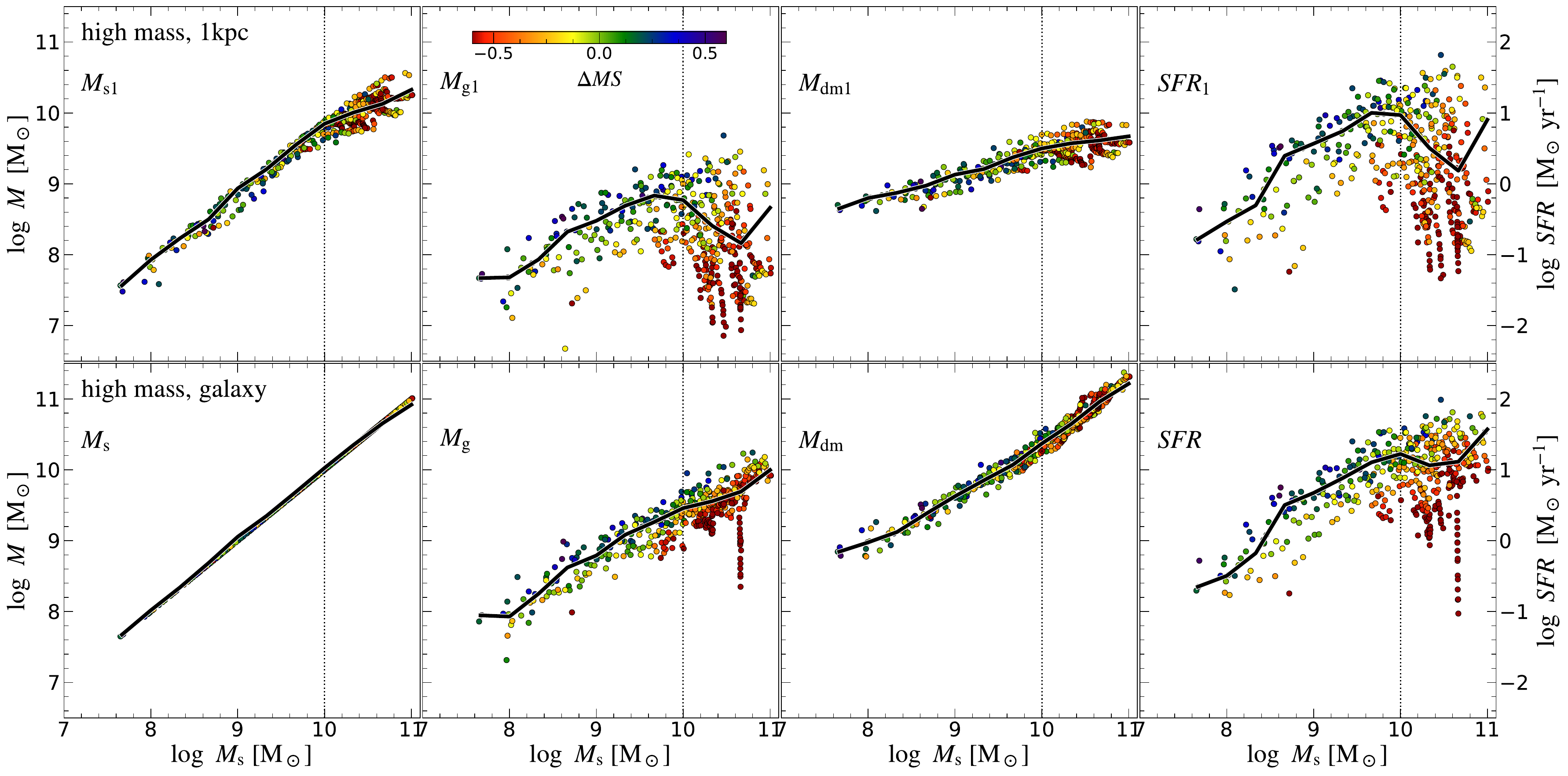}
\caption{
Same as \fig{M-a_hm} but as a function of stellar mass. The figures are qualitatively similar, reflecting the fact that the BN tends to occur at a characteristic mass $\simi10^{10}\msun$ {(bottom left panel is redundant, shown only to preserve the order of panels as \fig{M-a_hm})}. \FigApp{M-Ms_lm} shows the same for the low-mass subsample.
}
\label{fig_app:M-Ms_hm}
\end{figure*}

\begin{figure*} 
\includegraphics[width=0.95\textwidth]{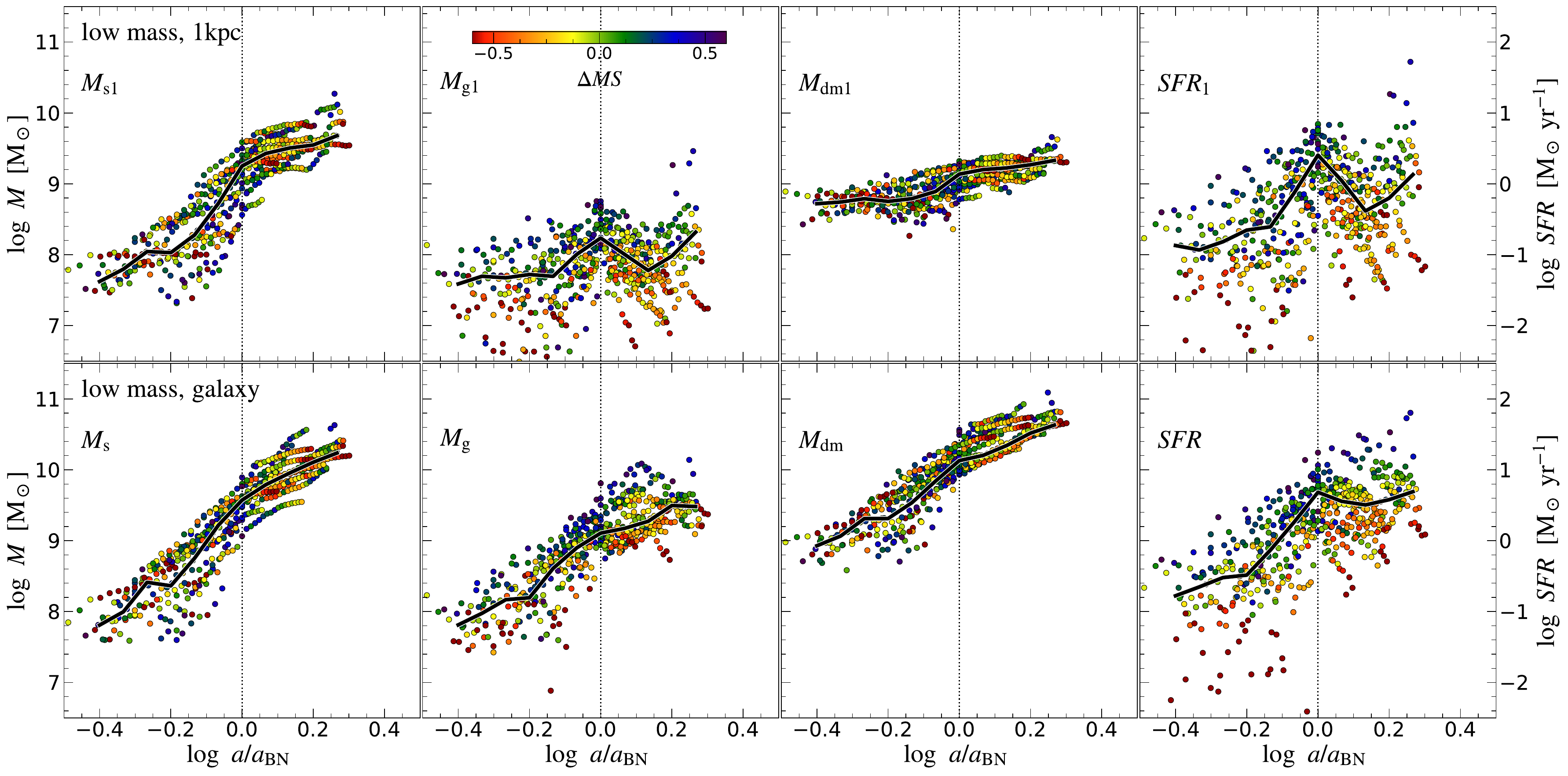}
\caption{
Similar to \fig{M-a_hm}, shown here for the low-mass subsample. 
}
\label{fig_app:M-a_lm}
\end{figure*}

\begin{figure*} 
\includegraphics[width=0.95\textwidth]{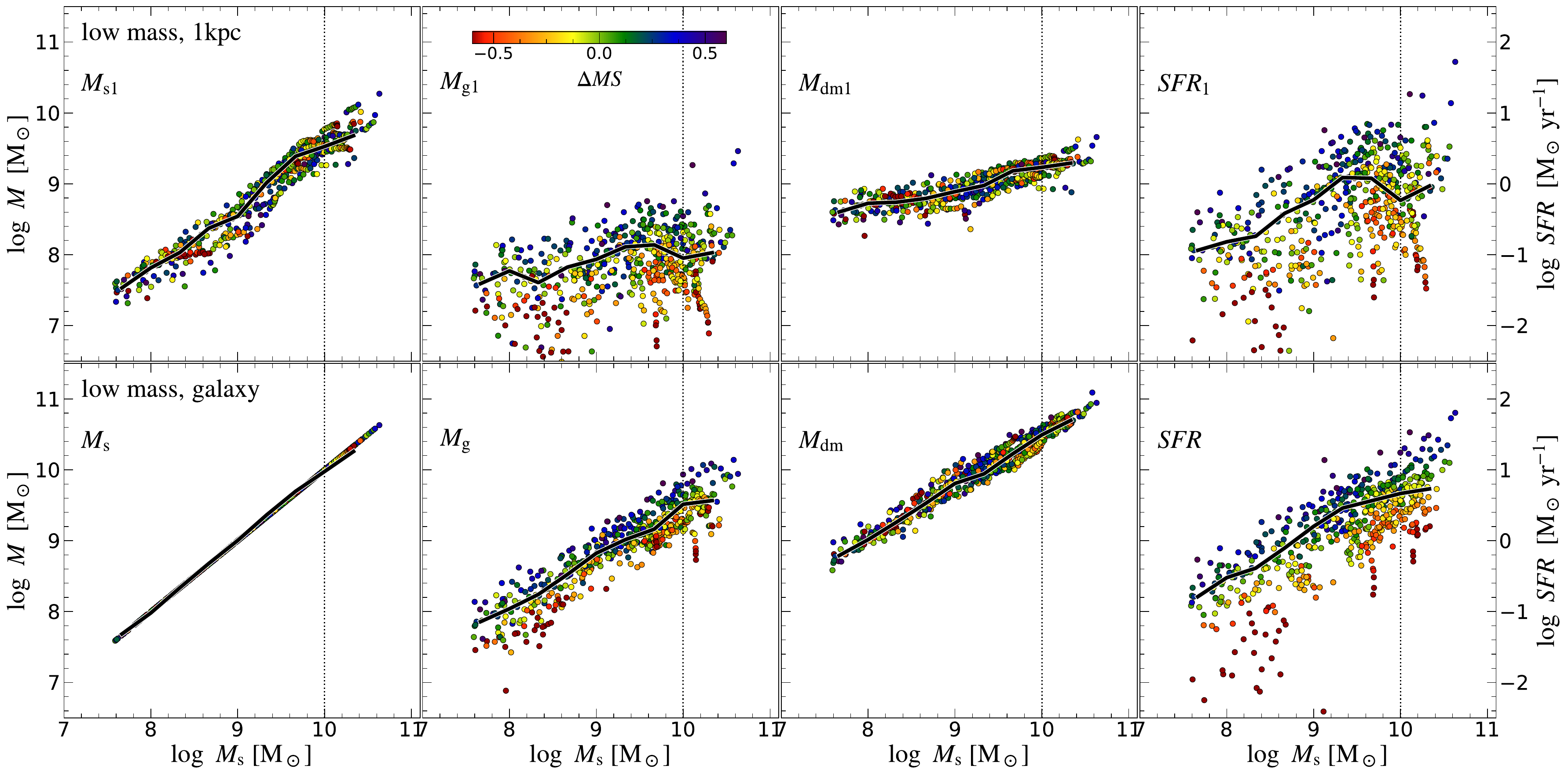}
\caption{
{Same as \figApp{M-a_lm} but as a function of stellar mass (bottom left panel is redundant, shown only to preserve the order of panels as \fig{M-a_hm}). \FigApp{M-Ms_hm} shows the same for the high-mass subsample.}
}
\label{fig_app:M-Ms_lm}
\end{figure*}

\begin{figure*} 
\includegraphics[width=0.8\textwidth]{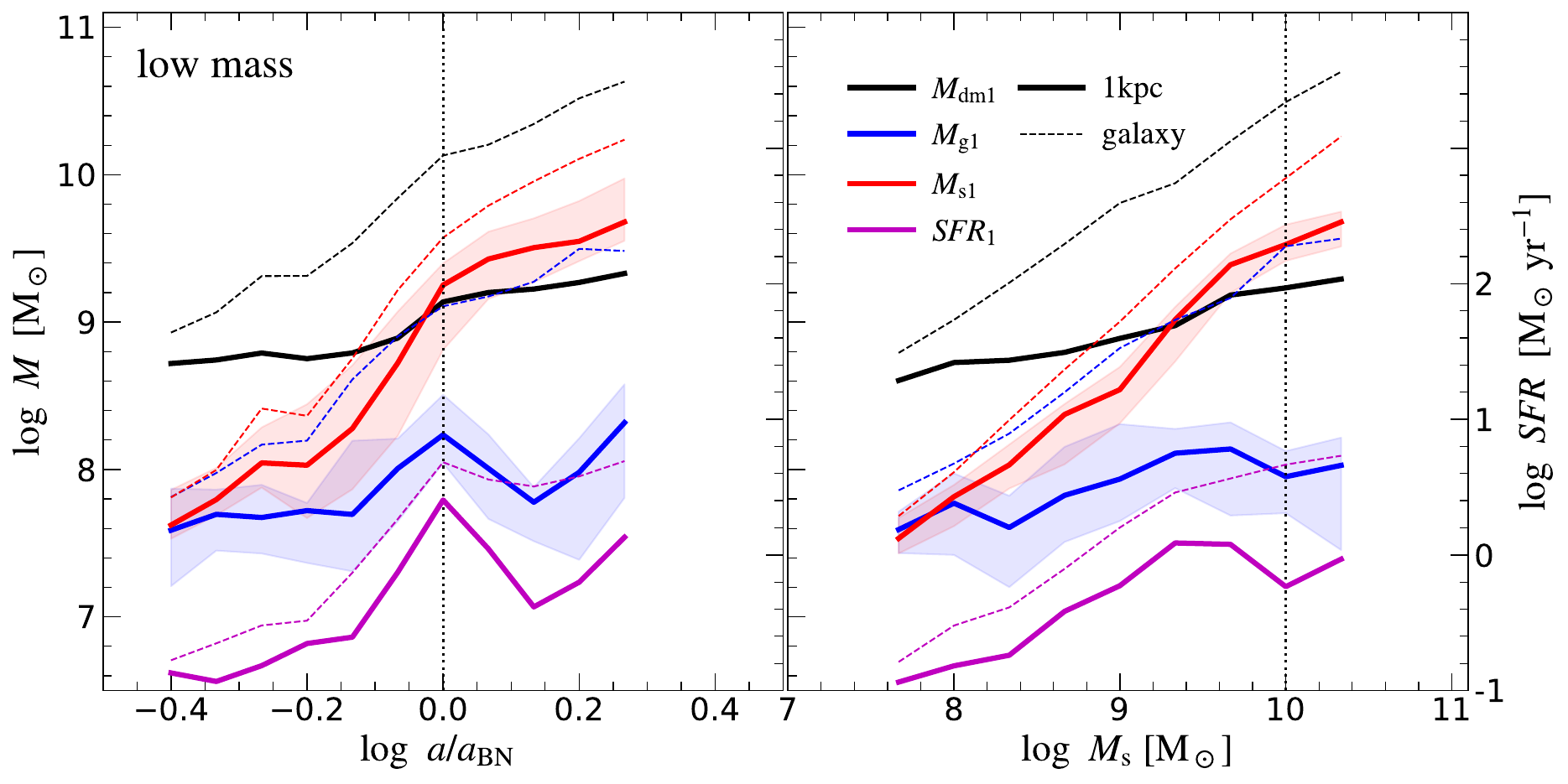}
\caption{
Same as \fig{M-a_med}, medians for the low-mass subsample
}
\label{fig_app:M-a_med_lm}
\end{figure*}

\begin{figure*} 
\includegraphics[width=0.8\textwidth]{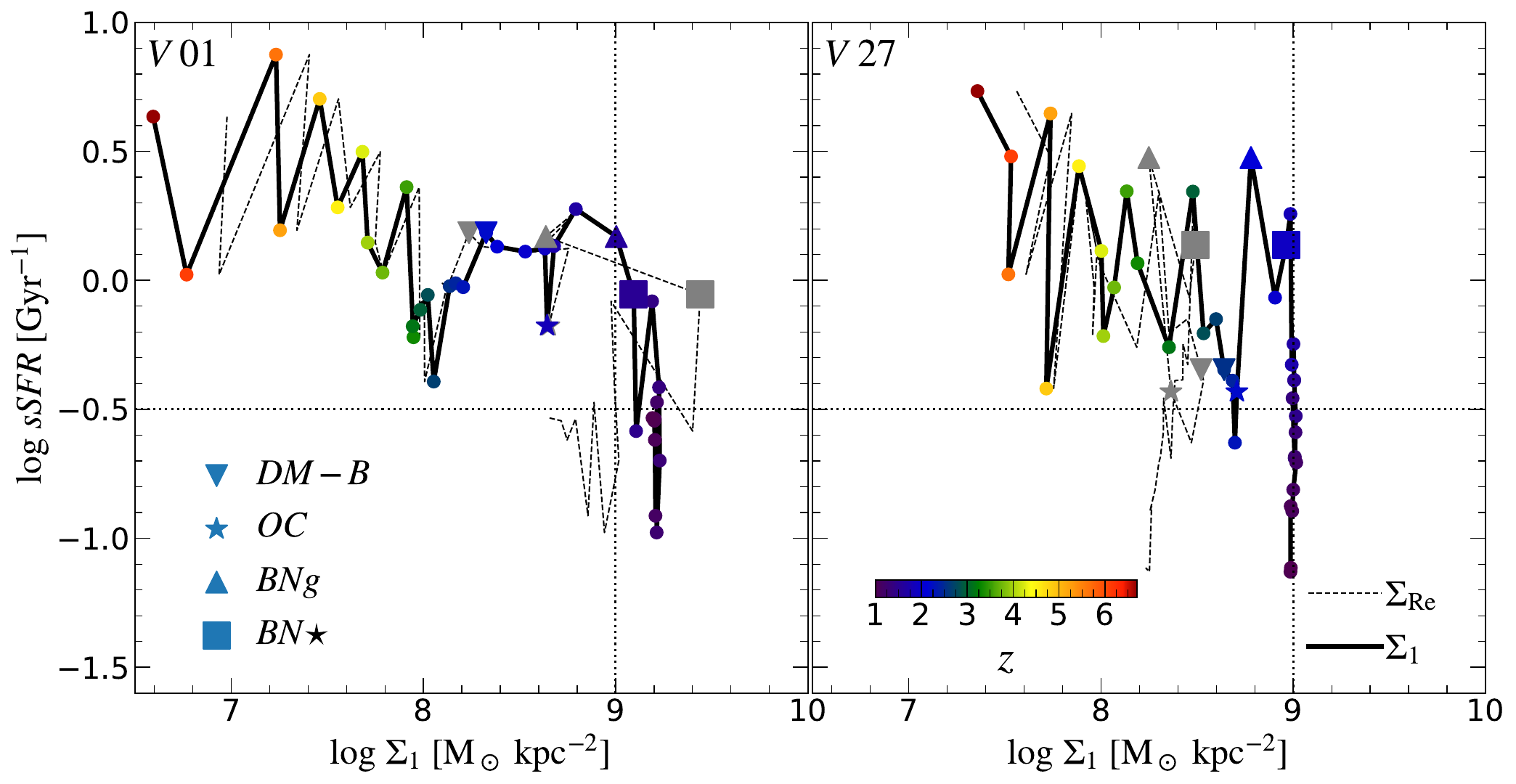}
\caption{
Same as \fig{L_V07_V12} but for two other simulated galaxies, V01 and V27, showing oscillations in sSFR due to repeating episodes of compaction and quenching attempts as long as the galaxy is below the critical mass.
}
\label{fig_app:L_V01_V27}
\end{figure*}


\begin{table*}
\centering
\begin{tabular}{@{}lcccc}
\hline
Quantity &    $\rm slope_{post-BN}$ & $\rp_{\rm post-BN}$ &  $\rm slope_{pre-BN}$ & $\rp_{\rm pre-BN}$\\
\hline
\hline\\
$\tdep $          & $-0.51\pm 0.07 $  & $0.71 $  & $-0.47\pm 0.03 $    & $0.67$   \\ 
$\fg $            & $+0.48\pm 0.04 $  & $0.69 $  & $+0.47\pm 0.03 $    & $0.69 $  \\ 
$\rm SFR_1 $      & $+1.22\pm 0.06 $  & $0.73 $  & $+1.10\pm 0.11 $    & $0.64 $  \\ 
\hline \\
$\M1s $           & $-0.01\pm 0.02 $  & $0.03 $  & $-0.04\pm 0.10 $    & $0.06 $  \\ 
$\Re $            & $+0.03\pm 0.04 $  & $0.08 $  & $-0.02\pm 0.03 $    & $0.01 $  \\ 
$n $              & $+0.03\pm 0.03 $  & $0.12 $  & $+0.05\pm 0.04 $    & $0.06 $  \\ 
\hline\\
$\vosg $          & $-0.31\pm 0.03 $  & $0.54 $  & $-0.21\pm 0.06 $    & $0.14 $  \\ 
$\lambdag $       & $-0.35\pm 0.04 $  & $0.62 $  & $-0.19\pm 0.03 $    & $0.29 $  \\ 
$\Resf $          & $-0.36\pm 0.09 $  & $0.36 $  & $-0.15\pm 0.06 $    & $0.15 $  \\ 
\hline\hline\\
$Z_{\rm3kpc,g} $  & $-0.08\pm 0.01 $  & $0.40 $  & $+0.10\pm 0.06 $    & $0.22 $  \\ 
$Z_{\rm1kpc,g} $  & $-0.48\pm 0.02 $  & $0.66 $  & $+0.00\pm 0.02 $    & $0.06 $  \\ 
$e_{\rm s,\Re} $  & $+0.15\pm 0.03 $  & $0.29 $  & $+0.04\pm 0.05 $    & $0.04 $  \\ 
$f_{\rm s,\Re} $  & $-0.05\pm 0.01 $  & $0.14 $  & $+0.01\pm 0.04 $    & $0.14 $  \\ 
\hline
\hline
\end{tabular}
\caption{
{Best-fitting slopes for gradients across the MS. Columns: (1) Galaxy property. (2) Slope of the gradient across the MS for post-BN phase, $\Ms\sgt10^{9.5}\Msun$ (solid black lines in \fig{gradients_q-dMS}). (3) Pearson correlation coefficient for post-BN phase (4) Slope of the gradient across the MS for the pre-BN phase, $\Ms\slt 10^{9.5}\Msun$. (5) Pearson correlation coefficient for pre-BN phase.}
}
\label{tab_app:gradients_dMS_app}
\end{table*}

\begin{figure*} 
\includegraphics[width=0.8\textwidth]{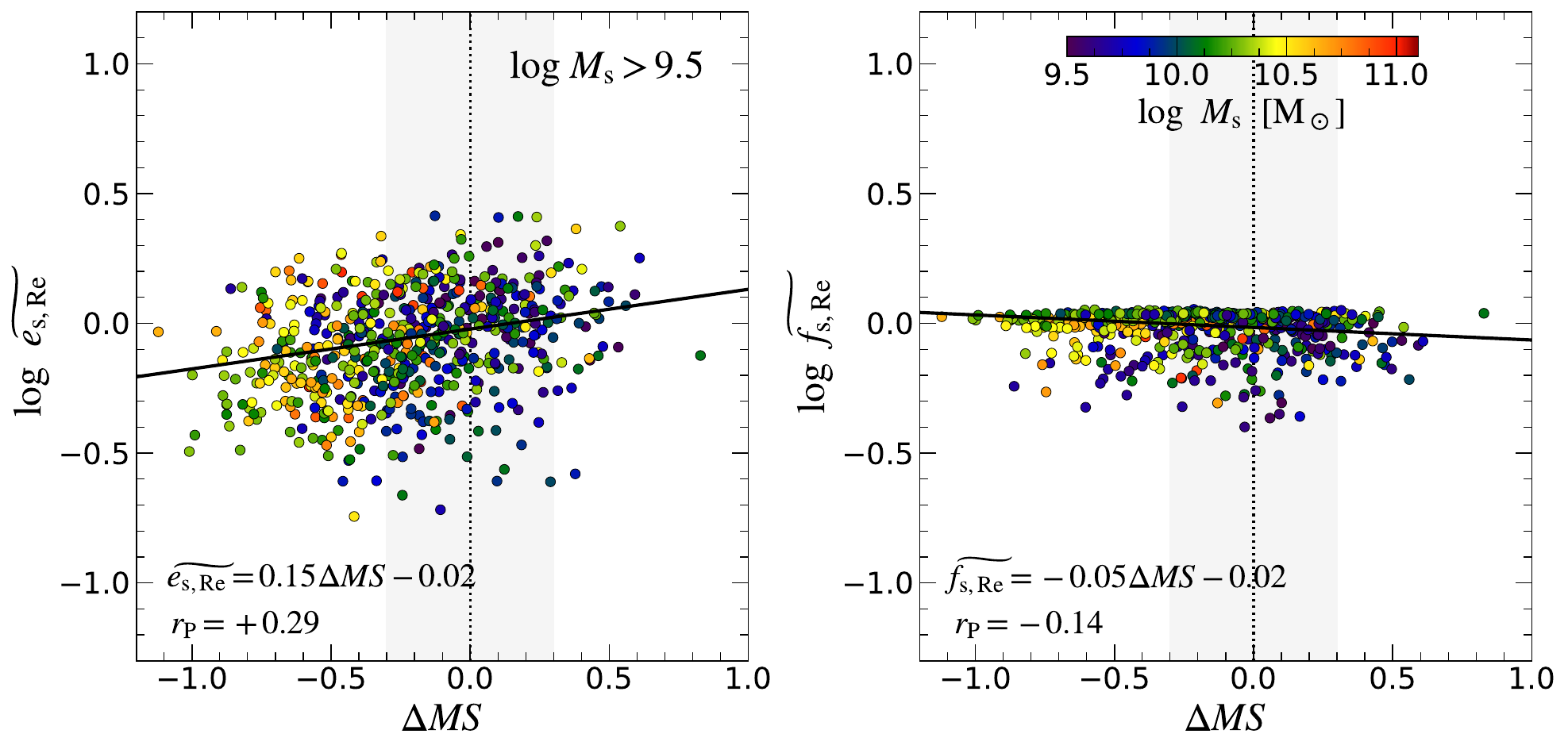}
\caption{
{Gradient across the MS for the shape parameters: elongation $e$, and flattening $f$ within $\Re$. Systematic redshift and mass dependence are removed. The stellar mass is indicated by colour.}
}
\label{fig_app:shape_dMS_app}
\end{figure*}

\begin{figure*} 
\includegraphics[width=0.95\textwidth]{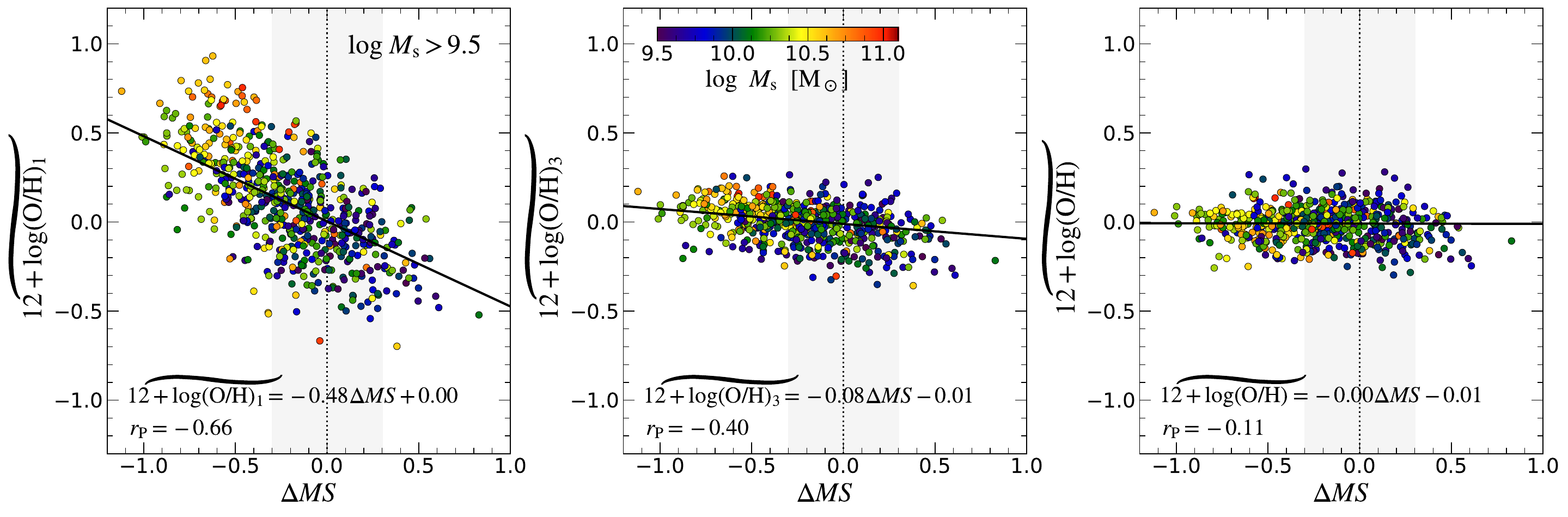}
\caption{
{Gradient across the MS for the gas metallicity. From left to right, measured within $1\kpc$, $3\kpc$ and $0.1\Rv$. Systematic redshift and mass dependence are removed. The stellar mass is indicated by colour.}
}
\label{fig_app:Z_dMS_app}
\end{figure*}

\begin{figure*} 
\includegraphics[width=0.95\textwidth]{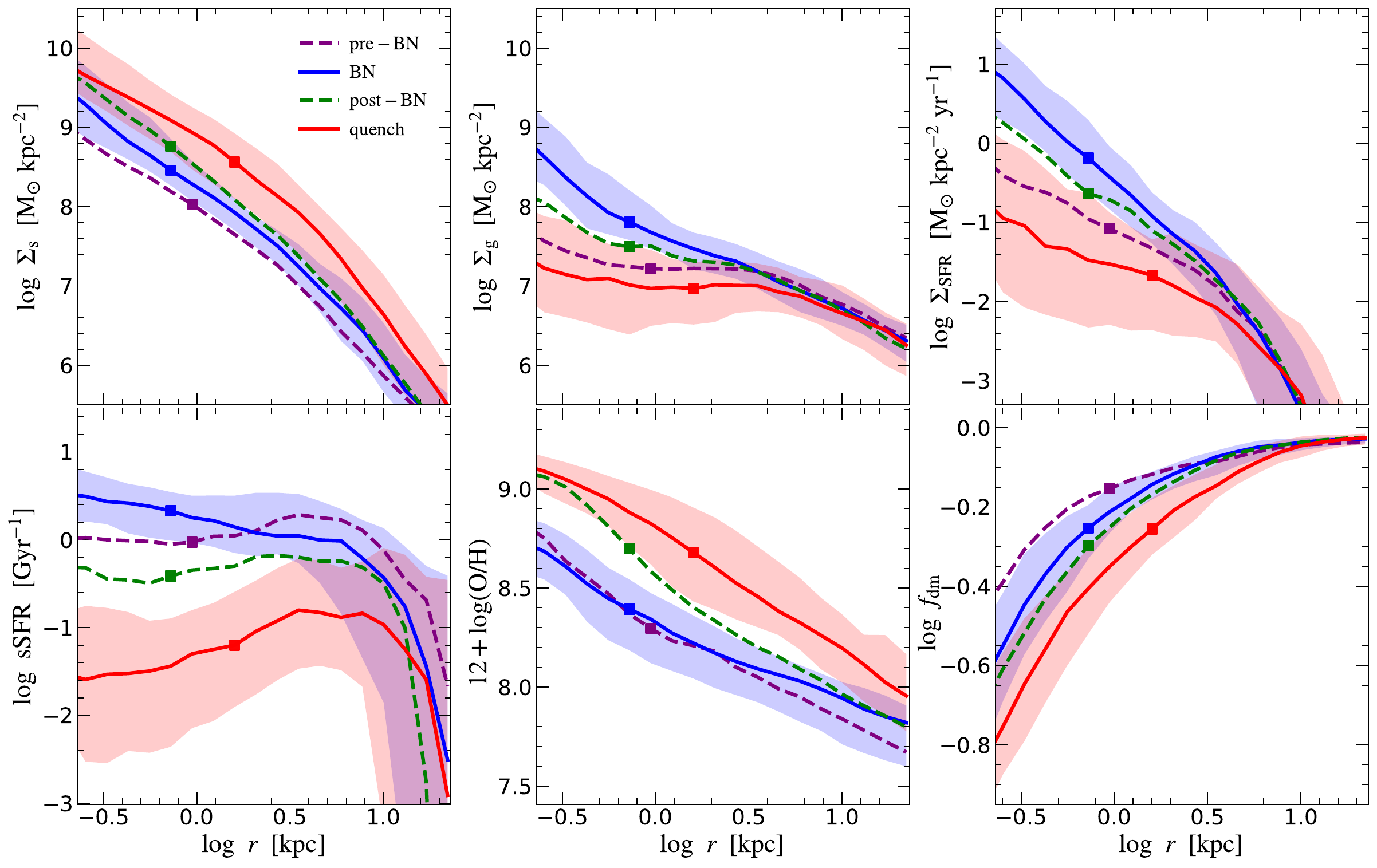}
\caption{
{Similar to \fig{profiles_re}. We show the raw medians of the profiles in each phase as a function of radius. Namely, without scaling the x and y-axis. The median of $\Re$ in each phase is marked with a square symbol.}
}
\label{fig_app:profiles_r}
\end{figure*} 